\pdfoutput=1
\RequirePackage{fix-cm}
\documentclass[smallextended]{svjour3}       
\smartqed  
\usepackage{graphicx}
\usepackage{txfonts}
\usepackage{longtable}
\usepackage{natbib}

\def\matisse{{\small MATISSE}}

\def\vlti{{\small VLTI}}

\def\HTO{\mbox{H$_{2}$O}}

\def\CO{$^{12}$CO}

\def\HII{H{\sc ii} }

\def\HC{HC~H{\sc ii}}
\def\UC{UC~H{\sc ii}}

\def\kms{\mbox{km~s$^{-1}$}}

\def\msolyr{\mbox{$M_\odot$\,yr$^{-1}$}}

\def\mic{\mbox{$\mu$m}}

\def\teff{\mbox{$T_{\rm eff}$}}

\def\g35{G35.03}
\def\farcs{\hbox{$.\!\!^{\prime\prime}$}}
\def\degr{\hbox{$^\circ$}}
\def\macc{$\dot M_{\rm acc}$}
\def\brg{Br$\gamma$}
\def\msol{\mbox{$M_\odot$}}
\def\goa{\mathrel{\mathchoice {\vcenter{\offinterlineskip\halign{\hfil
$\displaystyle##$\hfil\cr>\cr\approx\cr}}}
{\vcenter{\offinterlineskip\halign{\hfil$\textstyle##$\hfil\cr
>\cr\approx\cr}}}
{\vcenter{\offinterlineskip\halign{\hfil$\scriptstyle##$\hfil\cr
>\cr\approx\cr}}}
{\vcenter{\offinterlineskip\halign{\hfil$\scriptscriptstyle##$\hfil\cr
>\cr\approx\cr}}}}}

\begin{document}

\title{ACCRETION DISKS IN LUMINOUS YOUNG STELLAR OBJECTS
}

\titlerunning{Disks in young high-mass stars}        


\author{M.T. Beltr\'an \and 
  W.J. de Wit
}


\institute{   M.T. Beltr\'an \at
              INAF-Osservatorio Astrofisico di Arcetri\\
              Largo E.\ Fermi 5\\
              I-50125 Firenze\\
              Italy\\
              \email{mbeltran@arcetri.astro.it}\\
	      Senior Scientific Visitor at ESO Chile  \\
	              \and
           W.J. de Wit \at
           European Southern Observatory\\
           Alonso de C\'ordova 3107\\
           Vitacura, Casilla 19001\\
           Santiago de Chile\\
           Chile\\
           \email{wdewit@eso.org}
}

\date{Received: date / Accepted: date}

\maketitle

\begin{abstract}
An observational review is provided of the properties of accretion
disks around young stars. It concerns the primordial disks of
intermediate- and high-mass young stellar objects in embedded and
optically revealed phases. The properties were derived from spatially
resolved observations and therefore predominantly obtained with
interferometric means, either in the radio/(sub)millimeter or in the
optical/infrared wavelength regions.
We make summaries and comparisons of the physical properties,
kinematics, and dynamics of these circumstellar structures and
delineate trends where possible. Amongst others, we report on a
quadratic trend
of mass accretion rates with mass from T\,Tauri stars to the highest
mass young stellar objects and on the systematic difference in mass
infall and accretion rates.

\keywords{Accretion, accretion disks \and Techniques: high angular
  resolution \and Techniques: interferometric \and Stars: formation}
\end{abstract}

\section{Introduction}
\label{Sect:intro}

Circumstellar disks are an essential ingredient of the star and planet
formation process. They are found around young stars associated with
molecular clouds at a high incidence rate \citep{Hernandez2007}.  A
disk forms naturally from centrally infalling material under the
influence of gravity and the redistribution of specific angular
momentum by torques \citep{Turner2014}. The disk provides for
continued accretion of material onto the growing star and, as such, is
a key element of the accretion dynamics. In addition, the prevalent
physical conditions within disks are conducive to the formation and
growth of planetary bodies \citep{Williams2011}. The disks around
nearby solar-type stars have been studied to great extent and detail,
but the number of disk studies of more distant, massive stars (of
equivalent spectral type A and earlier) is comparatively small. Our
knowledge of the physical properties of massive star disks is
commensurately more uncertain. The aim of this review is to summarize
and put in perspective our current knowledge of accretion disks around
intermediate-mass (IM) and high-mass (HM) young stars. Our approach
makes use of spatially resolved observations whenever possible in
order to reliably identify and isolate circumstellar structures and
evaluate their role in the accretion process. Such data are often
obtained with interferometers. Resolved observations facilitate direct
comparisons with the more nearby low-mass stars for which disks are
readily spatially resolved. A review like the one presented here is
justifiable because in recent years interferometers in the
radio/(sub)millimeter (radio/mm) wavelength range, as well as those
that operate in the optical/infrared (optical/IR) wavelength range
have undergone substantial technical developments, which have resulted
in significant improvements of the data quality and observational
efficiency.
 
In the current era of the Atacama Large Millimeter Array (ALMA), 
(sub)milli\-me\-ter interferometry probes angular scales of
$\sim$10~milli-arcsecond at unprecedented sensitivity.  Although such angular 
scales were  already accessible with Very Long Baseline Interferometry
(VLBI) techniques for non-thermal phenomena (e.g. maser emission),
ALMA opens these scales for observing thermal processes. It unlocks thereby
a plenitude of diagnostics for the physical conditions in all kinds of astronomical
objects. Thermal processes are also accessed by optical/IR
interferometric arrays, like for instance the Very Large Telescope
Interferometer (VLTI, M\'erand et al. 2014\nocite{Merand2014}) and the Center
for High Angular Resolution Astronomy array (CHARA, ten
  Brummelaar 2005\nocite{tenBrummelaar2005}). Their continual 
improvement and the development of new beam combining
instrumentation have resulted in a strong increase of astronomical discoveries
and advancements at $\sim$1--10~milli-arcsecond scale. Indeed, some of the
current optical/IR facilities are now producing synthesized images in a similar way to
those produced by radio/mm interferometers \citep{Berger2012}. A
review of the astronomical progress from high angular resolution
techniques across wavelength regimes condenses an already large
literature volume on star and planet formation and is therefore a
necessary and timely exercise.

In the field of Galactic star and planet formation, the angular
resolution and sensitivity achieved by radio/mm and optical/IR
interferometry 
(OI)\footnote{From here on, we will use assume OI to stand for the technique that
  performs real-time, square-law interferometry with separate
  telescopes and 
  which covers both the optical and IR wavelength regimes. We will use the
  term radio/mm interferometry to indicate the heterodyne technique
  employing coherent amplification and product correlators (usually in
  the sub-mm, mm and cm wavelength regimes).}
allow us for the first time to investigate the phenomena that shape
the circumstellar environment less than 100~au from an
accreting star; a scale at which our understanding of the
star and planet formation process is least certain \citep{Armitage2011}. Here
important accretion disk processes are expected to take place and the interaction
between the disk and the star to become manifest, especially for
  stars in the high-mass regime \citep{Zinnecker2007}. Increased spatial
resolution is crucial for star formation in the
high-mass regime, where, due to the typical distances of the sources 
($>$1--2 kpc), the angular resolution achieved to date has proven to be insufficient
to probe the close by circumstellar structure of the  
(proto)star\footnote{We 
  will use the term {\it(proto)star} when referring to high-mass young 
  stellar objects to highlight the fact that most of them have
  already reached the zero age main sequence but are still actively
  accreting.}. 
In fact, and as we will summarize, the progress made in recent years by radio/mm and optical/IR
interferometers has made important breakthroughs in the field of disks
around IM and HM young stellar objects (YSOs). These have permitted to 
qualify and sometimes quantify disk properties and feed ideas
regarding the formation of high-mass stars.

This observational review focuses on the properties of circumstellar
disks of young high-mass stars, their evolution with time and their
dependency on star mass as derived from spatially resolved
observations. As most of the presented data is taken with
interferometers, we provide a brief description of the basic
principles behind astronomical interferometry in Sect.\,2. For
background purposes, we present a general vision of disks around young
stars as a function of stellar mass, from the low- to the high-mass
regime in Sect.\,3. We refrain from detailing the
fragmentation and core formation process in the earliest phases of
star and planet formation \citep{Bergin2007, McKee2007, Andre2014}, but
start the observational story of 
circumstellar 
disks from the Class\,0 stage until the pre-main sequence phase for
stars of early spectral type. Somewhat more technically, Sect.\,4 describes the
tracers and methods to derive the properties of disks with
interferometric techniques alongside published examples. The bulk of the review
in Sect.\,5 constitutes a detailed characterization of disks of
intermediate- and high-mass young stars in the early embedded phase
and during the pre-main sequence. This
collection of data forms the basis
to discuss the properties of the disk and the mass accretion rate as a
function of stellar mass and time, presented in Sect.\,6. Finally we
summarize our findings and  provide an outlook of future
instrumentation developments that will help to tackle the open issues
in this area. The desired outcome of this review is an
improved perspective of the evolution of young star circumstellar disks as
function of time and mass.

\section{Interferometry and synthesis imaging}
In this partly didactical section, we briefly clarify principles and
terminology related to astronomical interferometers and stellar
interferometry. The section is motivated to provide the general reader
with a better understanding of the reviewed interferometric data, as
interferometry is (still) less commonly used by 
astronomers than standard techniques like spectroscopy. 
Although the physical principle behind optical/IR and radio/mm
interferometry is the same, the respective implementations pose
different challenges with their own technical solutions as we will
briefly discuss next. We refer to Monnier \& Allen
(2012)\nocite{Monnier2012} for an extensive review on similarities and
differences between optical and radio interferometric detection
techniques.

\subsection{The physical basis}
Stellar interferometry works because individual (monochromatic)
photons from {\it a point source} produce identical fringe patterns
when the photon is interfered after propagating through two separate
apertures. Yet, the interference patterns produced by photons coming
from {\it an extended source} are shifted with respect to each
other. The overlap of shifted fringe patterns from an extended source
results in a lowering of the contrast (i.e.\  the visibility $V$)
between the light and dark bands. This manifestation is a direct
measure of the size of the luminous object, as first pointed out by
Fizeau (1868\nocite{Fizeau1868}). Therefore, any astronomical source seizes to
be point-like when the distance between the apertures is large
enough. The angular resolution $\theta$ of an instrument with a single
aperture is diffraction-limited by its diameter $D$,
$\theta\sim1.22\,\lambda/D$. On the other hand, the angular resolution
of an aperture array (interferometer) is equivalent to that reached by
an instrument the size of the entire array, $\theta\sim
1.22\,\lambda/b_{\rm max}$, where $b_{\rm max}$ is the maximum
baseline between the apertures in the array. Typical maximum baselines
range from a few 10~m to a few km for radio/mm interferometers, and
from a few 10~m to a few 100~m for optical/IR interferometers. Because
of that interferometers usually achieve a much higher angular
resolution than that obtained by a single aperture operating at the
same wavelength. In practice, astronomical interferometers are
designed to measure the amplitude and phase of the visibility function
$V(u,v)$. The observations aim to cover a range of baseline lengths and
angles, where each combination corresponds to a point in the $(u,v)$ plane.
The key notion of astronomical interferometry is that the
measured 2-D visibility function $V(u,v)$ corresponds to the Fourier transform
of the source's brightness distribution on the sky, $T(x,y)$. 

Astronomical interferometers are used across the light spectrum, from
optical to radio wavelengths making use of telescopes and antennas.
Aperture synthesis, which is the technique that produces astronomical images from
interferometric data, was first developed by radio
astronomers and has been successfully used since the 1950s (see e.g., Ryle \&
Hewish 1960\nocite{Ryle1960} and references therein). OI is technically quite
demanding and this has resulted in the fact that the technique has
been lifted from an experimental approach to a more common
astronomical tool only during the first decade of this century. 
It is only recently that 
aperture synthesis imaging has been achieved in OI with separate
apertures (Baldwin 1996\nocite{Baldwin1996}). 

\begin{figure*}
\centerline{\includegraphics[angle=-90,width=13cm]{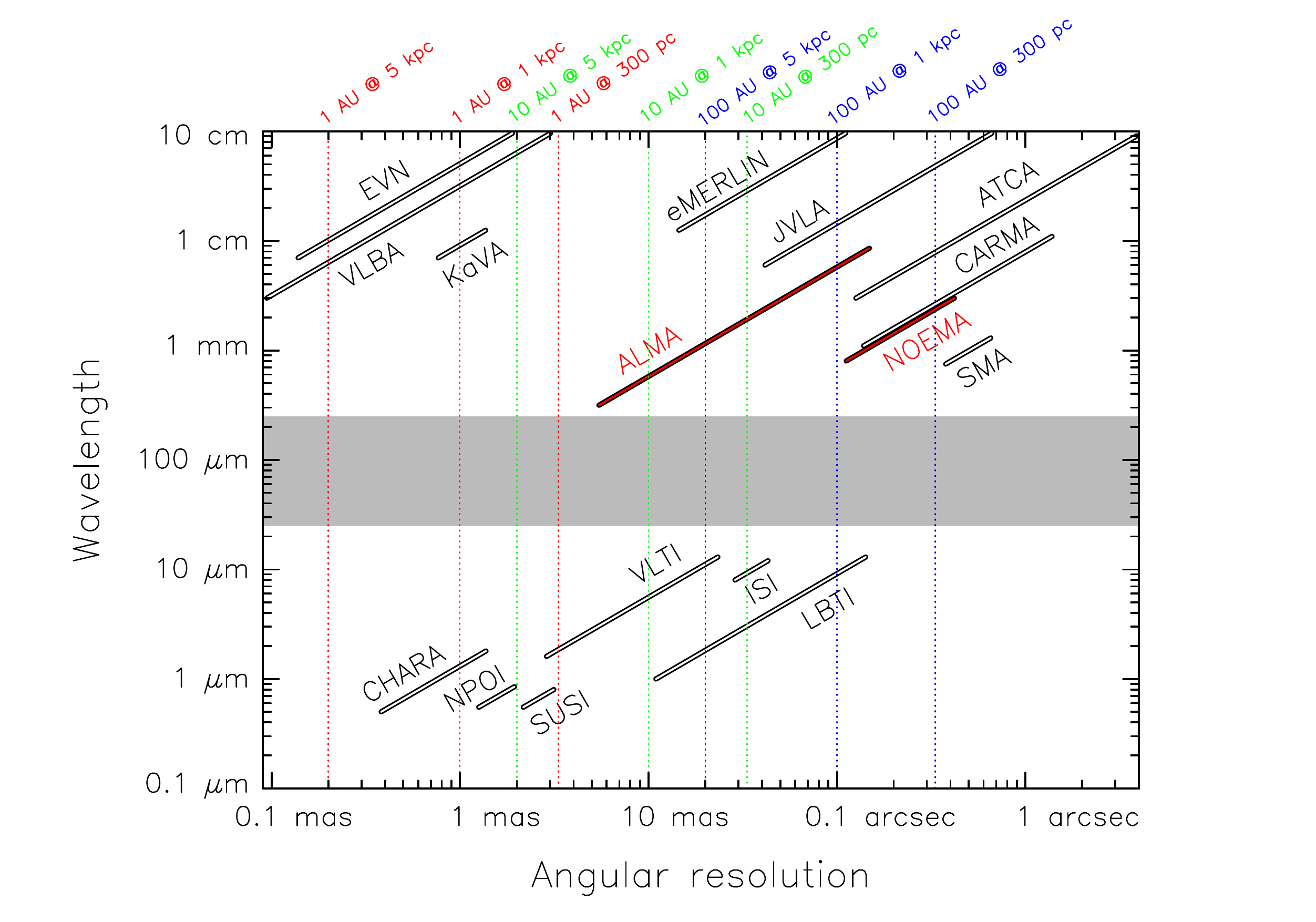}}
\caption{Wavelength coverage and maximum angular resolution provided by radio/mm and optical/IR 
interferometers. For ALMA and NOEMA, the wavelength coverage and angular resolution correspond
to those provided by the completed interferometers. The shaded area indicates
the range of wavelengths for which the atmosphere transmission is close to zero.
Dotted vertical lines indicate the angular
resolution needed to trace spatial scales of 1~au ({\it red lines}), 10~au ({\it green lines}), and 100~au 
({\it blue lines}), for sources located at 300~pc, 1~kpc, and 5~kpc.}
\label{fig-resolution}
\end{figure*}

\subsection{Comparing optical/IR with radio/mm interferometry} In astronomical
interferometry the angular resolution of an array is determined by the largest
baseline length. The image quality and the array sensitivity are strongly
dependent on the total number of independent baselines and thus on the number of
individual apertures (antennas or telescopes).  The number of baselines for an
array with $N$ apertures amounts to $N(N-1)/2$.  The extensive experience with
astronomical interferometers in the radio/mm range has resulted in the
construction and successful operation of large (sub)millimeter arrays over the
recent years. These arrays  are exquisitely suitable for aperture synthesis: (1)
the Plateau de Bure Interferometer (PdBI) has operated with 6 antennas and is
currently being upgraded to include 12 antennas and renamed to  NOEMA; (2) the
Submillimeter Array (SMA) located on Mauna Kea has 8 dishes; and (3) the
Combined Array for Research in Millimeter-wave Astronomy (CARMA) employs 23
antennas. The two most powerful (sub)\-millimeter arrays are the Jansky Very
Large Array (JVLA) with 27 antennas, which operates at centimeter and 7~mm
wavelengths and offers 351 baselines, and of course ALMA with 50 antennas that
deliver a total of 1225 baselines for a single, instantaneous observation. In
contrast, the current optical/IR arrays have only a small number of apertures
resulting in a much less efficient filling of the $(u,v)$ plane. The optical
array with the largest number of telescopes is currently the CHARA
interferometer with 6 elements. This array also offers the longest baselines,
delivering a spatial resolution below 1 milli-arcsecond (see
Fig.~\ref{fig-resolution}). The most productive optical/IR interferometer to
date is the VLTI which operates at NIR and MIR wavelengths. It is the only,
so-called, large aperture optical interferometer currently in use. The VLTI can
combine the light coming from either the four 8.2-m Unit Telescopes (UTs) or the
four 1.8-m Auxiliary Telescopes (ATs) rendering the array as the most
sensitive.  Optical interferometers make use of both Earth rotation and movable
telescopes to increase the $(u,v)$ coverage. Radio/mm interferometers rely on
the high number of apertures to produce high fidelity astronomical images.  On
the other hand, in OI, physical interpretation of the high angular resolution data is
mostly performed in Fourier space by fitting the visibility function to models
in OI.

Some of the differences in the technical implementation of astronomical
interferometry between radio/mm and optical/IR are caused by the differences in
atmospheric properties between these two wavelength regimes. The effects of a
turbulent atmosphere are much stronger at shorter optical wavelengths than at
the long wavelength, radio end. In particular, the coherence time at optical/IR
wavelengths is on the order of milliseconds and the Fried parameter $r_{0}$ (or
coherence length) is tens of cm. This means that in practice wavefront
restoration (e.g., by adaptive optics or monomode fibres) and stabilization of
fringe positions on the detector (by a fringe-tracker) are required for
exposures longer than milliseconds in optical/IR. The wavefront distortions are
not important in the radio/mm range where the coherence time is on the order of
minutes. The spatial scale of the atmospheric turbulence is also larger than the
antenna size, and therefore it is possible to make use of phase referencing by
switching between target and a nearby source with known fringe phase (i.e. the
phase calibrator). In the optical domain this is not possible as the phase is
subject to the atmospheric distortions. Yet, partial information on the fringe
phase is retrieved by means of the so-called closure phase. This quantity can be
extracted from the data when using three or more apertures. Otherwise, dual-beam
interferometry and an accurate metrology like for narrow-angle astrometry can be
employed \citep{Malbet2000}. Nonetheless, studies of the phase difference (or
differential phase) between continuum and atomic line transitions in
astronomical sources have proven to be powerful techniques in OI, allowing to
measure photocentric displacement and asymmetries on angular  scales of tens of
micro-arcseconds \citep[e.g.][]{Wheelwright2012b}.

In radio/mm interferometry, the actual detection of
the photon occurs at the antenna by means of the heterodyne
technique. In this technique, the in-coming signal is coupled to
an extremely stable reference signal and amplified. One records the
resulting amplitude and phase of the mixed wave. The sinusoidal signals of
each antenna are then digitized and cross-correlated by electronics in
the correlator. The resulting interference 
signal, which consists of the visibility amplitude and phase,
depends only on the geometry of the antenna pair in relation to the
source.
The nature of optical light is such that the signal cannot be stored
and amplified with conservation of the photon's phase (e.g., Oliver
1965\nocite{Oliver1965}). As a result, a heterodyne technique and a
post-detection, electronic cross-correlation process as implemented in
the radio/mm is simply not possible; the phase of the optical photon
is randomnized in the amplification process\footnote{The boundary
  wavelength between the techniques is $10\,\mu$m.}. In OI, the photon
that propagates through two (or more) telescopes needs to be recombined
in the same place (the detector) at the same time.  To achieve this,
one requires: (1) a light-path delay system in order to cancel out the
difference in path length between telescopes; and (2) a dedicated
instrument that performs 
the actual light-beam combination and final detection of the fringes.
The beam-combiner splits the incoming light from each telescope as each `photon'
can only be used once (unlike the radio case). For example, in a 
three telescope beam
combiner the light from each telescope is split in two to allow combination with the
other telescopes.

Figure~\ref{fig-resolution} shows the maximum angular resolution provided by the
different interferometers operating at the different wavelength domains. As seen
in this figure, the maximum angular resolution obtained at radio/mm and
optical/IR wavelengths is quite similar if one takes into account VLBI
techniques. In such a case, the angular resolution in the radio/mm regime can be
as high as 1 micro-arsecond.  Otherwise, the maximum angular resolution is 5
milli-arcsecond and is achieved by ALMA at the shortest wavelength (0.35\,mm;
Band 10) and largest configuration (15 km). This angular resolution is similar
to the angular resolution offered by, e.g. the VLTI. This complementarity is of
interest to high-mass star formation studies in, for example, uniting hot
accretion physics with the cold disk physics aiming for a complete picture of a
massive star accretion disk.

\section{Disks along the stellar mass sequence} 

This section provides the reader with background to the study of young
stars and their disks. It follows the established division of (young)
stars in three separate stellar mass intervals to which we
adhere. These are the low-, the intermediate- and the high-mass
  intervals. The dividing mass values that separate  
the three groups are 2 and 8\,M$_{\odot}$. The upper mass threshold
that identifies the high-mass stars follows
from (1) the
limiting mass for which a pre-main sequence (PMS) phase can be 
observationally identified and (2) the mass at which the effective
surface temperature produces ultraviolet photons able to ionize
significantly the young star's
environment. The lower mass threshold is somewhat arbitrary. It has some relation to
convective and radiatively stable stellar interiors during the PMS
contraction phase. The character of the stellar interior affects the topology of the
surface magnetic field which is an important ingredient in the accretion process. We will
review the properties of disks of intermediate-mass and high-mass
(proto)stars in detail in Section 5, while here we describe in broad
lines the characteristics of the young stars and their disks according
to mass.

\subsection{Disks in low-mass YSOs}	
	
\begin{figure*}
\centerline{\includegraphics[angle=0,width=8cm]{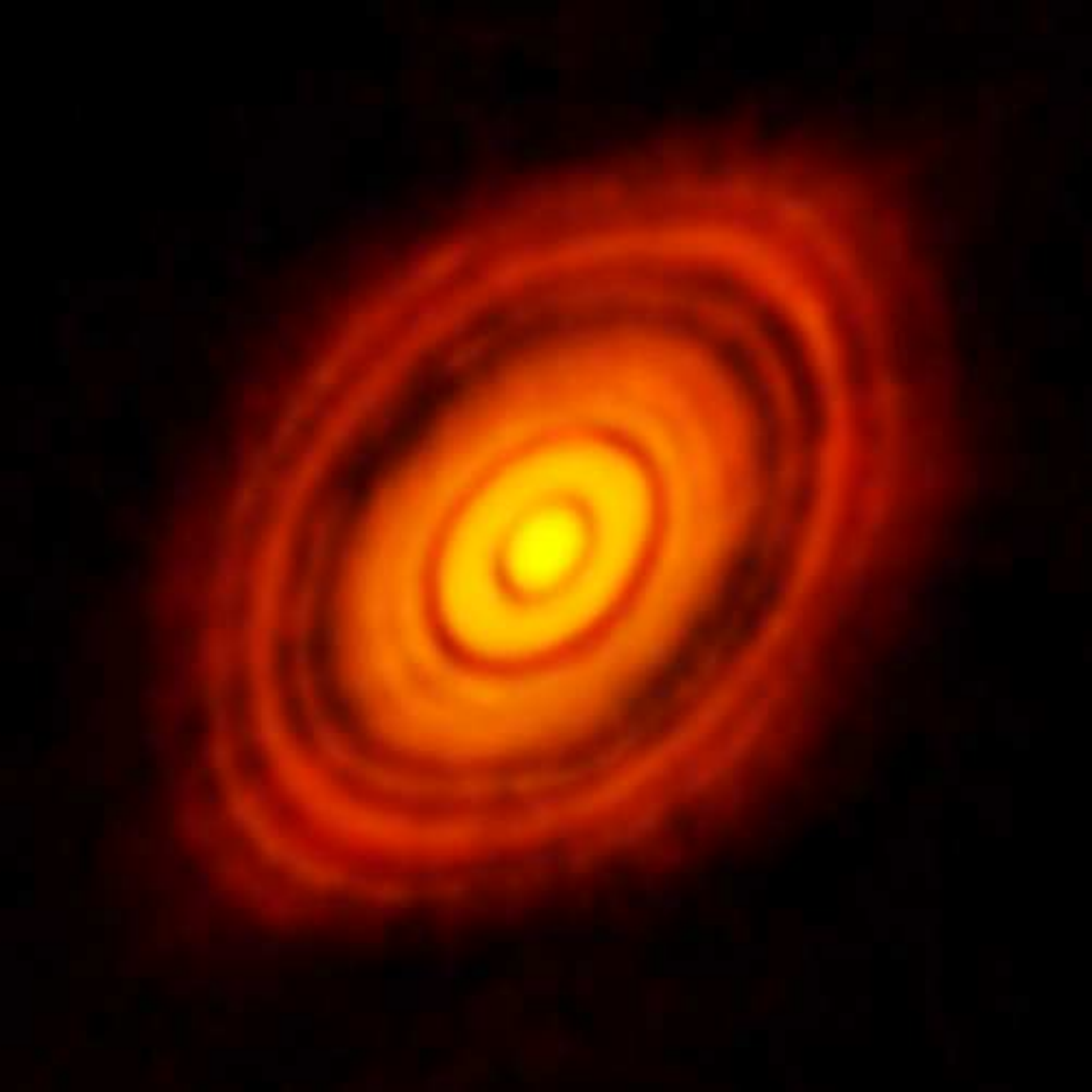}}
\caption{ALMA 1.3~mm image of the protoplanetary disk surrounding the young
low-mass star HL Tau. The gaps in the disk might have been created by forming
planets in the disk. Credit:
ALMA (ESO/NAOJ/NRAO).}
\label{fig-hltau}
\end{figure*}

Circumstellar disks of young stars were imaged for the first time by
the {\it Hubble Space Telescope}. Its superior spatial resolution
produced a detailed view of elongated structures and dark, dusty lanes
associated with the low-mass stars in M\,42 (O'Dell \& Wen,
1993). These high angular resolution observations in the optical
confirmed however the conclusions of a series of low resolution,
millimeter ( e.g., Beckwith et al.,
1990\nocite{Beckwith1990}) and infrared  (e.g., Cohen et al., 
1989\nocite{Cohen1989}) observations performed in the
80s. The strong IR excess discovered by the {\it InfraRed
    Astronomical Satelite (IRAS)} in non-mass losing stars like
Vega and $\beta$\,Pictoris, reinforced the notion of detectable proto-planetary systems
made from the prenatal cloud \citep{Aumann1984}. These studies indicated that
the shape of the dusty structures surrounding young stars is
necessarily  non-spherical. A relatively high mass measured for the dust structures combined with a low visual extinction seen towards these stars can only mean that the dust is
confined to a limited solid angle from the vantage point of the
central star \citep{Beckwith1990}. The angular resolution and
sensitivity of times past contrast starkly with the enormous wealth of information at hand in the
recent ALMA observations of HL\,Tau using the longest array baselines
(15~km). The ALMA Long Baseline Campaign confirms
  unequivocally the conclusions drawn in the 80s that 
  protoplanetary disks must be present. Courtesy of ALMA, the most 
detailed millimeter image of a protoplanetary disk around a young
solar-type star is depicted in Fig.~\ref{fig-hltau}. The disk is
characterized by an impressive, yet anticipated, series of rings and
gaps that could indicate the presence of forming planets (Partnership
ALMA~2015\nocite{Partnership2015}).

The evolution of the disk, its connection to the larger molecular
envelope and to the inner central star is cornerstone to our
understanding of star and planet formation. From a functional point of view, the
disk provides a way for potential energy to be converted into
radiation and kinetic energy and thereby powering dynamical phenomena
such as collimated jets and winds. The protoplanetary disk is the
vehicle with which matter is brought to the stellar surface as close
as possible and finally deposited on to it. Accretion of this
material is envisaged to be magnetically controlled. A structured
magnetosphere links the disk material to the star surface by accretion
funnel flows anchored in the disk at the radius where
the stellar magnetic field pressure is overcome by the ram pressure of
the inflowing material. This happens approximately at the co-rotation
radius. The material is lifted up from the disk and travels at nearly
free-fall velocity to the star, where it creates a shocked hot spot
(Hartmann, Hewett \& Calvet 1994\nocite{Hartmann1994}). This magnetic view of
accretion in protoplanetary disks finds support by many observations
as is overviewed in e.g., \citet{Bouvier2007}. 

Basic properties of the low-mass star disks such as size, mass and
velocity field have been determined from high-angular resolution
observations in the optical and the sub-millimeter
\citep[e.g.,][]{Beckwith1990,Burrows1996,Padgett1999}. In such
studies, the dust continuum and/or gas emission of the disk reveal
typical radii of 100--300~au. However, low-mass star disks have
been detected with radii as large as 700--800~au
\citep[e.g.,][]{Pietu2007,Andrews2007}. The disk masses traced by dust
continuum emission at millimeter and sub-millimeter wavelengths range
from 10$^{-3}$--10$^{-1}$~$M_\odot$ \citep{Williams2011}.
Interferometric molecular line emission observations of low-mass YSOs
have revealed that the velocity field of the circumstellar disks is
consistent with Keplerian rotation
\citep[e.g.,][]{Simon2000,Pietu2007}, even for the youngest Class~0
objects, as recently confirmed by CARMA and ALMA observations (L1527
IRS: Tobin et al., 2012\nocite{Tobin2012}; Ohashi et al.,
2014\nocite{Ohashi2014}; VLA 1623: Murillo et al.,
2013\nocite{Murillo2013}; HH~212: Lee et al., 
2014\nocite{Lee2014}; Codella et al., 2014\nocite{Codella2014}).

\subsection{Disks in intermediate-mass YSOs}
\label{disks-IM}

Arguably the best view of the formation and evolution of accretion disks in star
formation we owe to IM stars. This mass range covers stars with Zero Age Main
Sequence  (ZAMS) masses of 2 to 8~$M_\odot$, corresponding to bolometric
luminosities  of $L_{\rm bol}$$\sim$50--2000~$L_\odot$. The A and B-type stars
comprising the IM range carry a special importance as they form the transitional
group for various stellar astrophysical phenomena. For example, the IM range
covers the transition in star formation between isolated and clustered
formation, in stellar interiors between convective and  radiative interiors and
in stellar angular momentum evolution between slow and fast rotators. For disk
studies, the IM stars offer observational advantages as the stars are bright and
relatively close-by rendering the disks bright at most wavelengths while
subtending a large solid-angle on the sky. Importantly and similar to what has
been demonstrated for the solar-type stars, the existence of circumstellar disks
in the embedded protostellar phase has been established unambiguously by
observations  \citep{Zapata2007,Sanchez-Monge2010,VanKempen2012,Takahashi2012}.
The properties of these IM star disks are similar to those of their lower-mass
counterparts  \citep{Beltran2015}. However, whether they are in Keplerian
rotation is still a matter of debate  \citep{Sanchez-Monge2010}.

The disks persist well into the PMS phase
\citep[see e.g.,][]{Mannings1994,Mannings1997}, although it is believed
that the mass delivered to
the star in this phase is only a small fraction of the final star
mass. Yet the 
process of depositing material onto the stellar surface and the related phenomena like jet and disk-wind
generation, the structure of the disk and the question of how the disk disperses
can all be investigated in detail along the PMS evolution of IM
stars. Moreover, as the central stars can be age-dated by means of
evolutionary tracks, the time evolution of the disk can be determined. 

The best examples of IM PMS stars are the Herbig Ae/Be (HAeBe) stars
\citep{Herbig1960,Strom1972z,Hillenbrand1992}. They are relatively easily identifiable
by optical spectra, their spectral energy distributions (SEDs) with
near-IR excess by dust emission, and their association with dark or
scattering nebulae on scales of tens of arcsecond.  The HAeBe stellar
group \citep[e.g.][]{The1994} has been instrumental in establishing
detailed properties of disks. The dominance of circumstellar dust
emission is the base for the group I/II subdivision among this class
\citep{Meeus2001}. The sub-division finds a physical basis in the
structure of the disk, where group\,I sources have mid-IR bumps
produced by a flared disk and group\,II sources lack this emission and
are suspected of having geometrically flat disks. This SED
based idea is found to be consistent with resolved observations using
mid-IR interferometry (Leinert et al. 2004\nocite{Leinert2004}, Di Folco et
al. 2009\nocite{DiFolco2009}). Such spatially resolved mid-IR observations
  also provide  
an increase in the
number of HAeBe disks with evidence for gaps, i.e., near discontinuous
jumps in the radial density distribution
\citep{Panic2014,Matter2014}. A correlation is 
emerging between the presence of gaps and the flared geometry of the
disk \citep{Maaskant2013}. This is furthermore supported by near-IR scattered
light images of the outer disk, where the group\,II disks are rather
featureless, while the group\,I sources show the hallmarks of
ongoing dispersal, like multiple spiral arms, disk gaps, pericentre
offsets, and asymmetric shadowing of the outer disk favouring dynamical
clearing by sub-stellar objects \citep[see the overview by][]{Grady2015}.

\subsection{Disks in high-mass YSOs}
\label{Sect:diskinhmyso}
		
The picture and the role of accretion disks in young stellar objects
become increasingly unclear for the higher mass (proto)stars. The {\it high-mass}
adjective in our context applies to objects with ZAMS masses of
$8\,M_{\odot}$ and upwards and corresponds to OB-type stars with a
limiting spectral type B\,3V. Among this class of stars, the very
existence of disks during the formation phase has been under
debate. According to established PMS theory and the related idea of 
the stellar birthline \citep{Stahler2000}, stars more massive than
8~$M_\odot$ reach the ZAMS still deeply embedded in their molecular
surroundings.  Stellar winds and radiation pressure from the newly
formed early-type star may halt the infall of material (in case of
spherical symmetry), thus preventing further growth of the
protostellar nucleus \citep{Kahn1974,Wolfire1987}. In the past years,
different theoretical scenarios have been proposed as possible
solutions for the formation of OB-type stars, from non-spherical
accretion \citep{Nakano1995,Jijina1996} to stellar mergers
\citep{Bonnell1998}. However the theoretical ideas appear to converge
to a disk-mediated accretion scenario nonetheless. In fact, models suggesting
massive star-formation initiated by a monolithic collapse of a turbulent molecular core
\citep{McKee2002,Krumholz2009}, those based on competitive
accretion for core material \citep{Bonnell2006}, and those that propose
Bondi-Hoyle accretion for the growth of the star \citep{Keto2007} all
predict the existence of a circumstellar accretion disk
\citep[see][for a review of high-mass star-formation
  mechanisms]{Tan2014}. This is hardly surprising as the disk is a
natural mechanism to redistribute angular momentum. 

\begin{figure*}
\centerline{\includegraphics[angle=0,width=10cm]{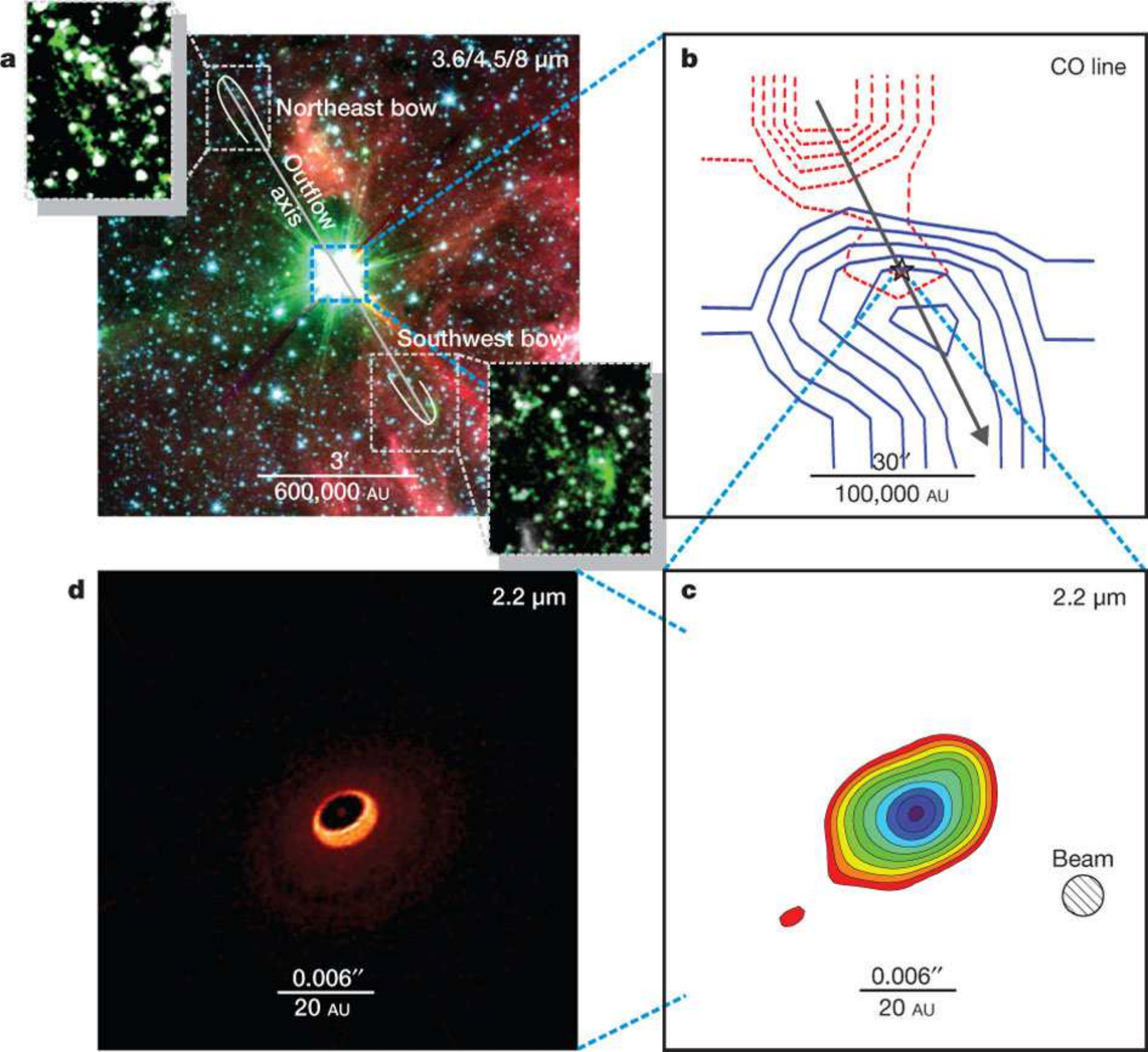}}
\caption{{\it a)} Three colour composite IRAC image (3.6~$\mu$m, blue, 4.5~$\mu$m,
red, 8.0~$\mu$m, green) of the high-mass star-forming region 
IRAS~13481$-$6124 taken with {\it Spitzer}. {\it b)} Blueshifted and redshifted 
CO~(3--2) emission of the molecular outflow associated with the high-mass YSO
observed with the single-dish antenna APEX. {\it c)}  VLTI/AMBER 2.2~$\mu$m aperture-synthesis
reconstructed image of the disk-like structure perpendicular to the outflow
direction. {\it d)} Best-fit radiative transfer model 2.2~$\mu$m image. From
\citet{Kraus2010}.}
\label{fig-kraus}
\end{figure*}

\begin{figure*}
\centerline{\includegraphics[angle=90,width=10cm]{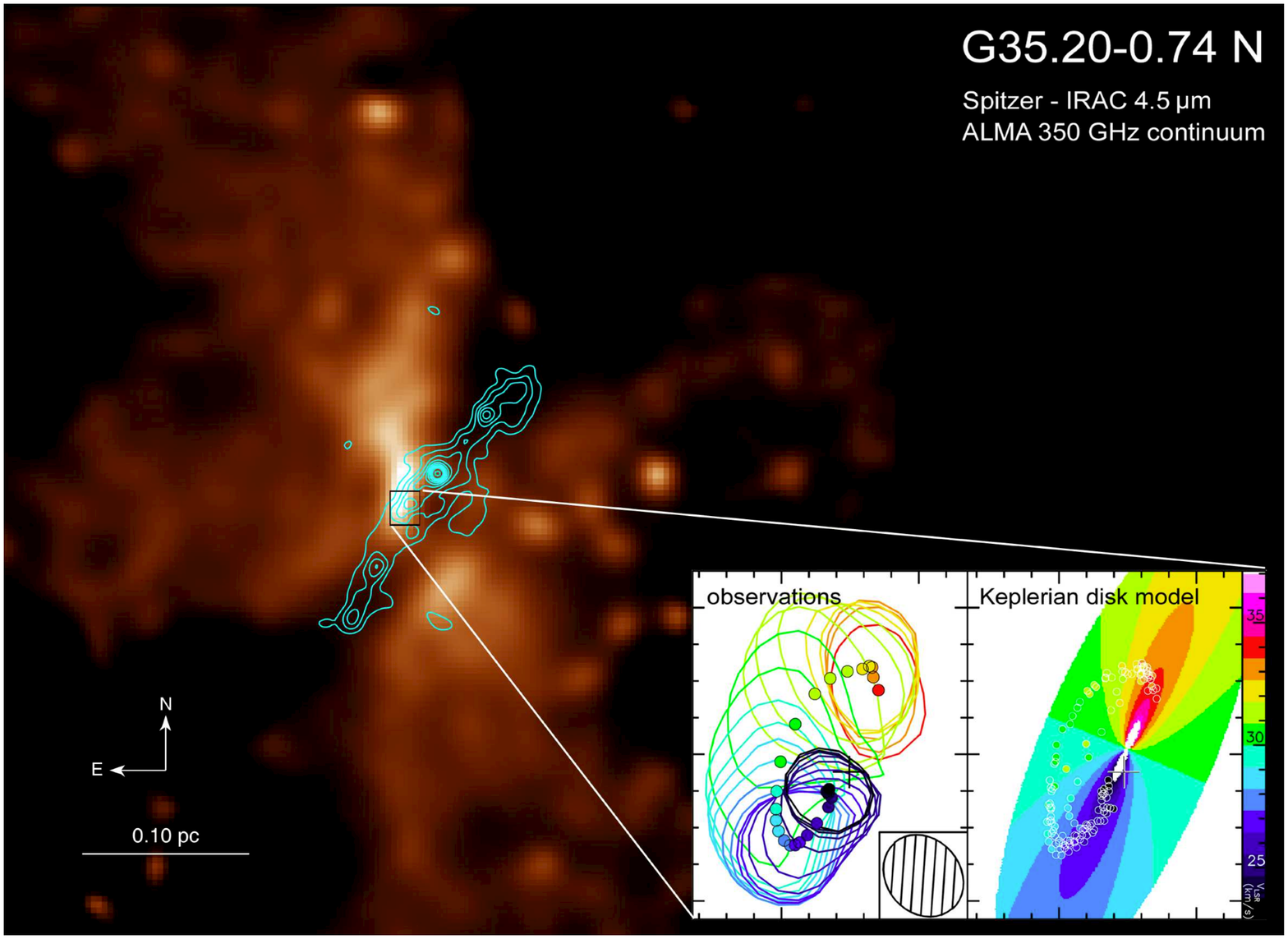}}
\caption{Large-scale {\it Spitzer} IRAC 4.5~$\mu$m image of the high-mass
star-forming region G35.20$-$0.74N, overlaid with the 850~$\mu$m continuum
emission ({\it cyan contours}) observed with ALMA. ({\it Left inset})
CH$_3$CN~(19--18) $K$=2 line emission peaks ({\it solid circles}) obtained with a
two-dimensional Gaussian fit channel by channel towards core B in G35.20$-$0.74N
observed with ALMA.
Open circles represent the 50\% contour level for each channel. ({\it Right inset})
Overlay of the velocity peaks of different high-density tracers ({\it solid circles}) and a velocity map of
the best-fit Keplerian disk model ({\it color map}). The velocity scale is on
the right. The comparison indicates a remarkable agreement between computed and 
observed velocities. From S\'anchez-Monge et al. (2013).}
\label{fig-g35}
\end{figure*}

From an observational point of view, a growing number of disk
candidates have been detected around high-mass YSOs in recent
years. It should however immediately be clarified that the accumulated
evidence for disks in young OB stars does not extend to stars beyond a
mass limit of 25--30~$M_{\odot}$ or $\sim$10$^5 L_\odot$. The disks of
these sources (of early-B and late O-type) have been spatially
resolved using line and continuum tracers from infrared to centimeter
wavelengths (e.g., IRAS 20126+4104: Cesaroni et al.,
2005\nocite{Cesaroni2005}, 2014\nocite{Cesaroni2014}; Cepheus A HW2:
Patel et al., 2005\nocite{Patel2005}; IRAS 13481$-$6214: Kraus et al.,
2010\nocite{Kraus2010}, see Fig.~\ref{fig-kraus}; CRL 2136: de Wit et
al., 2011\nocite{DeWit2011}). The kinematics of the high-density gas
towards B-type (proto)stars has been studied at high-angular
resolution ($\leq0\farcs5$) only for a handful of sources.  However,
for these few YSOs, molecular line observations reveal that the
velocity field, like for their low-mass counterparts, would be
consistent with Keplerian rotation (e.g., IRAS 20126+4104: Cesaroni et
al., 2005\nocite{Cesaroni2005}, 2014\nocite{Cesaroni2014}; AFGL 2591
VLA3: Wang et al., 2012a\nocite{WangK2012}; G35.20$-$0.74N B:
S\'anchez-Monge et al., 2013, 2014\nocite{Sanchez-Monge2013}, see Fig.~\ref{fig-g35};
G35.03+0.35: Beltr\'an et al., 2014\nocite{Beltran2014}). Moreover,
for those massive YSOs that are bright enough in K-band, the
kinematics has been investigated by means of the line profiles of CO
first overtone transitions in the NIR\citep{Bik2004,Blum2004}. The CO bandhead
profile is successfully reproduced by emission from material in
Keplerian rotation in the inner few astronomical units from the
central star \citep{Ilee2013}. That the CO molecules are in Keplerian
rotation around the star can also be concluded from the astrometric
drift of the photocenter as function of the velocity resolved bandhead
emission. The drift was measured on sub-milliarcsecond angular scales
\citep{Wheelwright2010} and provides strong evidence for the presence
of disks on linear scales of a few au near embedded early-B and late-O type 
stars.

For stars of the highest stellar mass ($>30~M_{\odot}$), the existence
of circumstellar disks has remained elusive up to now. This
observational result is probably unsettling for theory and simulations
that show that radiation pressure does not prevent disk accretion to
form stars up to 140~$M_\odot$
\citep{Krumholz2009,Kuiper2010}. However the problem maybe more
serious in that no genuine (proto)star is currently known that would
constitute an accreting hydrostatic object with a mass over this
limit. Nonetheless, huge ($\sim$0.1~pc), dense ($n_{\rm H_2} \gtrsim
10^7$~cm$^{-3}$), massive (a few 100~$M_\odot$), rotating cores have
been detected around early O-type 
(proto)stars in studies performed at spatial resolutions attainable
before the advent of ALMA. These objects are in all likelihood
non-equilibrium structures enshrouding young stellar clusters and not
merely individual massive stars (see Cesaroni
2007\nocite{Cesaroni2007}; Beltr\'an et al., 2011\nocite{Beltran2011a} and references
therein). These rotating structures are referred to as toroids, so as
to distinguish them clearly from accretion disks in Keplerian
rotation which they clearly cannot be
\citep[e.g.,][]{Beltran2005}. The open question is whether higher angular 
resolution observations with ALMA will demonstrate the presence of
circumstellar (or circumcluster) disks interior to the rotating toroids.

Our knowledge of the formation of the most massive stars is incomplete
and it is uncertain whether these stars conform to the disk-mediated
accretion scenario. Milli-arcsecond resolution studies will be able to
shed more light on the accretion dynamics of the high-mass
stars. In particular and under the assumption that accretion onto a
low-mass object be universal, a growing object should produce a
jet-driven outflow very soon in its evolution. Could the launching of
the ouflow be quenched initially by the massive envelope or the
prevalent strong gravitational forces? What would be the influence of
the irradiation of the disk by a high-mass star and how tenacious is
the disk in withstanding this additional eroding force? Because of the
accumulating evidence for disks in stars with  masses less than
$\sim$30~$M_\odot$ and the lack of evidence of disks for stars above 
this threshold, we will discuss these issues and related questions for
both groups separately in Sect.\,\ref{obs-prop}.

\section{Deriving properties of YSO disks} 
\label{deriving-prop}

Disks around IM and HM (proto)stars are deeply embedded in the innermost part of
the surrounding infalling envelopes. As a result, identifying candidate
accretion disks, especially around embedded HM (proto)stars, requires a careful
selection of targets and tracers to overcome the problems related to the nature
of massive star-forming regions. The most important of these are the large
distances, the high dust extinction factors, the high
  multiplicity that characterizes these stars, and the complexity of  the
environment typical of the clustered mode of high-mass star formation. The
onset of the clustered mode of star formation becomes  manifest among the
intermediate-mass A and B-types \citep{Testi1999} and only few O-type stars may
form in isolation \citep{DeWit2005}. The increasingly shorter time-scale of
formation compounded with the shape of the IMF results in low probabilities of
finding a Class\,0 massive (proto)star or massive starless cores.  Moreover, the
clustered formation mode makes it very difficult to trace back the physical
characteristics of the parent cloud and to derive the initial conditions for
massive star formation.  To overcome these problems, high-sensitivity and
high-angular resolution observations are needed to study  the earliest phases of
IM and HM (proto)stars. True accretion disks are probably spawned at the radius
where the centrifugal force of the infalling envelope is balanced by gravity
\citep{Terebey1984}. Resolved observations of Keplerian disks near IM young
stars with ALMA show outer radii of a few times 100\,au (e.g. Pineda et al. 
2014\nocite{Pineda2014}). Resolving such structures around HM (proto)stars
requires sub-arcsecond angular resolution already for the Orion molecular cloud,
the nearest high-mass star-forming (HMSF) region at $\sim$410\,pc. The situation
is slightly more favourable for IM protostars, but even for them, minimal
typical distances are $\gtrsim$300~pc. In Sect. \ref{disk-tracers} we discuss
the tracers used to identify disks around IM and HM (proto)stars from near-IR to
centimeter wavelengths.


\begin{table*}
 \begin{center}
\caption{Rotating disks in intermediate-mass embedded protostars}
\label{IMprop}
\begin{tabular}{lccccccccc}
\hline
&\multicolumn{1}{c}{$d$} &
\multicolumn{1}{c}{$L_{\rm bol}$} &
\multicolumn{1}{c}{$M_{\rm gas}^{\rm OH94\,\,a}$} &
\multicolumn{1}{c}{$R$} &
\multicolumn{1}{c}{$M_{\star}^b$} &
\multicolumn{1}{c}{$\dot M_{\rm out}^c$} &
\\
\multicolumn{1}{c}{Core} &
\multicolumn{1}{c}{(kpc)} &
\multicolumn{1}{c}{($L_\odot$)} &
\multicolumn{1}{c}{($M_\odot$)} &
\multicolumn{1}{c}{(au)}&
\multicolumn{1}{c}{($M_\odot$)} &
\multicolumn{1}{c}{($M_\odot$/yr)} &
\multicolumn{1}{c}{Refs.$^{d}$} & \\
\hline

Serpens FIRS 1  	  &0.31     &  46   & 0.1   &  65   &2.5     &---  	&1 \\
OMC-2/3--MM6 Main	  &0.414    &  60   & 0.32  & 500   &2.6     &2.4$\times10^{-5}$  &2 \\
L1641 S3 MMS1		  &0.465    &  70   & 7.6   & 400   &2.7     &---  	&3 \\
Cepheus E     	          &0.73     &  78   & 1.4   & 440   &2.8     &2.0$\times10^{-5}$	&4, 5 \\
OMC-2/3--MM7 SMM	  &0.414    &  99   & 1.3   & 370   &3.0     &---  	&6, 7 \\ 
IRAS 20050+2720 A	  &0.70     & 280   & 5.6   &1650   &4.0     &2.7$\times10^{-4}$     &8, 9 \\
IRAS 22198+6336 MM2	  &0.76     & 370   & 0.95  & 200   &4.2     &4.9$\times10^{-5}$     &10, 11 \\ 
IC1396N BIMA2		  &0.75     & 440   & 0.36  & 110   &4.4     &7.5$\times10^{-5}$     &12, 13 \\ 
NGC 2071 A (IRS 3)         &0.422    & 440   & 0.27  & 100   &4.4     &---  	&3, 14 \\
NGC 2071 B (IRS 1)        &0.422    & 440   & 0.23  & 100   &4.4     &---  	&3, 14 \\
H288			  &2.0      & 480   & 12.4  &3000   &4.4     &2.2$\times10^{-3}$     &15 \\
IRAS 00117+6412 MM1	  &1.8      & 500   & 2.0   &1700   &4.5     &4.5$\times10^{-5}$     &16 \\
NGC7129 FIRS 2		  &1.25     & 500   & 2.7   & 400   &4.5     &3.8$\times10^{-5}$     &10, 17, 18 \\
IRAS 00117+6412 MM2	  &1.8      & 600   & 1.6   &1800   &4.7     &---   	&16 \\
IRAS 22172+5549 MM2	  &2.4      & 800   & 2     & 750   &5.1     &7.7$\times10^{-4}$	&14, 19 \\
OMC-2 FIR4		  &0.42     & 920   & 5.0   &2000   &5.4     &---       &20, 21, 22  \\
CB3-1			  &2.5      & 930   & 0.66  & 600   &5.4     &---       &23, 1 \\
CB3-2			  &2.5      & 930   & 0.25  & 330   &5.4     &---       &23, 1 \\
S235 NE--SW core A	  &1.8      &1000   & 30    &6600   &5.4     &2.0$\times10^{-3}$     &24, 25 \\
IRAS 20293+3952 A	  &2.0      &1050   & 8.8   &1800   &5.5     &7.0$\times10^{-4}$     &26, 27 \\
L1206			  &0.91     &1200   & 12    &1600   &5.6     &5.5$\times10^{-5}$     &28 \\
IRAS 05345+3157 C1-a	  &1.8      &1400   & 2.0   &1850   &6.0     &8.7$\times10^{-4}$     &29 \\
IRAS 05345+3157 C1-b	  &1.8      &1400   & 0.35  &1650   &6.0     &---       &29 \\
IRAS 05345+3157 C2	  &1.8      &1400   & 1.6   &1340   &6.0     &---       &29 \\ 
IRAS 23385+6053 	  &4.9      &1500   &28.5   &2700   &6.1     &2.3$\times10^{-3}$     &30, 31 \\
IRAS 20343+4129 IRS 1	  &1.4      &1500   & 0.73  &2800   &6.1     &8.7$\times10^{-5}$     &32 \\
AFGL 5142 MM1		  &1.8      &2300   & 4.4   & 520   &7.2     &4.7$\times10^{-4}$     &14, 33 \\
AFGL 5142 MM2		  &1.8      &2300   & 3.2   & 320   &7.2     &---       &14 \\
OMC1-S 139$-$409 	  &0.460    &1$\times10^4$$^e$   & 0.14$^f$  & 120$^f$   &12      &---       &34 \\
OMC1-S 134$-$411	  &0.460    &1$\times10^4$$^e$   & 0.20$^f$  & 80$^f$   &12      &---       &34 \\
\hline				 
\end{tabular}
\end{center}
\vspace{1mm}
 \scriptsize{
{\it Notes:} \, 
 $^a$ Masses estimated assuming the dust opacities of \citet{Ossenkopf1994}; \\
 $^b$ Masses estimated from the simulated clusters (see \S~\ref{mass-star}); \\
 $^c$ Outflow parameters corrected for opacity and assuming a mean inclination angle of the outflow of 32.7\degr;  \\ 
 $^d$ References for the core parameters: 
1: \cite{Fuente2007};
2: \cite{Takahashi2012};
3: \cite{VanKempen2012};
4: \cite{Moro-Martin2001};
5: \cite{Froebrich2003};
6: \cite{Takahashi2008}; 
7: \cite{Takahashi2013}; 
8: \cite{Beltran2008}; 
9: \cite{Froebrich2005};
10: \cite{Sanchez-Monge2010}; 
11: \cite{Palau2011};
12: \cite{Neri2007}; 
13: \cite{Beltran2002};
14: \cite{Palau2013}; 
15: \cite{Gueth2001};
16: \cite{Palau2010};
17: \cite{Fuente2001}; 
18: \cite{Fuente2005};
19: \cite{Fontani2004a}; 
20: \cite{Kim2008};
21: \cite{Crimier2009}; 
22: \cite{Lopez-Sepulcre2013};
23: \cite{Launhardt1997};
24: \cite{Felli2004}; 
25: \cite{Saito2008};
26: \cite{Palau2007a}; 
27: \cite{Beuther2004};
28: \cite{Beltran2006b};
29: \cite{Fontani2009};
30: \cite{Fontani2004b}; 
31: \cite{Molinari1998};
32: \cite{Palau2007b};
33: \cite{Zhang2007};	 
34: \cite{Zapata2007};  \\	
$^e$ Bolometric luminosity of the whole OMC1-S region. \\
$^f$ $M_{\rm gas}$ and $R$ estimated with CH$_3$CN at 1.3~mm. At 7~mm, \citet{Zapata2007} have resolved the 1.3~mm
continuum emission towards both IM protostars OMC1S 139--409 and 134--411 into two
compact binary systems with radii $R$ of 20--25~au and masses $<$0.1~$M_\odot$.\\ 
      
} 
\end{table*}
 
\begin{table*}
 \begin{center}
\caption{Rotating disks and toroids in high-mass embedded (proto)stars}
\label{OBprop}
{\scriptsize
\begin{tabular}{lccccccccccc}
\hline
&\multicolumn{1}{c}{$d$} &
\multicolumn{1}{c}{$L_{\rm bol}$} &
\multicolumn{1}{c}{$M_{\rm gas}^{\rm OH94\,\,a}$} &
\multicolumn{1}{c}{$R$} &
\multicolumn{1}{c}{$V_{\rm rot}$} &
\multicolumn{1}{c}{$M_{\star\,{\rm Lyman}}^b$} &
\multicolumn{1}{c}{$M_{\star\,{\rm cluster}}^c$} &
\multicolumn{1}{c}{$\Delta V$} &
\multicolumn{1}{c}{$\dot M_{\rm out}^d$} &
\\
\multicolumn{1}{c}{Core} &
\multicolumn{1}{c}{(kpc)} &
\multicolumn{1}{c}{($L_\odot$)} &
\multicolumn{1}{c}{($M_\odot$)} &
\multicolumn{1}{c}{(au)}&
\multicolumn{1}{c}{(\kms)} &
\multicolumn{1}{c}{($M_\odot$)} &
\multicolumn{1}{c}{($M_\odot$)} &
\multicolumn{1}{c}{(\kms)} &
\multicolumn{1}{c}{($M_\odot$/yr)} &
\multicolumn{1}{c}{Refs.$^{e}$} & \\
\hline
IRDC 18223$-$1243	 & 3.7     &1$\times10^{2 \,f}$     & 47     &14000	&1.5  &---   &---   & 1.8   &5.5$\times10^{-3}$   &1 \\
NGC6334 I(N) SMA1b	 & 1.3     &1$\times10^3$     & 4.3    &  800	&3.5  &---   & 5.5  & 8.8   &  ---		  &2, 3  \\
AFGL 490		 & 1.0     &2$\times10^3$     & 4.1    & 1600	&1.3  &  8--10$^g$   &  7   & 3.0   &  ---		  &4, 5 \\
G192.16$-$3.82  	 & 2.0     &3$\times10^3$     & 11     & 2100	&3.0  &  8   &  8   & 1.5   &3.8$\times10^{-4}$   &6, 7, 8, 9 \\
IRAS 04579+4703 	 & 2.5     &4$\times10^3$     & 8      & 5000	&1.0  &  7   & 8.5  & 3.6   &1.7$\times10^{-4}$   &10, 11, 12 \\
GH2O 92.67+3.07 	 & 0.80    &4.7$\times10^3$   & 12$^h$     & 7200	&1.2  &  6$^i$   &  9   & 3.0   &2.7$\times10^{-4}$   &13 \\
G35.03+0.35 A		 & 3.2     &6.3$\times10^3$   & 0.75   & 2200	&2.0  & 11   & 10   & 8.5   &  ---		  &14 \\
IRAS~20126+4104 	 & 1.7     &1$\times10^4$     & 0.9    & 3600	&1.3  & 7--10$^g$   & 12   & 3.0   &1.3$\times10^{-3}$   &15, 16, 17 \\
G23.01$-$0.41		 & 4.6     &1$\times10^4$     & 41     & 6000	&0.6  & 18   & 12   & 8.3   &2.0$\times10^{-4}$   &18, 19, 20 \\
NGC7538S MM2		 & 2.7     &1.5$\times10^4$   & 5.0    & 1000	&1.0  &---   & 13   & 4.0   &  ---		  &21 \\
AFGL 2591 VLA3  	 & 1.0     &2$\times10^4$     & 0.41   &  400	&2.2  & 16   & 14   & 1.5   &  ---		  &22, 23 \\
IRAS 18162$-$2048 MM1	 & 1.7     &2$\times10^4$     & 4.9    &  800	&2.0  &---   & 14   & 5.5   &  ---		  &24 \\
IRAS 18151$-$1208	 & 3.0     &2$\times10^4$     & 43     & 5000	&2.0  & 15   & 14   & 1.9   &  ---		  &25, 26, 27 \\
G16.59$-$0.05            & 4.7     &2$\times10^4$     & 75     &  600	&7.2$^j$  &---	& 14   & 5.6   &3.8$\times10^{-3}$   &19, 28, 29 \\
S255IR SMA1              & 1.59    &2$\times10^4$     & 3.8    & 1850	&1.0  &---	& 14   & ---   &6.2$\times10^{-4}$   &30 \\
Cepheus A HW2		 & 0.725   &2.5$\times10^4$   & 2.2    &  360	&3.5  & 15   & 15   & 4.0   &1.7$\times10^{-3}$   &31, 32, 33, 34 \\
G35.20+0.74 N A 	 & 2.19    &3$\times10^4$     & 1.0    & 1500	&1.5  &---   & 16   & 4.5   &  ---		  &35 \\
G35.20+0.74 N B 	 & 2.19    &3$\times10^4$     & 0.9    & 2600	&1.0  & 18$^g$   & 16   & 2.8   &  ---		  &35 \\
IRAS 18089$-$1732	 & 3.6     &3.2$\times10^4$   & 68     & 3600	&3.0  &---   & 16.6 & 6.0$^k$ &  ---		  &36, 37, 38 \\
G240.31+0.07		 & 5.3     &3.2$\times10^4$   &133     &10000	&2.5  &---   & 16.6 & 1.7   &6.4$\times10^{-3}$   &39, 40, 41 \\
G24.78+0.08 C		 & 7.7     &3.3$\times10^4$   & 95$^l$     & 3500$^l$	&0.5  &---   & 17   & 3.5   &4.7$\times10^{-4}$   &42, 43, 44 \\
G24.78+0.08 A1  	 & 7.7     &3.3$\times10^4$   &130$^m$     & 4600$^m$	&1.5  & 18   & 17   & 7.0   &  ---		 &42, 43, 44 \\
G24.78+0.08 A2  	 & 7.7     &3.3$\times10^4$   & 87$^n$     & 3500$^n$	&0.75 &---   & 17   & 7.0   &2.3$\times10^{-3}$   &42, 43, 44 \\
IRAS 16547$-$4247	 & 2.9     &6.2$\times10^4$   & 22     & 1500	&1.7  &---   & 21   & 7.6   &  ---		  &18, 45, 46 \\
IRAS 16562$-$3959	 & 1.7     &7$\times10^4$     & 7.6    & 3000	&2.2  &---   & 22   & 5.0   &  ---		  &47 \\
NGC7538 IRS 1		 & 2.7     &8$\times10^4$     & 18     & 1000	&3.0  & 30   & 23   &10.0   &  ---		  &21, 48, 49 \\
IRAS 18566+0408 	 & 6.7     &8$\times10^4$     & 263    & 7000	&3.0  & 25   & 23   & 8.3   &2.3$\times10^{-4}$   &18, 50, 51 \\
IRAS 23151+5912 	 & 5.7     &1$\times10^5$     & 11     & 2150	&3.0  &  8   & 25   & 6.2   &3.6$\times10^{-4}$   &52, 53, 54 \\
W33A$-$MM1 Main 	 & 3.8     &1$\times10^5$     & 16     & 1900	&3.0  &---   & 25   & 6.8   &5.6$\times10^{-3}$   &55 \\
NGC6334I SMA1 Main	 & 1.7     &1$\times10^5$     & 37     &  280	&5.1  &---   & 25   & 8.0   &  ---		  &2, 56 \\
IRAS 18360$-$0537 MM1	 & 6.3     &1.2$\times10^5$   &124     &22000	&1.5  &---   & 27   & 3.4   &1.1$\times10^{-2}$   &12, 57 \\
W3 IRS 5 		 & 2.0     &2$\times10^5$     & 13.4   & 2000	&2.5  &---   & 32   & 4.5   &  ---		  &58, 59 \\
G28.87+0.07		 & 7.4     &2$\times10^5$     & 78     & 6000	&0.5  &---   & 32   & 9.1   &1.1$\times10^{-2}$   &19, 28 \\
G31.41+0.31		 & 7.9     &3$\times10^5$     &653     & 3600	&2.5  &25$^o$    & 38   & 8.0   &  ---		  &42, 43, 60, 61, 62 \\
G10.62$-$0.38		 & 3.5     &4$\times10^5$     & 80     & 3400	&2.1  &---   & 42   & 6.0   &  ---		  &63 \\
G20.08$-$0.14N  	 &12.3     &6.6$\times10^5$   & 28     & 5000	&3.5  & 25   & 52   & 8.2   &3.9$\times10^{-3}$   &64, 65 \\
G29.96$-$0.02		 & 6.2     &8$\times10^5$     & 88     & 4000	&1.6  & 33   & 56   & 9.0   &7.8$\times10^{-4}$   &63, 66 \\
W51e8			 & 5.4     &1.5$\times10^6$   &116     & 8200	&$4.0^p$  &---   & 75   & 9.0   &  ---		  &18, 67 \\
W51e2 $-$E		 & 5.4     &1.5$\times10^6$   &157     & 8000	&2.5  & 15   & 75   & 8.0   &1.6$\times10^{-3}$   &18, 67, 68, 69 \\
G19.61$-$0.23		 & 12.6    &2.2$\times10^6$   &401     & 6300	&1.0  &---   & 80   &10.0   &3.9$\times10^{-3}$   &63 \\
W51 North		 & 6.0     &3$\times10^6$     & 248    &20000	&1.5  &---   & 87   & 8.0   &1.7$\times10^{-1}$   &70, 71, 72 \\
\hline

\end{tabular}
}
\end{center}
\vspace{1mm}
 \scriptsize{
{\it Notes:} \, 
 $^a$ Masses estimated assuming the dust opacities of \citet{Ossenkopf1994}; \\
 $^b$ $M_\star$ estimated from the Lyman-continuum photons (see \S~\ref{mass-star}); \\
 $^c$ Masses estimated from the simulated clusters (see \S~\ref{mass-star}); \\
 $^d$ Outflow parameters corrected for opacity and assuming a mean inclination angle of the outflow of 32.7\degr;  \\ 
 $^e$ References for the core parameters: 
1: \cite{Fallscheer2009};
2: \cite{Beuther2008}; 
3: \cite{Hunter2014};
4: \cite{Schreyer2002}; 
5: \cite{Schreyer2006};
6: \cite{Shepherd2001}; 
7: \cite{liu2013}; 
8: \cite{Shepherd2004};
9: \cite{Shepherd1998};
10: \cite{Sanchez-Monge2008}; 
11: \cite{Xu2012};
12: \cite{Molinari1996};
13: \cite{Bernard1999};
14: \cite{Beltran2014};
15: \cite{Cesaroni2005}; 
16: \cite{Shepherd2000};
17: \cite{Cesaroni2014};  
18: \cite{Hernandez-Hernandez2014};
19: \cite{Furuya2008};
20: \cite{Sanna2014};
21: \cite{Beuther2012};
22: \cite{WangK2012}; 
23: \cite{VanderTak2005};
24: \cite{Fernandez-Lopez2011};
25: \cite{Fallscheer2011}; 
26: \cite{Watt1999}; 
27: \cite{Tackenberg2014};
28: \cite{Rosero2013};  
29: \cite{Sanna2010};
30: \cite{WangJ2011};
31: \cite{Patel2005}; 
32: \cite{Torrelles1996}; 
33: \cite{Comito2007}; 
34: \cite{Gomez1999};
35: \cite{Sanchez-Monge2014};
36: \cite{Beuther-Walsh2008}; 
37: \cite{Beuther2005}; 
38: \cite{Beuther2002b};
39: \cite{Qiu2009}; 
40: \cite{Qiu2014}; 
41: \cite{Chen2007};
42: \cite{Beltran2004}; 
43: \cite{Beltran2005}; 
44: \cite{Beltran2011b};
45: \cite{Rodriguez2008}; 
46: \cite{Franco-Hernandez2009};
47: \cite{Guzman2014};
48: \cite{Campbell1984};
49: \cite{Mallick2014};
50: \cite{Zhang2007}; 
51: \cite{Carral1999}; 
52: \cite{Beuther2007}; 
53: \cite{Rodriguez-Esnard2014}; 
54: \cite{Qiu2007};
55: \cite{Galvan-Madrid2010};
56: \cite{Hunter2006};
57: \cite{Qiu2012};
58: \cite{Rodon2008}; 
59: \cite{WangK2013};
60: \cite{Girart2009}; 
61: \cite{Cesaroni2011};
62: \cite{Osorio2009};
63: \cite{Beltran2011a};   
64: \cite{Galvan-Madrid2009}; 
65: \cite{Yu2013}; 
66: \cite{Beltran2013};
67: \cite{Klaassen2009}; 
68: \cite{Gaume1993}; 
69: \cite{Shi2010};
70: \cite{Zapata2008}; 
71: \cite{Zhang1997}; 
72: \cite{Zapata2009};	 \\    
$^f$ The bolometric luminosity is low because this object is an IRDC and therefore in a very early evolutionary phase.; \\
$^g$ $M_\star$ estimated from the fitting of the velocity field assuming Keplerian rotation; \\
$^h$ The mass of this core has been estimated from CS \citep{Bernard1999}; \\ 
$^i$ $M_\star$ estimated from the mass infall rate and the dynamical age of the outflow; \\
$^j$ $V_{\rm rot}$ estimated from CH$_3$OH maser emission; \\
$^k$ $\Delta V$ estimated from single-dish observations \citep{Beuther2002b}. \\
$^l$ \citet{Beltran2011b} estimated a mass of 22--48~$M_\odot$ (for a dust opacity of 1\,cm$^2$\,g$^{-1}$) and a radius of
1500~au with PdBI and SMA higher angular resolution ($\sim0\farcs5$) observations. \\
$^m$ \citet{Beltran2011b} estimated a mass of 13~$M_\odot$ (for a dust opacity of 1\,cm$^2$\,g$^{-1}$) and a radius of
1800~au with PdBI and SMA higher angular resolution ($\sim0\farcs5$) observations. \\
$^n$ \cite{Beltran2011b} estimated a mass of 18~$M_\odot$ (for a dust opacity of 1\,cm$^2$\,g$^{-1}$) and a radius of
2700~au with PdBI and SMA higher angular resolution ($\sim0\farcs5$) observations. \\
$^o$ $M_\star$ estimated from SED modeling. \\
$^p$ Upper limit because estimated from the OCS velocity gradient that could be contaminated by outflow emission \citep{Klaassen2009}. \\
}
 \end{table*}

Spatially resolved observations characterize YSO disks according to a
number of basic quantities: (1) inner and outer radius; (2) radial
profiles of scale height and surface density; (3) total mass; and (4) mass
accretion rate. These quantities change and evolve with stellar mass,
envelope mass, the evolutionary stage or YSO class, and bolometric
luminosity.  Our knowledge of the evolution of disks, especially
around young, embedded IM and HM (proto)stars, in terms of the above
mentioned four properties is currently quite sparse, hampered by
insufficient resolution and sensitivity. It is not clear yet when in
the star-formation process the disk forms and becomes the vehicle of
mass and angular momentum transport. The disk could already be
established during the Class\,0 phase of a low-mass object. The
properties of optically revealed pre-main sequence disks of low- and
intermediate-mass stars are better known. Yet for the HM (proto)stars,
an extensive disk census combined with detailed disk views 
needs to be established with forthcoming observations. In
Sect. \ref{density-height} to \ref{macc} we describe the predominantly
interferometric methods employed that allow to derive disk
properties.

\begin{figure*}
\centerline{\includegraphics[angle=0,width=12.5cm]{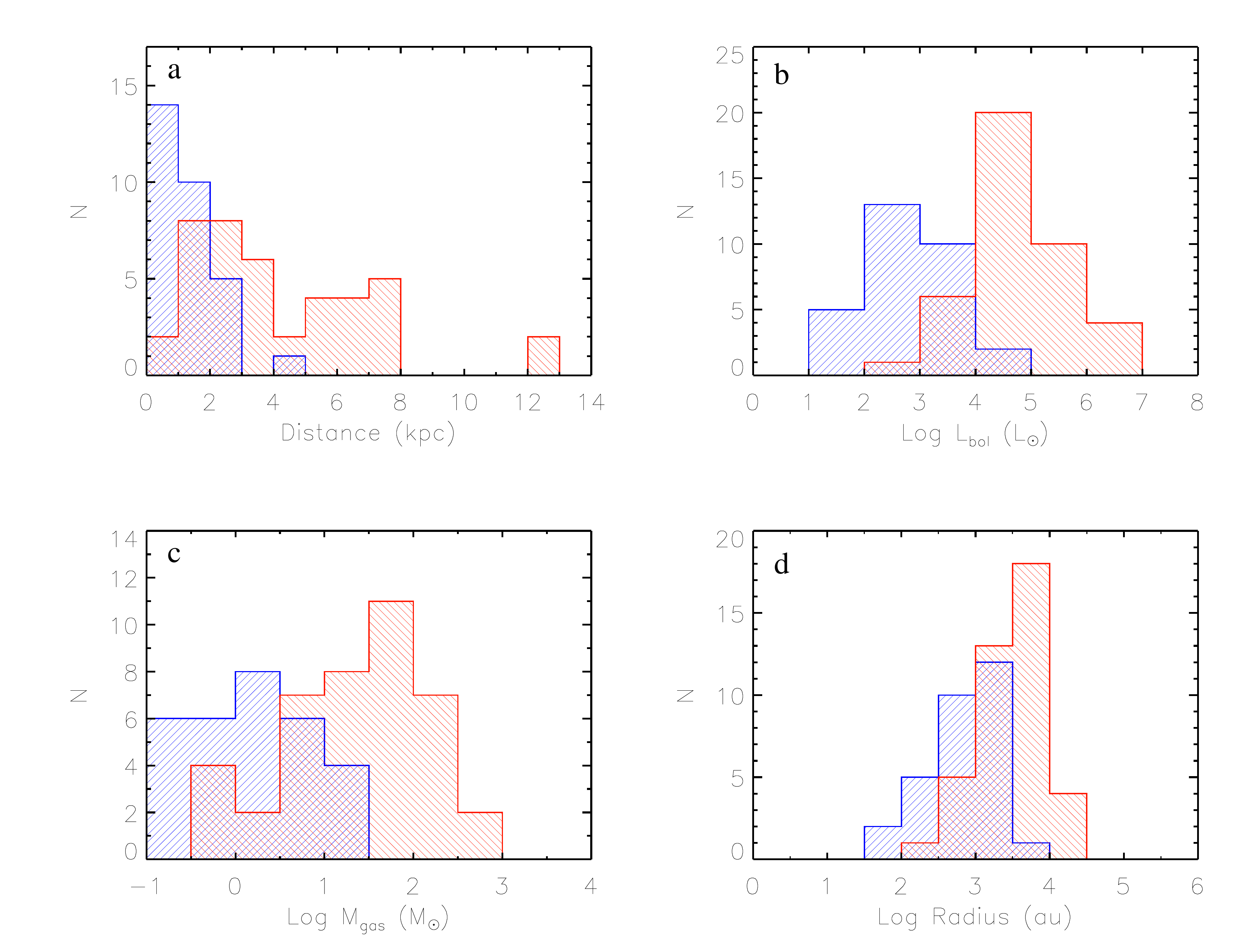}}
\caption{Histograms of {\it a)} distance, {\it b)} bolometric luminosity,   
 {\it c)} gas mass of the circumstellar structure, and {\it d)} outer radius  of
 the circumstellar structure of the IM ({\it blue}) and HM ({\it red})
 (proto)stars in Tables \ref{IMprop} and \ref{OBprop}.}
\label{histos}
\end{figure*}

To analyze the properties of disks around massive embedded
(proto)stars in a statistical way, we have compiled a list of IM and
HM disk candidates from different studies based on (sub)millimeter
interferometric observations. 
As explained
before, because of the distance and complexity of IM and HM
star-forming regions, high-angular resolution observations are needed
to distinguish the emission of different embedded sources and to
better trace the properties of their surrounding structures. What is
more, interferometric observations have the advantage of limiting the
contribution of the surrounding envelopes to the estimated quantities,
because some (most) of the extended envelope emission is being
filtered out by the interferometer if short-spacing information is not
available. Table~\ref{IMprop} shows the distance $d$, bolometric
luminosity $L_{\rm bol}$, mass $M_{\rm gas}$, radius $R$, and mass of
the central star $M_{\star}$ for the rotating structures found around
IM disk candidates.  Table~\ref{OBprop} shows the same properties and
includes also the rotation velocity $V_{\rm rot}$ and line width
$\Delta V$ for the HM disk candidates. Figure~\ref{histos} shows the histograms
of  $d$, $L_{\rm bol}$, $M_{\rm gas}$, and $R$ for the IM and HM (proto)stars
in the tables. Table~\ref{OBprop} also shows the stellar mass of
  the HM sources estimated in two different ways: from the Lyman-continuum
photons when available, $M_{\star\,{\rm Lyman}}$, and from simulated clusters,
$M_{\star\,{\rm cluster}}$ (see Section\,\ref{mass-star}).

We have searched the literature for interferometric observations of molecular
outflows for the candidate disk IM and HM sources. High-angular resolution
observations are necessary to disentangle the emission of possible different
outflows powered by different sources in these very clustered environments.
Because for some molecular outflows the parameters have been corrected for
inclination angle and for others the inclination angle is not known, we have
decided to assume a mean inclination angle of 32.7$\degr$ with respect to the
plane of the sky \citep{Bontemps1996} for all the outflows.
Therefore, the parameters of all the outflows were re-calculated assuming this
inclination angle. Regarding the opacity of the emission, we applied a
correction factor of 3.5 \citep{Cabrit1992} to the parameters of those outflows
observed with (partially) optically thick tracers, like for example $^{12}$CO,
and not corrected for opacity. Tables~\ref{IMprop} and \ref{OBprop} show the
corrected values of the mass loss rate $\dot M_{\rm out}$ for HM and IM disk
candidates, respectively.

\subsection{Disk tracers} 
\label{disk-tracers}

When searching for disks around IM and HM (proto)stars at millimeter wavelengths,
optically thin and high-density tracers are required because the structure is
relatively hot and dense. \citet{Cesaroni2007} discuss the main techniques used to
search for disks in HM YSOs and categorize them in: continuum emission, thermal
line emission, and maser emission. Based on the fact that the dust in the disk
emits as a grey-body peaking at $\sim$100~$\mu$m, continuum observations from IR to
(sub)millimeter wavelengths have been performed to search for disks around IM and HM
(proto)stars. Because of the high visual extinctions of the HM star-forming regions,
observations in the NIR (and sometimes even in the MIR) may not attain sufficient
depth to detect the disk. The sources which are detected at IR wavelengths are
either somewhat advanced in their early evolution or have a favourable orientation
where the line-of-sight intersects the envelope cavities created by the outflow, in
both cases lowering the total extinction and revealing the innermost parts.

Optical/infrared interferometry reaches typical spatial resolutions of
1\,mas in the
NIR and 10 mas in the MIR (see Fig.~\ref{fig-resolution}). This technique provides therefore
access to the inner disk at linear scales of 1-10\,au and to the
interaction zone between envelope and outflow at $\sim$100\,au for
objects located at 1\,kpc.  The observed flux in this wavelength range
is dominated by thermal dust emission conveniently at sublimation
temperature of 1500\,K (``hot'') in the NIR or at a typical
temperature of 300\,K (``warm'') in the MIR.  Access to velocity
fields is provided by those beam combining instruments employing a
dispersive element of at least intermediate spectral resolution
($R>1500$). The technique of spectro-interferometry can resolve
therefore both spatially and spectrally the atomic and molecular gas
in the inner disk region. The infrared interferometric instruments
deliver visibility and phase data. Using spectral differential phase,
the nominal spatial resolution of the interferometer can be superseded
by a factor $\sim$100 when probing the photo-centre shifts encoded in
the differential phase signal between spectral line and continuum
emission. Differential phase precisions of 30\,$\mu$as have been
reached for measurements of the \CO~ first overtone bandhead emission
arising in a Keplerian disk \citep{Wheelwright2012b}. The inner disk
physics of most intermediate mass PMS stars are within reach of OI,
yet most of the deeply embedded sources (mostly the high-mass stars)
and the intrinsically faint sources (the low-mass stars) are currently
too faint even for the sensitivity offered by optical interferometers
using 8m-class telescopes (ESO's VLT-I and the LBT-I) and only a few
exceptionally bright sources of these classes have been studied in
detail.

Millimeter line emission from high-energy levels of
low-abundance molecular species are used to probe the velocity field
of the bulk of the disk mass. Complex organic molecules, such as
CH$_3$OH, CH$_3$CN, and HCOOCH$_3$, including isotopomers and
vibrationally excited transitions, with excitation energies $>$1000~K
and pre-biotic molecules, such as CH$_2$OHCHO, are abundant in HM
star-forming regions and also in a few IM star-forming regions
\citep[see][and references therein]{Beltran2015}. The large
bandwidth of current radio/mm interferometers allows us to study the
physical properties and velocity field of these embedded (proto)stars
in many different species, covering a broad range of excitation
conditions, as recently demonstrated by ALMA observations at
850~$\mu$m of B-type (proto)stars: G35.20+0.74N\,A and B
\citep{Sanchez-Monge2014} and G35.03+0.35\,A \citep{Beltran2014}. 

Alternative strategies to search for massive YSO disks were proposed
by \citet{Hoare1994} and \citet{Hoare2006} for the massive YSO sources S\,106 and
S\,140-IRS1 by means of centimeter emission. It potentially traces ionized
equatorial winds driven by the radiation pressure of the central star and inner
disk acting on the surface layers of the disk. Indeed, these
observations have revealed elongated disk-like structures 
perpendicular to molecular outflows that have been interpreted as
produced by such equatorial disk winds. Such winds can also be traced
using IR hydrogen recombination lines accessible to OI and exploited
for example in the study of the disk of the young B\,1.5V star MWC\,297 \citep{Malbet2007}.
Circumstellar disks can also be studied with maser
emission, as nicely demonstrated by the multi-epoch SiO maser emission VLBI
observations of the HM (proto)star  Orion Source I (Matthews et al.,
2010\nocite{Matthews2010}). Because of the strength and compactness of their emission, masers can
be observed with VLBI techniques that can provide angular resolutions of a few
mas. The advantage of using masers to study circumstellar disks is that they are
very bright, the drawback is that their emission is sensitive to the
excitation conditions, which makes it hard to estimate the physical
parameters of the disks. In addition, the interpretation of maser emission is
complicated and ambiguous, and sometimes the same maser species have been claimed
to trace disks and outflows simultaneously. For example, \citet{Norris1998} and
\citet{Pestalozzi2004} interpret CH$_3$OH masers as originating in disks rotating
about massive YSOs, while \citet{DeBuizer2003} finds evidence that methanol
masers are associated with outflows. Therefore, any possible evidence of disks
from maser lines has to be confirmed by the presence of a jet/molecular outflow
perpendicular to the distribution of the masers.

\subsection{Estimating inner and outer radii}
\label{density-height}
Disk outer radii are estimated from either resolved dust continuum
emission or from line emission of high-density tracers. Such
observations are performed at long wavelengths, given the low
temperatures of the outer disk. This strategy can be applied to
embedded sources and to PMS sources by means of (sub)millimeter
interferometers (e.g., \"Oberg et
al. 2011\nocite{Oberg2011}). Importantly, the radii estimated from
molecular tracers are often found to be larger than those estimated
from dust continuum emission.  One possible explanation is that the
gaseous emission of the envelope is difficult to disentangle from that
of the disk. Alternatively, the gaseous disk could be intrinsically
larger than the dusty one. Tables~\ref{IMprop} and \ref{OBprop} list
the outer radii for the rotating structures around IM and OB-type
(proto)stars estimated from line emission when possible.  Otherwise,
the value was estimated from the dust continuum emission.

The inner disk boundary is a crucial parameter for a number of
reasons. The inner radius is required for estimating the accretion
energy released in the disk with respect to that which is released on
the star. The inner disk also represents the zone where the
interaction between disk and star is strongest and where processes
take place by which the material actually lands on the star
(magneto-spheric or boundary layer accretion) and where disk erosion
and planet formation are manifest. Direct spatial information on the
inner disk and its associated accretion physics requires a linear
resolution between 10\,au and approximately the stellar radius.  For
most stars, such scales are only indirectly accessible through
spectroscopic techniques like high resolution spectroscopy,
spectro-astrometry and spectro-polarimetry, i.e.\  techniques that have
proven their efficacy in the optical and near-IR wavelength range
\citep[see e.g.][]{Meynet2015}.  Such spatial scales and even smaller ($\sim$1~au)
can also easily achieved by maser emission VLBI observations.
To actually resolve the inner radii of disks, an angular resolution of
approximately 35~mas for the embedded IM protostars, and better than 10~mas for HM 
ones is required. ALMA can achieve angular resolutions of $\sim$25~mas in Band 6
($\sim$1~mm) and of $\sim$40~mas in Band 7 ($\sim$0.85~mm), with
baselines of 10 and 5~km, respectively bringing the inner radii of
embedded IM stars within reach. Resolutions beyond these scales can in principle
be obtained in Bands 9 ($\sim$0.45~mm) and 10 ($\sim$0.35~mm) at
baselines of 10--15~km, but it is not clear if the array sensitivity
is enough to detect the inner disk emission. An additional
complication could be introduced by the optical thickness of the
dust emission at those wavelengths that could prevent the detection of line
emission.

On the other hand, OI provides the means to resolve spatial scales of
$\sim$10~au and disk regions even closer to the central star
($\sim$1~au), accessing the thermal processes related to the accretion
dynamics (see Fig.~\ref{fig-resolution}).  Given that the average dust
sublimation temperature is $\sim$1500\,K, a temperature for which the
corresponding blackbody emission peaks at 2~$\mu$m, OI can probe
directly the angular size of the dust sublimation region.  For a
substantial sample of young stellar objects, especially the HAeBe
stars, these measurements can be obtained, depending of course on the brightness
of the source and the distance. Often however  the uv-plane coverage
of OI measurements is rather sparse. Hence, the measured visibility and phase
measurements are compared to model geometries in order to translate visibilities to
angular scales.  The assumption in this approach is that the adopted model is an
idealization of the actual shape of the emission region on the sky. It
makes sense to assume simple models initially when only a limited
number of visibility measurements are available. 
Popular geometries to which visibilities are compared are rings, uniform
brightness disks, and Gaussians.  These mathematical functions have
relatively simple Fourier transforms and can be readily compared to the
visibility data. The derived sizes can be referred to as {\it
  equivalent ring size}, i.e., the size (in milli-arcsecond) if the emission region
were identical to a ring. Difference in visibility function of the
simple geometrical models become only clear in the second lobe, where
for example, the power of a ring starts to diverge significantly from
that of a uniform disk \citep[for graphical examples see
  e.g.,][]{Tycner2006} or a ring without structure to one with
structure, c.f., the discussion regarding the rounded and straight
inner dust rims in HAeBes (see Sect.\,\ref{sect412}). 
The mathematical functions can be found in textbooks related
to interferometry schools like the Michelson summer schools or the VLTI
schools \citep[e.g.,][]{Perrin2003}.

A slight sophistication in modelling the visibility functions of the
inner disk is possible when OI observations are performed employing a
dispersive element. Most beam-combining instruments deliver spectrally
dispersed visibility and phase. The OI data can then be interpreted in terms
of simple geometries at different black-body temperatures. This makes
physical sense if the disk emission is optically thick. For a disk
without accretion heating, i.e., a passive, geometrically flat,
irradiated dust disk, it can be shown that the resulting radial
temperature gradient will have a power law with exponent $-3/4$
\citep{Adams1986}, the same exponent as for a steadily accreting, thin
disk \citep{Shakura1973}. The temperature gradient becomes less steep
(exponent of $-0.5$) for flared disks \citep{Kenyon1987} as it
intercepts more stellar radiation and is consequently
hotter. Temperature gradient models are in common use to approximate the inner rim of the
disk and the degree of flaring \citep[e.g.][]{Eisner2007,Kreplin2012}.

A next level of sophistication in deriving physical
parameters of disks, OI studies often adopt
radiative transfer (RT) modelling aimed at reproducing the SEDs and
interferometric observables. The geometries of the RT models are
characterized by many parameters which could complicate the analysis.
Popular RT codes publicly available and suitable for young star disk studies are RADMC
\citep{Dullemond2011}, HoChunk \citep{Whitney2013},
MC3D \citep{wolf2003}, MCFOST \citep{Pinte2006}. Model-to-data comparison yields constraints on the free 
parameters of the chosen model, like
scale height and flaring index, inner rim distance, inclination and
P.A. For example, in \citet{DiFolco2009} the flaring of the disk of the famous
A0\,V PMS star AB\,Aur is determined thanks to spectrally and
spatially resolved MIR N-band observations.

\subsection{Estimating scale height profiles}
\label{estimate-height}

To investigate whether IM and HM circumstellar disks are geometrically
thin like those around low-mass YSOs \citep[e.g.,][]{Padgett1999}, one
can measure the hydrostatic scale height $H$. 

For the younger embedded
sources, in most cases $H$ cannot be estimated directly from
millimeter maps because the interferometric observations do not
resolve the emission.  However, according to \citet{Cesaroni2014}, it
can be estimated following
\begin{equation}
\label{height}
H = \frac{FWHM}{\sqrt{8 \ln 2}}\sqrt{\frac{R^3}{G\,M_\star}},
\end{equation}
where $FWHM$ is the line width of the high-density tracers, $R$ is the
outer radius of the disk, and $M_\star$ is the mass of the central
star. As these authors state, this expression is obtained by replacing
$c_{\rm s}$ with $FWHM/\sqrt{8 \ln 2}$ in Eq.\ (3.14) of
\citet{Pringle1981}, which calculates the local structure of a
geometrically thin disk in hydrostatic equilibrium and isothermal in
the $z$ direction. 

\subsection{Estimating disk masses}
\label{sect32}

The mass of the disk determines its dynamical state, a critical
property for the evolution of the disk. The state
can be typified as either self-gravitating and in non-Keplerian rotation or
centrifugally supported. The bulk of the disk material is cold and its
total mass can be effectively estimated by the thermal emission from
dust. Under the assumption that the emission at (sub)millimeter wavelengths
is fully optically thin one can use:
\begin{equation}
M_{\rm gas}=\frac{g\,S_\nu\,d^2}{k_\nu\,B_\nu(T_{\rm d})}.
\label{eq-mgas}
\end{equation}
The parameters in this equation are the following: $S_\nu$ is the flux
density, $d$ is the distance to the source, 
$k_\nu$ is the dust opacity coefficient, $g$ is the gas-to-dust mass ratio,
and $B_\nu(T_{\rm d})$ is the Planck function with temperature 
$T_{\rm d}$. The above conversion relation from flux to $M_{\rm gas}$ is
subject to uncertainties mainly in $g$, $k_\nu$, and $T_{\rm  d}$. We briefly
step through the systematics of these three parameters.

The gas-to-dust mass ratio $g$ in a disk is usually assumed to be 100, the typical
value for the interstellar medium (ISM). In other words, only 1\% of the
disk mass is in dust particles. The total mass of a disk is 
practically the same as the gas mass. However, the physical conditions
dominating the disk are likely to be much unlike those found in the
ISM. The gas-to-dust ratio could actually be lower in
circumstellar disks due to gas removal. Dust related processes such as 
grain growth and dust settling could result in a predominantly gaseous
disk atmosphere that would be subject to photo-evaporation leading to gaseous
disks winds or to the selective accretion of gas onto the central
star \citep[e.g.,][]{Williams2014}. These processes would result in an
overestimate of the disk mass based on dust emission.

An estimate of the dust temperature $T_{\rm  d}$ can be obtained by
fitting the Spectral Energy Distribution (SED) with black-body
curves. Alternatively, by assuming local thermodynamic equilibrium (LTE),
one could adopt the rotation temperature of the high-density molecular
tracers.  However, if dust and gas are not perfectly coupled, then
$T_{\rm d}$~can differ by a few degrees from the temperature  
estimated of the high-density tracers \citep{Goldsmith2001,Juvela2011}. 

The principal source of uncertainty in Eq.~\ref{eq-mgas}  is however the dust
opacity coefficient $k_\nu$. It depends on the composition and
properties of the dust particles themselves in particular on the dust
opacity power-law index $\beta$ ($k_\nu\propto\nu^\beta$). Different
studies have computed different dust opacity laws
\citep[e.g.,][]{Hildebrand1983,Draine1984,Ossenkopf1994} that yield
disk mass estimates differing by a factor 4--5
\citep{Beuther2002b}. We underline that the disk masses presented in
our compiled sample of (proto)stars are re-calculated assuming a dust
opacity of 1~cm$^2$\,g$^{-1}$ at 1.4~mm \citet{Ossenkopf1994}. The dust opacity
power-law index $\beta$ used is 2, except for those sources for which
an accurate determination of this coefficient has been made
\citep[e.g., G192.16$-$3.82:][]{Shepherd2001}. All mass estimates are
based on dust emission, except the source GH2O 92.67+3.07,
for which \citet{Bernard1999} estimated $M_{\rm  gas}$ from CS observations.

Apart from the systematic uncertainties in the input parameters, an additional
source of uncertainty on the disk mass worth mentioning, relates to the achieved spatial
resolution and the nature of the sources. Insufficient angular resolution
makes it very difficult to 
disentangle the disk emission from that of the envelope and would lead
to serious overestimates of the disk mass. Interferometers have the
advantage of spatially filtering out the most extended envelope
emission, and decrease the bias on the disk mass estimates.

\subsection{Estimating star masses}
\label{mass-star}
Stellar masses are ideally estimated through dynamical
considerations. Modelling the rotation in a Keplerian disk could however be 
achieved only for a small number of HM (proto)stars in the
(sub)millimeter \citep[e.g., IRAS 20126+4104:
][]{Cesaroni2005,Cesaroni2014} and in the near-IR \citep[e.g.,
  W33A:][]{Davies2010}.  At the youngest stage, $M_\star$ is mostly  
estimated from the bolometric luminosity and a ZAMS assumption. In 
later stages, young OB star masses relate to the emitted UV
radiation. This radiation rapidly ionizes the surroundings of the 
forming star and creates an \HII region. 
Assuming that the centimeter continuum emission comes from an
optically thin \HII region, one can deduce the Lyman-continuum photons
emitted per second ($N_{\rm Ly}$)
\citep[e.g.,][]{Mezger1967} and relate it to the spectral type
\citep{Martins2005, Lanz2007, Davies2011}.  
We caution however that free-free emission is also produced by shock
ionization in thermal radio jets \citep{Curiel1989} associated with both 
IM and HM (proto)stars (e.g.,IRAS 21391+5802: Beltr\'an et al. 2002;
G35.20$-$0.74N: Gibb et al., 2003\nocite{Gibb2003}; IRAS\,16562-3959:
Guzm\'an et al., 2011\nocite{Guzman2011}; W75N(B): Carrasco-Gonz\'alez
et al., 2015\nocite{Carrasco-Gonzalez2015}). A distinction can be made
between the two phenomena by means of the spectral index $\alpha$. The
radio jets are partially optically thick producing $\alpha<1.3$,
whereas the \HII region has typical index values of 2. Moreover, a jet
is generally fainter than an \HII region. 

We chose to obtain an estimate of $M_\star$ from $L_{\rm bol}$ for
the sources in the compiled sample taking into account the presence of
stellar clusters. This was motivated by the partial
availability of centimeter observations for the HM (proto)stars and 
because $M_\star$ for IM YSOs cannot be estimated from the ionized emission. 
In brief, we assumed that $L_{\rm bol}$ was consistent with that of a stellar
cluster and estimated $M_\star$ from the simulations of a large collection
($10^6$) of clusters with sizes ranging from 5 to 500000 stars each
(L.\ Testi, priv.\ comm.). The distribution of the number of clusters
$N_{\rm cl}$ with of a given number of stars $N_{\rm st}$ is d$N_{\rm
  cl}$/d$N_{\rm st} \propto N^\gamma_{\rm st}$ with $\gamma=2$. Each
cluster is populated assuming a randomly sampled
Chabrier~(2005\nocite{Chabrier2005}) initial mass function with masses ranging
from 0.1 to 120~$M_\odot$. For each simulated cluster, the total mass,
maximum stellar mass, bolometric luminosity and Lyman continuum are
computed. For each IM protostar of Table~\ref{IMprop}, we
assumed that its $M_\star$ corresponds to the maximum stellar mass of the 
simulated cluster.
For the HM (proto)stars, we estimated $M_\star$ in two different ways (Table~\ref{OBprop}):
one from the Lyman-continuum photons, when continuum centimeter observations of the
associated \HII region are available, and the other from the simulated cluster
by assuming that $M_\star$ is the maximum stellar mass of the cluster. Taking into
account the uncertainties of both methods, the agreement between the estimated
values of $M_\star$ is quite good for most of the sources (Table~\ref{OBprop}).

Stellar mass estimates for IM PMS stars present their own set of difficulties.
Masses are estimated from comparison with evolutionary tracks or directly from
the observed spectral type. The main uncertainties in spectral type
determination for young stars lie with  the extinction correction and the
associated total-to-selective dust extinction law. The superposed emission line
spectra of the circumstellar environment hides the crucial photospheric lines.  
Moreover, for stars in the A and B-type spectral range, the  photospheric lines
are only few and often compounded by fast stellar rotation. This can lead to
uncertainties in spectral type of 5 sub-types and in some cases it can be even
more uncertain \citep{Hernandez2004,Alecian2013}. The spectral types and
luminosities of HAeBe stars that are included in the analysis of the present
work are taken from \citet{Fairlamb2015}. 

Finally, the embedded and optically revealed sources share inaccurate
distance estimates leading to a systematic 
uncertainty in the bolometric luminosity. Only for the nearest PMS stars,
Hipparcos distances are available \citep{VandenAncker1998} and
spatial co-location with young stellar clusters and star-forming regions would
entail reasonable distances. The situation regarding the PMS star distances
should strongly improve with the data from the GAIA satellite. 

\subsection{Estimating rotation velocities}
Velocity field information is crucial in the selection process of good Keplerian
disk candidates, especially when searching for the youngest disks in complex embedded
environments. Particular emphasis in this process is on the existence of
velocity gradients perpendicular to molecular outflows as they are indicative of
rotation.  The rotation velocity ($V_{\rm rot}$) for the embedded disks is a
quantity that is used to refer to the rotation velocity at the outer disk
radius. It gives access to parameters like the rotational timescale $t_{\rm
rot}$ and the dynamical mass $M_{\rm dyn}$ of the system (see Fig.\,3). In
practice, $V_{\rm rot}$ is often estimated as  half the range of the velocity
gradient measured over the resolved disk of a high-density molecular tracer
\citep[e.g.,][]{Beuther2008,Beltran2011a}. For well resolved objects, Keplerian
rotation can be established by means of the position-velocity (PV) diagram  when
along the major axis of the rotating structure a ``butterfly" pattern appears,
characterized by low-velocity ``spurs" of emission and high-velocity ``spikes"
towards the position of the central object. These curves can be fitted with a
Keplerian rotation curve  \citep[e.g.,][]{Beltran2014}. Finally, a more accurate
way to estimate the kinematics is by radiative transfer modeling of the
molecular line emission, like  e.g.\ in the case of the HM source
AFGL~2591 VLA3, where \citet{WangK2012} find a velocity field consistent with
sub-Keplerian rotation with radial expansion.

For the embedded HM sources in Table~\ref{OBprop}, the values of $V_{\rm rot}$ 
are those quoted in the literature. However, when $V_{\rm rot}$ is not
reported, we have estimated it from the maps of the 
velocity gradient shown in the papers. 
This has left us with a quite homogeneous sample to
derive overall trends and properties: all HM sources have been observed at
high-angular resolution and for all of them kinematical information
regarding possible rotating structures is available. For the IM
protostars sample (Table~\ref{IMprop}) we had to limit our research to
dust continuum emission only. Unfortunately, only for a few cases high-angular
resolution and high sensitivity line observations are available and the kinematics of these
disks has been studied in detail \citep[e.g., IRAS
  22198+6336:][]{Sanchez-Monge2010}. Finally, for direct comparison between 
  disk properties among embedded and PMS objects, we have estimated $V_{\rm
  rot}$ from the observed velocity gradient for HAeBe stars for which
  resolved millimeter observations exist (see Sect.\,\ref{disk-evolution}).

\subsection{Estimating accretion rates}
\label{macc}

The mass accretion rate (\macc) is a central quantity in star
formation.  It provides the timescale for star birth and constrains
the physics involved in the process of mass, energy and angular
momentum transport from the envelope to the disk and from the disk to
the star. According to recent models of HM star formation,
non-spherical accretion through a disk and onto the stellar surface is
characterized by high accretion rates
\citep[$>10^{-3}\,\msolyr$, also inferred from observations,
  e.g.,][]{Nakano2000, Testi2010} that are 
needed to overcome the radiation pressure of the newly
formed OB-type star \citep[see e.g.,][]{McKee2003, Kuiper2014}.  The
adjective {\it high } is in relation to the derived mass accretion rates in
the low-mass regime and various studies aim to demonstrate that a
correlation between $M_\star$ and \macc\, exists
\citep[e.g.,][]{Ercolano2014}.  This aspect will be explored in
detail in \S\,\ref{evolution}.

In the low-mass regime, instantaneous mass accretion rates are derived
within the framework of magneto-spheric accretion, in which the
accreting material is channeled from the circumstellar disk onto the
stellar surface along the stellar magnetic field lines. Accretion
luminosities can be derived from the flux released when the material
impacts the stellar surface, which appears as excess flux in relation to
the photospheric one. The strong shocks create accretion hotspots in the
photosphere with temperatures around  10\,000\,K and optically thin
pre-shock material can reach temperatures up to 20\,000\,K.
This potential energy release produces hydrogen continua and emission in
various atomic transitions \citep[e.g.,][]{Hartmann1998,Calvet1998}. The
extra continuum emission is easily distinguishable from the cool
atmospheric emission of late-type stars \citep{Herbst1994} and it can be
identified in the weakest  magneto-spheric accretors \citep{Herczeg2008} by means of
excess UV and optical light. It becomes however increasingly harder to
detect towards earlier spectral types. A correlation between accretion
luminosity and line luminosity is however apparent for many
transitions \citep{Herczeg2008,Antoniucci2014} and these can be made
to good use for dust extincted and embedded sources
\citep[e.g.,][]{Natta2006} in particular
the higher mass young stars. Special attention in this framework
receives the \brg\, 
transition as it is known to be powered by accretion energy along the mass
sequence from the T\,Tauri stars \citep{Muzerolle1998}  up to $4\,\msol$
stars \citep{Calvet2004}. Note however that the latter work investigates
IM PMS stars which are still fully or partially convective as their
Hertzprung-Russell diagram (HRD) positions place them on top of the Hayashi
tracks. A continuity in magneto-spheric accretion related properties between these IM objects and
the low-mass ones can be expected based on the fact that the incidence of
magnetic fields drops with \teff\, among the PMS stars, i.e., when the
stars lose their convective outer-layer \citep{Hussain2014}. Nonetheless,
observed Balmer and sodium line profiles of PMS A-type stars can be
reproduced by the magneto-spheric accretion model \citep{Muzerolle2004}.

Systematic studies determining \macc\ in the disks of HAeBe stars
have so far used either the \brg\ line \citep[][with a sample
  restricted to A-type stars]{GarciaLopez2006} or the Balmer
continuum excess \citep{Donehew2011,Mendigutia2011,Fairlamb2015}.  The
latter studies include early Herbig Be stars and an extension of the
relation between \brg\, and accretion luminosity is provided albeit
with a large scatter in the data \citep{Mendigutia2011}.  These surveys
demonstrate that, as a group, the Herbig stars are accretors. They
also show that the magneto-spheric accretion model may not be applicable to early B-types
($\teff>12\,000\,K$), because the required surface area of the
magnetic footprint needs to be larger than the entire stellar
surface. 
A large number of OI studies targeted the
\brg\,transition in HAeBe stars in order to determine the exact origin
of the emission within the accretion environment, probing the presumed
transition in accretion physics (see Sect.\,\ref{sect412}).

The above methods cannot be applied to estimate
$\dot M_{\rm acc}$ for the  embedded IM and HM (proto)stars. The very
high extinction due to obscuring 
dust, typically many tens of magnitudes, causes young massive stars to be
essentially invisible at optical, near-IR and for some sources even at mid-IR
wavelengths. Nonetheless, an indirect method to derive an estimate of global
protostellar \macc\,  (i.e., not instantaneous) is given by the position
of PMS stars in the HRD. The protostellar mass accretion rate determines the
luminosity and \teff\, of an accreting object and therewith the locus of the
stellar birthline in the HRD. The upper limit to the distribution of
PMS stars in the HRD was shown to be well matched by
a \macc\, of $10^{-5}\,M_\odot$\,yr$^{-1}$ \citep{Palla1993}.

Information on the instantaneous \macc\, during the main accretion
phase and therewith the properties of the accretion disk can be
accessed in the mid-IR N-band, under favourable lines of sight. In the
mid-infrared, the extinction due to obscuring dust is much lower than
at shorter wavelengths, whereas the SED of HM YSOs exhibits a
strong increase. Observations at these wavelengths allow us to access the earliest phases, crucial for
high-mass star formation. \citet{Rigliaco2015} demonstrate
that the mid-IR hydrogen recombination lines are accretion tracers
although the exact origin of these lines is not yet established.

At centimeter and (sub)millimeter wavelengths, a very effective method
of identifying the presence of infalling material in HM star-forming
regions is to observe red-shifted absorbed line profiles. At
centimeter wavelengths, the red-shifted absorption has been observed
against the bright continuum emission of an embedded hypercompact \HII
region, as for G10.62$-$0.02 \citep{Keto1988,Sollins2005}
 and G24.78+0.08 A1 \citep{Beltran2006c}, and at (sub)millimeter wavelengths, against
the bright optically thick continuum emission from the core center, as
for W51 North \citep{Zapata2008}, G19.61$-$0.23 \citep{Wu2009,Furuya2011,Beltran2011a}, and G31.41+0.31
\citep{Girart2009}. Since the continuum source is very
bright ($\sim$$1\,000$~K), especially when observed with an
interferometer (because the emission does not suffer severely from beam
dilution), it is easy to observe the colder molecular gas
($\sim$$100$~K) in absorption against it.  The absorption is observed
at positive velocities relative to the stellar velocity if the
material surrounding the (proto)star is infalling onto the central
star.  Following \citet{Beltran2006c}, the infall
rate $\dot M_{\rm inf}^{\rm \, red-abs}$ in a solid angle $\Omega$ can be estimated from the expression
\begin{equation}
\label{eq-minf}
\dot M_{\rm inf}^{\rm \, red-abs}= (\Omega/4\pi)\,
2\pi\,m_{\rm H_2}\,N\,R_0^{1/2}\,R^{1/2}\,V_{\rm inf}, 
\end{equation}
where $N$ is the gas column density, $R_0$ is the radius of the
continuum source, $V_{\rm inf}$ is the infall velocity, and $R$ is the
radius at which $V_{\rm inf}$ is measured. The main caveat of this
method is the uncertainty on $R$. In fact, the radius at which $V_{\rm
  inf}$ is associated is not known, and usually the size of the
infalling core is given as an upper limit. In addition, what
red-shifted absorbed profiles are tracing is the infall of material in
the inner part of the core or in the disk, but not actual accretion
onto the central star.

Infall has also been inferred from the modeling of PV plots of
high-density tracers with models including both rotation and infall
\citep[e.g.,][]{Nakamura1991,Bernard1999}, and from the SED modeling
\citep[e.g.,][]{Osorio1999}.

For HM YSOs for which rotation has been detected, assuming that  the infall
velocity $V_{\rm inf}$ is equal to the rotation velocity \citep{Allen2003},
 the infall rate $\dot M_{\rm inf}^{\rm \, v_{rot}}$ can be estimated from the expression 
\begin{equation}
\label{eq-mvrot}
\dot M_{\rm inf}^{\rm \, v_{rot}}= 2\,\pi\,\Sigma\,R\,V_{\rm inf}, 
\end{equation}
where $\Sigma=M_{\rm gas}/\pi\,R^2$ is
the surface density.  

Finally, a rough method to estimate the infall rate $\dot M_{\rm inf}^{\, ff}$ is to assume that all
the material in the rotating structure will collapse in a free-fall time $t_{ff}$
and use the expression 
\begin{equation}
\label{eq-mfreefall}
\dot M_{\rm inf}^{\, ff} = M_{\rm gas}/t_{ff}. 
\end{equation}
However, like for the red-shifted absorption method, what one is
measuring here is the infall in the core or in the disk and not the
accretion onto the central star. 

A method to estimate the mass accretion rate \macc\, onto the central
star is to use the mass loss rate of the associated outflow $\dot
M_{\rm out}$ \citep[e.g.,][]{Beuther2002a}. Assuming
conservation of the momentum rate of the outflow and of the internal
jet entraining the outflow, then $\dot M_{\rm out}$ is related to the
mass loss rate of the internal jet $\dot M_{\rm jet}$ jet as $\dot
M_{\rm out}=\dot M_{\rm jet}\,V_{\rm jet}/V_{\rm out}$. Assuming furthermore 
a ratio between the jet velocity $V_{\rm jet}$ and the molecular outflow
velocity $V_{\rm out}$ of $\sim$20 \citep{Beuther2002a} and a ratio between $\dot M_{\rm jet}$ and
the mass accretion rate onto a low-mass protostar $\dot M_{\rm acc}$ of
approximately 0.3 \citep{Tomisaka1998,Shu1999}, one finds that 
\begin{equation}
\label{eq-mout}
6\,\dot M_{\rm acc}=20\,\dot M_{\rm jet}=\dot M_{\rm out}. 
\end{equation}
Properties of accreting stars by parametrizing 
the \macc\, following the above argumentation were calculated by
\citet{Behrend2001}. The growing \macc\,rates with stellar mass have a
significant effect on the stellar structure where a (proto)star starts
to bloat being able to reach radii of $>100\,R_{\odot}$ under influence of the high
\macc. This results in low stellar effective temperatures and
consequently little UV luminosity. For the quoted values of
$10^{-3}\,\msolyr$, the accreting (proto)star settles on the ZAMS only
once it has accreted a mass of $\sim 30$\,\msol\,
\citep{Hosokawa2010}.

\section{Observed properties of circumstellar disks}
\label{obs-prop}

In this section we gather and analyze the observed properties of the
circumstellar structures. For the IM
YSOs, we describe the disks in the early embedded phase and in 
the later pre-main sequence phase separately. For the HM YSOs,
we discuss the disk properties during the embedded phase only as
optically revealed accreting HM YSOs are not known. We discuss
separately the early-B/late-O type (proto)stars and the early-O type 
(proto)stars as
the properties of the circumstellar structures found for these two
groups are different, as touched upon in Sect.\,\ref{Sect:diskinhmyso}.

\subsection{Disks around intermediate-mass stars}

\subsubsection{Disks in embedded IM protostars} 
\label{disks-IM-prop} 

\begin{figure*}
\centering
\includegraphics[angle=0,width=11cm]{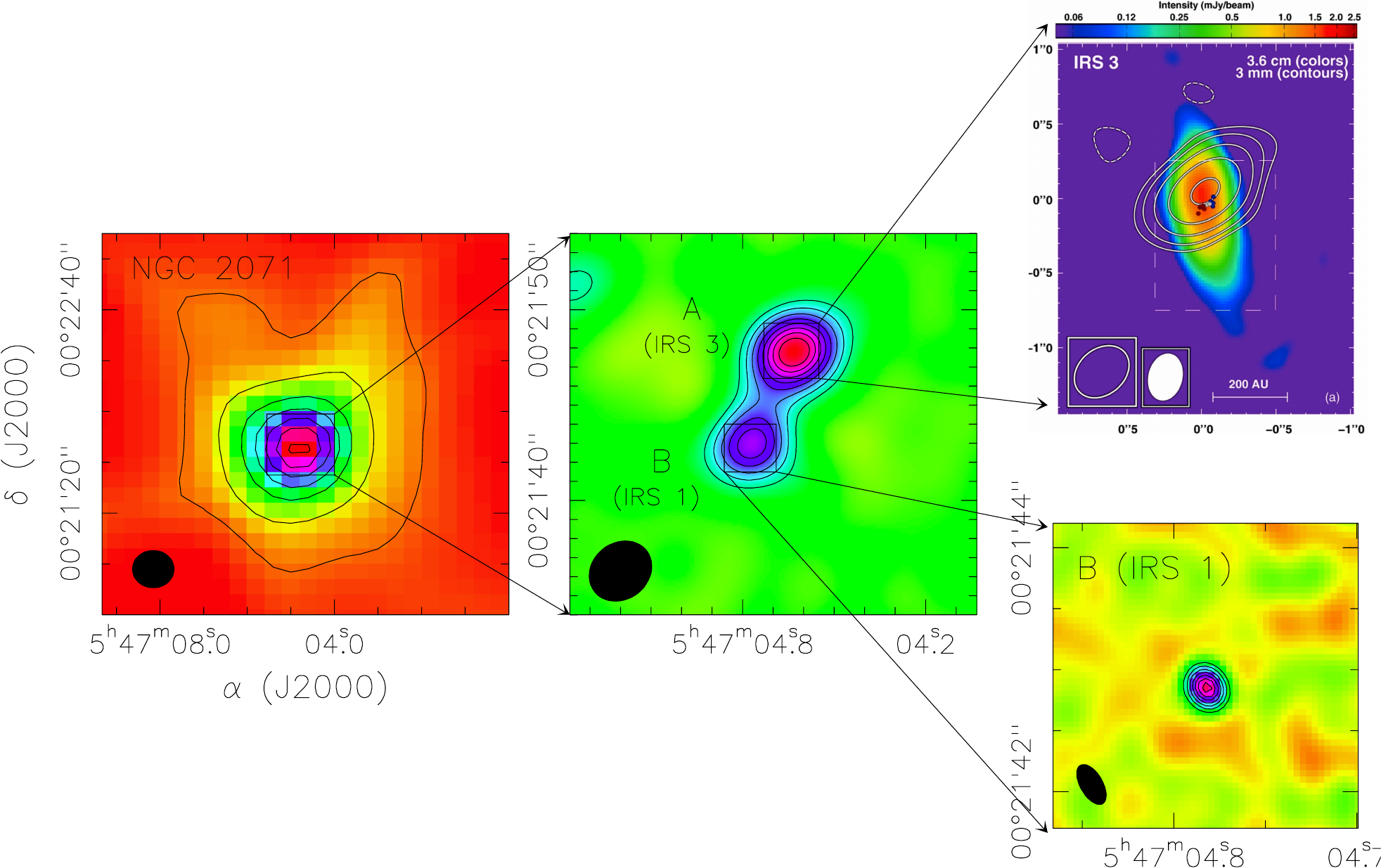}
\caption{{\it Left and middle panels:} Dust continuum emission towards the IM core NGC 2071 mapped with
SCUBA at 850~$\mu$m ({\it left}) and with the SMA in compact configuration at 1.3~mm
({\it middle}). The interferometric observations resolve the emission
of the NGC 2071 core into two cores, named A (or IRS 3) and B (or IRS 1). Adapted from \citet{VanKempen2012}. {\it Top right panel:} 
Overlap of the 3~mm dust continuum emission observed with
CARMA ({\it white contours}) on the 3.6~cm continuum emission observed with 
the VLA ({\it color map}) towards NGC 2071 A (IRS 3). The centimeter continuum
emission tracing the thermal radio jet is perpendicular to the circumstellar
disk. Water masers are marked with red, gray, and blue
circles.  From \citet{Carrasco-Gonzalez2012}.  {\it Bottom right panel:} Dust 
continuum emission observed with the eSMA (SMA combined with the JCMT and CSO telescopes)
at 840~$\mu$m ({\it right})  of the circumstellar disk embedded in core NGC 2071
B (IRS 1). Adapted from \citet{VanKempen2012}.
}
\label{fig-disksIM}
\end{figure*}

Clear detections of circumstellar disks of embedded IM
protostars are provided by (sub)millimeter interferometric
observations of the dust continuum. As bona-fide examples can be
considered the following sources: OMC1-S
139$-$409 and 134$-$411 (Zapata et al., 
2007\nocite{Zapata2007}), IRAS~22198+6336\,MM2 (S\'anchez-Monge et al.,
2010\nocite{Sanchez-Monge2010}), L1641\,S3, NGC\,2071\,A and B (van
Kempen et al., 2012\nocite{VanKempen2012}) (see Fig.~\ref{fig-disksIM}), and MMS\,6/OMC3 (Takahashi
et al., 2012\nocite{Takahashi2012}). The bolometric luminosity of the
sources extends up to a few times
$10^{3}\,L_{\odot}$, covering the whole IM protostar range and suggesting that 
there is no limit with stellar mass for the occurrence of a disk in
this evolutionary phase. 

The embedded dusty disks have a typical size of a few hundred
au. For a few cases however much larger sizes are measured with radii
of up to 1000--2000~au (see Table~\ref{IMprop}). We consider that
disks with outer radii of over 1000\,au are the product of
insufficient angular resolution. In fact, disks observed at increased
angular resolutions of less than $0\farcs4$~are found to have much
reduced radii of $\lesssim$100~au. Examples in our compiled sample are
Serpens FIRS\,1 with a deconvolved size of approximately 65~au and
NGC~2071~A (IRS\,3) with a deconvolved size of
$\sim$120$\times$50~au. These smaller dusty disks
appear as elongated structures oriented perpendicular to molecular
outflows and/or radio jets. The source NGC~2071~A (IRS\,3) is in this
respect a tellingly clear example (Carrasco-Gonz\'alez et al.,
2012\nocite{Carrasco-Gonzalez2012}, and see Fig.\,\ref{fig-disksIM}).

A particularly interesting case is presented in Zapata et al.,
(2007)\nocite{Zapata2007}. They observed IM
protostars in OMC1-S with the JVLA at 7~mm.  At a linear resolution of
$\sim$20~au, the continuum emission towards two sources ({\it viz.}
139--409 and 134--411) resolves in compact binary systems, each system
surrounded by a circumbinary ring of radius $\sim$100~au.  In turn,
each of the four individual objects might be surrounded by a compact
circumstellar disk of radius 20--25~au with a mass $<$0.1~$M_\odot$.
The picture that is suggested from these observations is that IM disks 
in young binary systems can be physically rather small for they cannot
extend beyond the circumbinary ring.

The angular resolution of the (sub)millimeter observations of IM protostars
performed to date has not been sufficient to resolve disks in the vertical
direction. From such observations it is clearly not possible to assess the
geometry of disks directly. However, Eq.~\ref{height} (see
Sect.\,\ref{density-height}) does offer a handle on the scale height
$H$. We applied this estimate for the compiled sample making the
assumption that a line width of $\sim$2~\kms~is typical (e.g.,
Beltr\'an et al., 2006b\nocite{Beltran2006b}, 
2008\nocite{Beltran2008}; Fuente et al.,
2009\nocite{Fuente2009}). We find that the hydrostatic scale height is
in  most cases $>$20--30\% of the disk radius but can be 
as high as 60\%. This simple estimate indicates that the disks of
embedded IM protostars are probably geometrically thick. For the
smaller disks observed at a resolution below $0\farcs4$~the scale 
height is only 10\% of the disk radius.  
This may indicate that the inner disk geometry is much thinner than
the outer disk suggesting an overall flared geometry. More observations are
clearly required to confirm this finding. 

\begin{figure*}
\centerline{\includegraphics[angle=0,width=11cm]{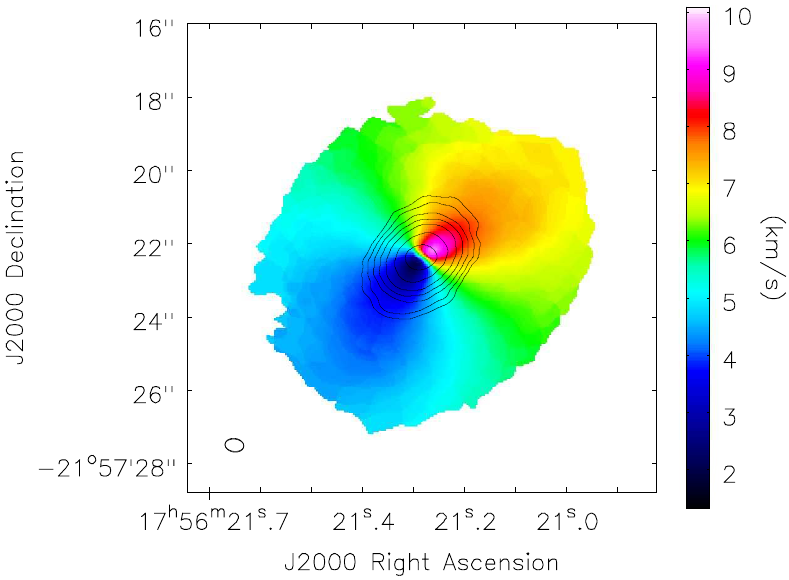}}
\caption{Overlay of the 850~$\mu$m continuum emission ({\it contours}) on 
the CO~(3--2) velocity map ({\it colours}) of the Keplerian disk around the Herbig Ae star HD 163296.
From \citet{DeGregorio-Monsalvo2013}.}
\label{fig-hd163296}
\end{figure*}

\begin{figure*}
\centerline{\includegraphics[angle=-90,width=9cm]{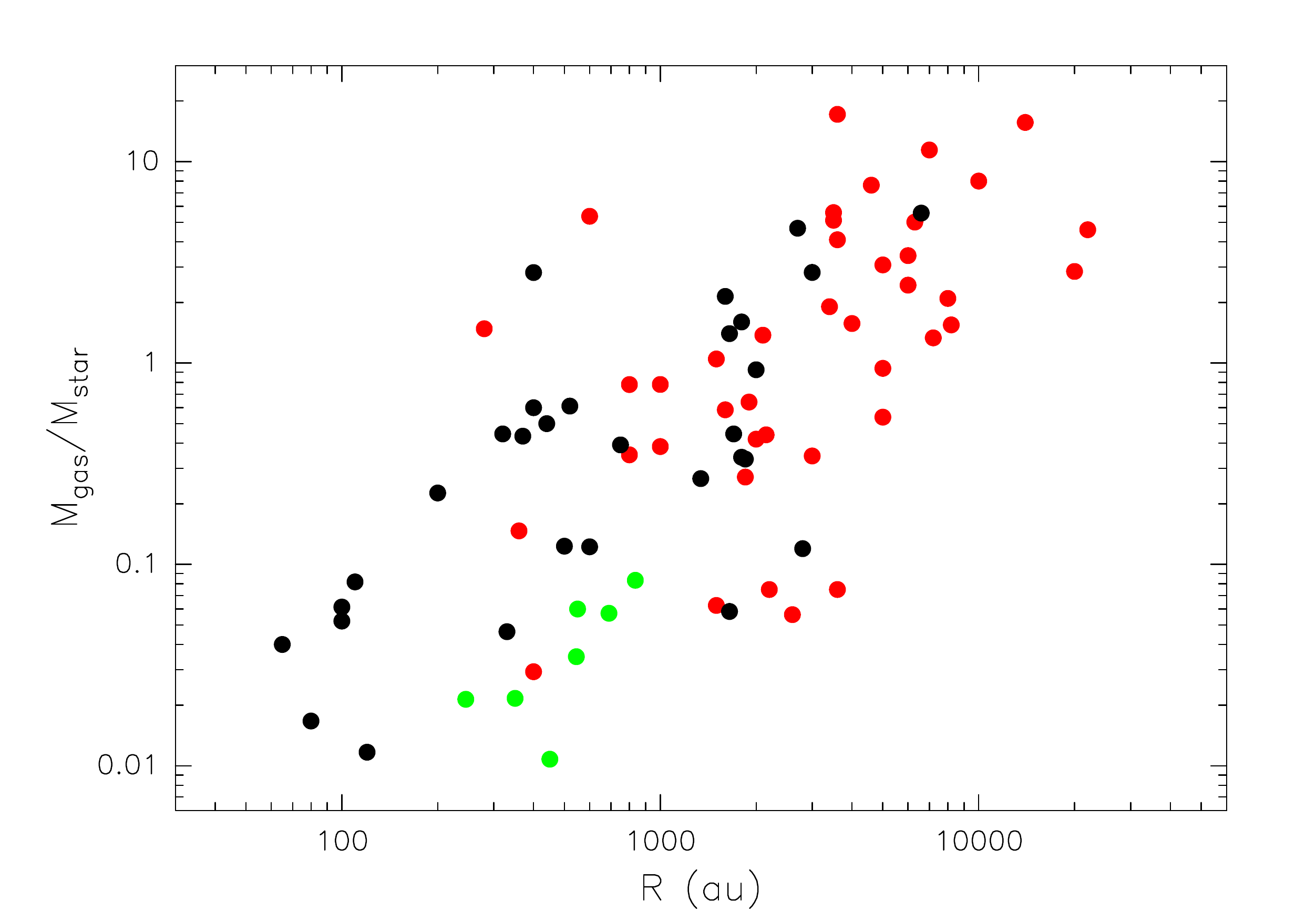}}
\caption{{\it Left panels:} Gas mass to stellar mass ratio $M_{\rm gas}/M_\star$ as a
function of the radius $R$ of the circumstellar structures for the IM (black
dots) protostars (Table~\ref{IMprop}) and the HM (red dots) (proto)stars 
(Table~\ref{OBprop}). Green dots correspond to Herbig Ae stars (see
\S~\ref{sect412}).} 
\label{fig-rad}
\end{figure*}

The disk masses $M_{\rm disk}$ of the embedded IM YSOs are typically
less than $2\,M_\odot$, which is smaller than that of the central
star. This finding suggests that IM circumstellar disks are likely to
be in Keplerian rotation like those of their low-mass counterparts. Direct
evidence for rotation is available for a few cases, where the
kinematics are probed with high-density tracers such as CH$_3$CN
(e.g., IRAS~22198+6336: S\'anchez-Monge et al.,
2010\nocite{Sanchez-Monge2010}; OMC1-S 139--409 and 134--411: Zapata et al.,
2007\nocite{Zapata2007}). They reveal velocity gradients
perpendicular to the direction of the corresponding molecular
outflows, which suggests rotation.  Rotation has also been inferred
from water maser emission (e.g., NGC~2071 A: Torrelles et al.,
1998\nocite{Torrelles1998}). However, the nature of the detected
velocity field (Keplerian or other) could not be established in any of
these case studies.  On the other hand, Keplerian rotation has been
clearly demonstrated towards the more evolved Herbig Ae/Be PMS stars,
like for example towards the Herbig Ae star HD~163296 (Qi et al.,
2013\nocite{Qi2013}; de Gregorio-Monsalvo et al.,
2013\nocite{DeGregorio-Monsalvo2013}; see Fig.~\ref{fig-hd163296}), as
unveiled by ALMA Science Verification observations at 850~$\mu$m and
1.3~mm.

For a few outliers though, the disk mass is found to surpass the star
mass as can be seen in Fig.\,\ref{fig-rad}. The explanation that we
prefer for such high masses would attribute the cause to (again)
insufficient angular resolution to separate the contributions by
envelope emission from that of the disk. This explanation is also
supported by the correlation seen in Fig.\,\ref{fig-rad}. The
figure displays the gas to stellar mass ratio $M_{\rm gas}/M_\star$
as a function of the radius $R$. The outlying IM protostars follow the
distribution of the more distant HM (proto)stars. The figure shows a
clear correlation between the distance-independent $M_{\rm
gas}/M_\star$ ratio and the radius of the circumstellar structure. This
correlation holds along the mass range of protostars and also for the
more evolved Herbig Ae stars.

\subsubsection{Optically revealed disks around Herbig Ae/Be stars}
\label{sect412}

It has become clear from spectroscopy that accretion continues after
the protostellar phase into the pre-main sequence phase for the A-type
\citep{Muzerolle2004,GarciaLopez2006} and B-type
objects \citep{Donehew2011, Mendigutia2011, Fairlamb2015}. Early
theoretical estimates of detecting young star (geometrically flat)
accretion disks by means of OI were sobering. \citet{Malbet1995}
expressed the minimal need for 100\,m telescopic baselines at
2.2\,$\mic$ (or $\sim$5~mas) for strong accretors ($10^{-6}\,\msolyr$)
located a little beyond Taurus (150\,pc). The first systematic studies
aimed at resolving the inner disk regions of young stars delivered the
pleasant surprise of finding much larger typical sizes than
predicted. The sizes were found to be in a linear range
between 0.1 and 10\,au. \citet{Monnier2002} demonstrated that the
H-band and K-band 
equivalent ring sizes $R$ (see Sect.\,\ref{density-height} for the meaning
of this term) determined for a sample of Herbig Ae/Be stars
show an apparent correlation with the bolometric luminosity of the
star with a functional dependency of $R \propto \sqrt{L_{\rm bol}}$, a
relation insensitive to distance errors. Such a relation suggests
physics related to dust sublimation and it inspired revision and
adaptations of the idea related to the inner disk structure: the
geometric {\it ring} was 
promoted to a physical dust sublimation {\it rim}.  It became an inflated dust
structure at the sublimation radius, puffed up by stellar irradiation
and the source for the poorly understood near-IR ``bump'' apparent in the SED
of HAeBe stars. An image 
using the technique of aperture masking interferometry of the
enigmatic Herbig Be star LkH$\alpha$\,101 delivered a resolved image
at $\sim$20~mas resolution in K-band. It demonstrated the evacuated
inner zone surrounded by a torus of hot dust at 3.4\,au radius \citep{Tuthill2001} and
provided justification for the use of the geometrical ring model in
order to estimate the dust sublimation radius. The ``puffed-up'' inner
rim scenario suggests that the dominant fraction of the disk emission
is reprocessed stellar irradiation and not the release of accretion energy. This
interpretation is valid in case the region interior to the dust rim
remains largely  optically thin to stellar irradiation such that the
rim can be illuminated by the star. Many of the OI studies in the past
decade are directed towards understanding the 
nature of the disk located interior to the dust sublimation
radius. This requires an angular resolution of the order 1--10~mas. A
comprehensive discussion of the physics and the problems of 
the disk inner rim of HAeBe stars and an overview of the advances achieved during the
first decade of this century can be found in \citet{Dullemond2010}. 

The most recent accretion rates determined for HAeBe
stars  render the zone interior to the dust
sublimation radius however optically thick \citep{Fairlamb2015}. For a fraction
of 20-30\% of their sample, \macc\ supersedes $10^{-6}\,\msolyr$, generating
accretion luminosities between 50\% and 80\% of the stellar bolometric
luminosity. Such \macc~rates would result in optically thick inner
disks \citep{Hartmann1993} and in smaller typical radii than expected
from dust sublimation physics \citep{Muzerolle2004}. 
These
findings for the accretion rate strongly indicate that 
the one-to-one identification of the geometrical ring 
with the innermost radius where dust exists cannot be the complete
solution to the origin of the near-IR emission in the Be nor the Ae types. This
finding is also supported by spectroscopic evidence where the strength
of the emission lines correlates with near-IR excess. As the emission
lines trace accretion activity, it would argue in favour of inner
disks that are not entirely passive \citep{Manoj2006}. 

\begin{figure*}
\centerline{\includegraphics[angle=0,width=10cm]{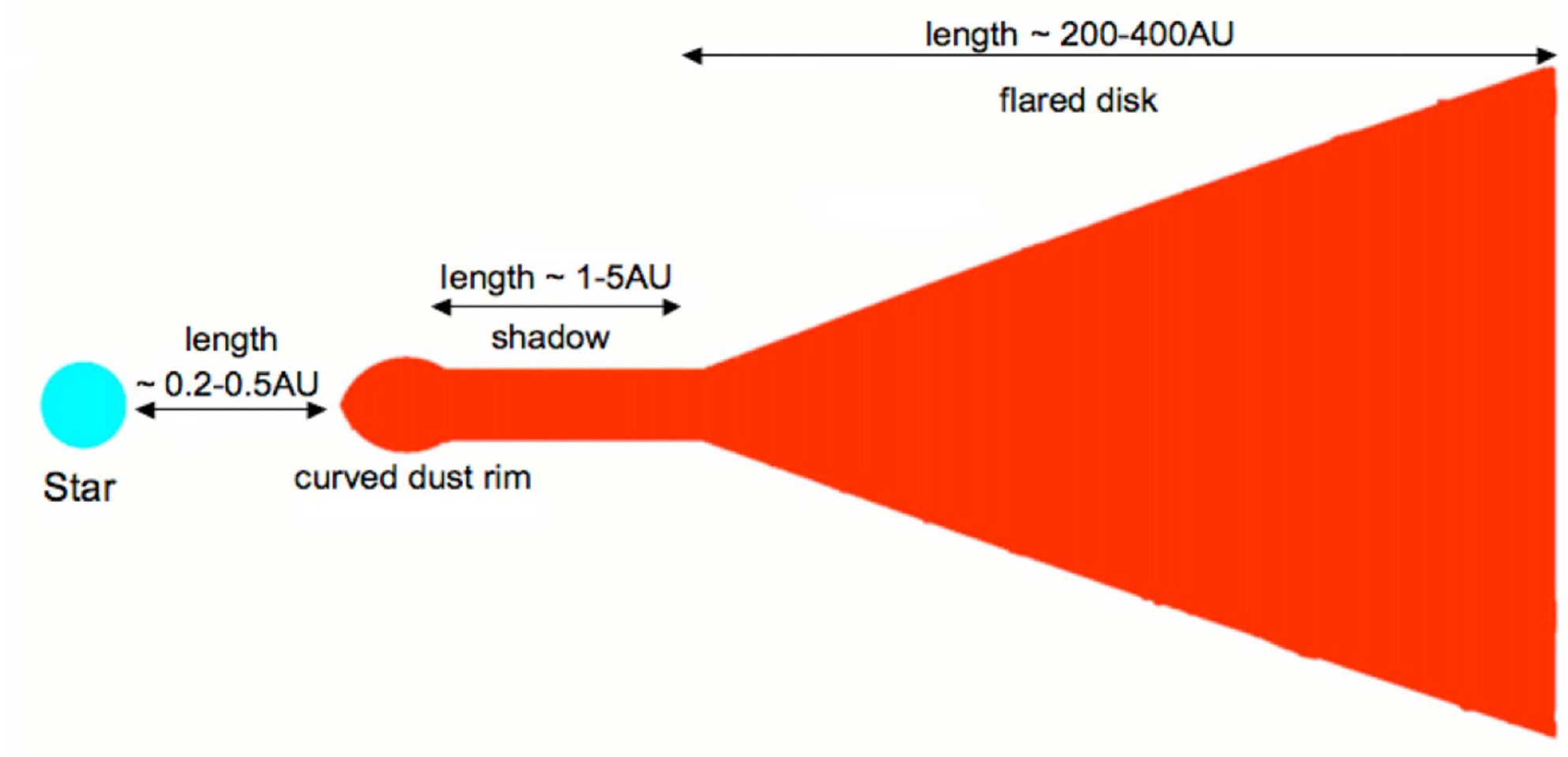}}
\centerline{\includegraphics[angle=0,width=10cm]{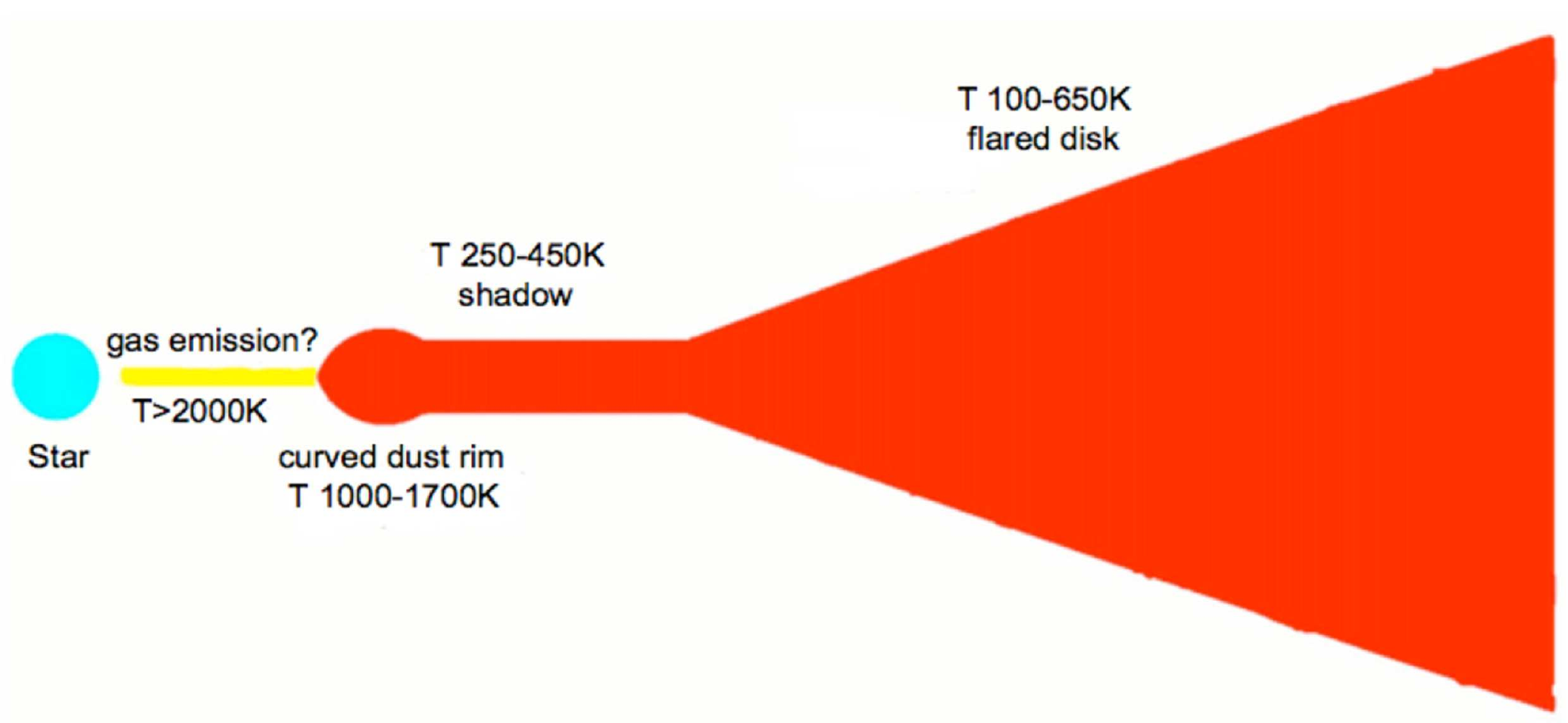}}
\caption{Schematic models of disks around Herbig Ae/Be stars. ({\it Top:}) Standard rim model
with a flared disk and a curved dust rim.  
({\it Bottom:})  Same model as above but with an additional smooth emission component within the dust
sublimation radius. From \citet{Tannirkulam2008}.}
\label{fig-tannirkulam}
\end{figure*}

The detailed properties of near-IR line and continuum emission require material
interior to the dust sublimation region. This advancement is
thanks to OI studies using facilities allowing increased data
production and more uniform filling of the uv-plane, the employment
of longer baselines, and the capacity to do high spectral resolution
interferometry up to R=12000. \citet{Kraus2008} find a satisfactory fit
to spectral and near/mid IR spatial data only when adopting an optically
thick gas disk interior to the dust rim for the $\sim$6\,$M_{\odot}$
Herbig Be star MWC\,147. This gaseous inner disk was confirmed and
found to be a rotation dominated structure by means of double-peaked
[O{\scriptsize{I}}] forbidden line emission \citep{Bagnoli2010}. In a number of
Herbig Ae cases, the standard inner rim scenario is not able to fit
long-baseline ($>100$\,m) interferometric data. Inner rim models
produce power (high visibilities) after the first null, a feature
which is not seen in the data. The modeling
in \citet{Tannirkulam2008} (see Fig.~\ref{fig-tannirkulam}) of two A-type stars favour the presence of
a smooth emission component inside the rim producing $\sim$50\% of
the emission. Similar findings are present for other A-type stars, for
which e.g. \citet{Benisty2011} argue that a large fraction of the
near-IR emission finds its origin interior to the dust sublimation
ring. The nature of this material could not be constraint apart from
the fact that it needs to withstand temperatures above 2000\,K. Based
on close to a total of 1500 visibility
measurements, \citet{Benisty2010} come to a similar conclusion for the
Herbig A1e MWC 275 (HD\,163296), where the {\it dominant} contribution
in H and K-band originates within the dust sublimation rim.

Given these OI results of HAeBe inner disks, the historical question regarding the nature of the near-IR SED bump 
 \citep{Hillenbrand1992} has therefore returned.  The expected temperatures and 
densities would predict strong molecular emission like the \CO\
and \HTO\ \citep[see the discussions
in][]{Muzerolle2004,Benisty2010}. \citet{Brittain2007} detect \CO\
ro-vibrational transition in 100\% of their sources with optically
thick inner disks, as defined as $K-L>1$. Also \citet{VanderPlas2015}
show the ubiquity of \CO\ fundamental ro-vibrational line emission in
HAeBe stars. On
the contrary, a general absence of the first-overtone emission of this
molecule is the result of a large spectroscopic survey (Ilee et
al. 2014\nocite{Ilee2014}), yet demonstrating that the few examples with \CO\
first overtone are all of mid to early B-type. High spectral resolution \CO\
first-overtone data can be fitted with Keplerian rotation with
corresponding velocities that place the molecular emission interior to
the dust sublimation region, but outside the corotation radius.
Spectro-interferometric observations with extended uv-coverage
corroborate this result in the case of a Herbig B5e star HD\,85567 where
the \CO\ emission is spatially resolved to be from within the
rim \citep{Wheelwright2013}. The reasons behind the \CO~ detection
differences can be related to temperature (accretion heating) or
column density, where fundamental transition is excited at the disk
surface covering radii up to 10\,au in case the disk is flared 
\citep{VanderPlas2015}. This
could imply that the B-type stars demonstrate much denser disks
interior to the continuum emission consistent with the \macc\ rates
 derived from the Balmer excess.

Spatially and (simultaneously) spectrally resolved observation of the
inner parts of the disk provide important information on the nature of
the inner disk. The first description of spatially resolved Br$\gamma$
emission can be found in \citet{Malbet2007} studying the enigmatic
young Herbig B0.5e MWC\,297 at linear scales of 1\,au. The
recombination emission in this star is found to be exceptional to that
seen in other HAeBes, as it is found at {\it larger} spatial scales
than the continuum emission. The near-IR continuum can be interpreted with a
standard accretion disk (a geometrically thin but optically thick, viscous
disk). The Br$\gamma$ emission is found to be larger than the
continuum emission and is modelled to arise in a disk
wind \citep[see also][]{Weigelt2011}.  In the majority of the HAeBe cases however, the
Br$\gamma$ emission is found to be at similar or at smaller spatial
scales than the dust continuum, where
sizes could be so small as to be consistent with magneto-spheric
accretion \citep{Kraus2008,Ragland2009}.  Garcia-Lopez et
al. (2015\nocite{GarciaLopez2015}) and \citep{Mendigutia2015} spatially resolve the Br$\gamma$ emission
at high spectral resolution ($\rm R=12\,000$). The velocity resolved Br$\gamma$ emission
emanates from a region smaller than the near-IR ontinuum and provides a
view of the inner gas disk dynamics. The inner gas disk is rotating
and can be modelled by Keplerian rotation \citep{Mendigutia2015} or a
rotating disk wind \citep{GarciaLopez2015}, whereas an even smaller
component of the Br$\gamma$ traces possibly the magneto-sphere itself.

Thanks
to high resolution near-IR spectrographs, it has also become apparent
that HAeBes are active accretors and capable of generating
jets \citep{Reiter2013, Ellerbroek2011}. 
Although the exact nature of the accretion onto the star remains
unknown, spectroscopic and spectro-polarimetric evidence has been
presented that the process by which the Ae stars form is analogous to
that by which solar-type stars form. This includes the idea of magneto-spheric accretion onto the stellar surface \citep{Costigan2014} albeit with smaller magnetospheres
than the T\,Tauri stars \citep{Cauley2014}. On the other hand, the
accretion process among the B-types is shown to be different and these
stars may accrete matter making use of a boundary
layers \citep{Vink2002,Mottram2007,Cauley2014}. This division in the
nature of accretion between Ae and Be is supported by OI data,
where \citet{Vinkovic2007} show a systematic difference in the nature
of the inner au of the disk with a break at a luminosity of
1000\,$L_{\odot}$, corresponding to spectral type B3V or 6\,\msol. The
ultraviolet flux falls precipitously in early B-stars and stars of
later spectral type do not emit significant Lyman continuum photons. 

The differences in accretion between Ae and Be type PMS stars, and the shared
properties of HAe and T\,Tauri stars may reveal the transition in star formation
from low-mass to high-mass. Regarding the overall disk structure,
\citet{Alonso-Albi2009} demonstrated that the masses of circumstellar disks
around HBe stars are a factor 5--10 lower than those of disks around HAe stars and that crystalline dust is found
to be more prominent in HAe disks \citep{Acke2004}. This difference is
attributable to the shorter life-time ($\sim$$10^{5}$\,yr) of the disk caused by
the increasingly stronger stellar radiation field with mass. The geometry of the
HAeBe disk, as captured in the Meeus group\,I and II  (see Sect.\,\ref{disks-IM}
for a description of the groups), changes too with mass. There are fractionally
more Herbig Be stars in group\,I (flared) than in group\,II (flat)
\citep{Acke2005}.  Consistency is found between disk geometry inferred from the
SED shape and spatially resolved OI data in the mid-IR
\citep{Leinert2004,Fedele2008,DiFolco2009,Ragland2012}.  

Estimates of the outer radius of HAeBe disks by resolving the outer parts
has been achieved for some notable cases by means of (sub)millimeter interferometry.
Mostly, in the pre-ALMA epoch the disks remain unresolved. The resolved disks
are found to have outer radii which are a couple of 100s of au in radius
(Guilloteau et al., 2013\nocite{Guilloteau2013}) and they are often found to be
in Keplerian rotation, allowing dynamical estimates of the central star mass
(e.g. for MWC\,480 and HD\,163296: Simon et al., 2001\nocite{Simon2001}; Isella
et al., 2007\nocite{Isella2007}; Qi et al., 2013\nocite{Qi2013}; de
Gregorio-Monsalvo et al., 2013\nocite{DeGregorio-Monsalvo2013}; see
Fig.~\ref{fig-hd163296}). The sub-Keplerian velocity field in the
outer-disk of the typical Herbig A0e AB\,Aur poses an interesting exception to
this (Mannings et al., 1997\nocite{Mannings1997}; Tang et al.,
2012\nocite{Tang2012}). The outer-disk measurements using molecular lines give
rise to a complication because of the physical conditions in early-type star
disks. The UV photons penetrate deeper into these disks and photo-dissociate the
molecules. Additionally, in the disk midplane freeze-out onto dust grains occurs
lowering the molecular abundances. Spatially resolved line emission and chemical
analysis have therefore been achieved mostly by making use of simple molecules
(e.g., Pietu et al., 2007\nocite{Pietu2007}; Dutrey et al.,
2014\nocite{Dutrey2014}).  In Sect.\,\ref{disk-evolution} we will use the disk parameters determined from
resolved observations in order to make a direct comparison between the embedded
and optically revealed disks. So far, the resolved outer disks are found
surrounding the later type Herbig Ae stars, {\it viz.} MWC 758 (Mannings et al.,
2000\nocite{Mannings2000}), HD\,142527 (\"Oberg et  al. 2011\nocite{Oberg2011}),
HD~34282 \citep{Pietu2003}, an exception the late B star HD\,100546
\citep{Pineda2014}. We also refer to the compilation of HAeBe disk masses in
the study of the cold dust emission by \citet{Alonso-Albi2009}. Like we do for
the  $M_{\rm gas}$ of the compiled sample of IM and HM (proto)stars
(see Sect.\,\ref{sect32}), the Herbig  Ae disk gas masses have been estimated using the
dust opacities of \citet{Ossenkopf1994}. 

In summary, the early-type PMS objects constitute good examples of young IM stars
surrounded by disks. The inner parts and the overall geometry of these
disks change character with mass. This is reflected by the information
obtained by high resolution techniques discussed in the previous
paragraphs. The group also provides direct confirmation of disks
surrounding PMS stars up to a mass of  
10\,\msol, namely: (1) R\,Mon - for 
which \citet{Fuente2006} found a Keplerian gaseous disk with outer
radius $<150$\,au from millimeter line observation; (2) HD200775
- \citet{Okamoto2009} reported a disk from direct mid-IR imaging; (3)
MWC\,1080; and (4) MWC\,297 have both been shown to possess inner
structures corresponding to hot dust
emission \citep{Malbet2007,Eisner2010} and with mm detections of cold
dust \citep{Alonso-Albi2009}. These disks can be considered as
descendants of the type of accretion disks expected among the embedded
early-B and late O-type stars. The accretion environment of the early-type 
Herbig Be stars shows similarities with the embedded ones, for
example in near-IR line emission of the CO
bandhead \citep{Wheelwright2012,Ilee2014} and the mid-IR dominance by
disk and outflow cavities \citep{DeBuizer2006, Okamoto2009,
DeWit2010}. To what extent these similarities hold, we will discuss in
the next section.

\subsection{Disks around high-mass stars}

\subsubsection{Disks around early B-type and late O-type (proto)stars}
\label{Bprop}
The current evidence for young stars with circumstellar disks is
limited to objects with masses up to 25--30\,$M_\odot$ or
$\sim$10$^5$\,$L_\odot$. The group of stars discussed in this section
contains the most massive objects for which unambiguous detections of
circumstellar disks exist. 
The bolometric luminosities correspond to ZAMS stars with temperatures
of early B to late O spectral type (see Davies et al., 2011\nocite{Davies2011} and Mottram et al.,
2011\nocite{Mottram2011}, and references therein), a regime at the onset of 
strong photospheric UV emission. Young, disk-bearing stars in this group are
deeply embedded in their birth environment; for very few 
sources, if any, the envelope component of the parent cloud disperses before the disk does. A situation which would allow an
unobscured view of the disk.  The disks have been traced in dust and
molecular line emission from centimeter to (sub)millimeter
wavelengths.  Line emission observations, in particular, have been
crucial in identifying these disks because in most cases they reveal
velocity gradients perpendicular to the molecular outflows powered by
the embedded HM (proto)star \citep[e.g.,
IRAS~20126+4104:][]{Cesaroni2005,Cesaroni2013, Cesaroni2014}.

Among the sources in this group, we
find typical disk radii of a few thousands of au, although radii as
small as 300--400~au have been reported (see Table~\ref{OBprop}). 
The largest disks of the HM
(proto)stars could therefore be 
larger than those of the IM protostars as is depicted in panel d of Fig.\,5.
At a maximum angular resolution of a few $0\farcs1$, the smallest spatial
scales traced to date are of the order of 100~au for a distance 
of $\sim$1\,kpc (as seen in  Table~\ref{OBprop}, HM (proto)stars are typically 
located at distances $>$1~kpc). Such
spatial scales belong to the outer regions of the disk, where
effects of material infall from the envelope to the disk may be
apparent.
Aiming for the inner regions of the disk using the JVLA in the
extended configuration, observations of dust emission at 7~mm can
reach an angular resolution of $0\farcs05$ or $<$100~au. However, 7~mm
emission may be significantly 'contaminated' by free-free emission
from thermal radio jets as demonstrated for the prototypical B-type
disk candidate IRAS~20126+4104 \citep{Hofner2007}. Also \UC\ regions
may affect the emission at this particular wavelength. Therefore,
multi-wavelength observations are mandatory to study the inner disk of
these sources, in order to separate the cold dust from the ionized
gas.

\begin{figure*}
\centerline{\includegraphics[angle=0,width=6.7cm]{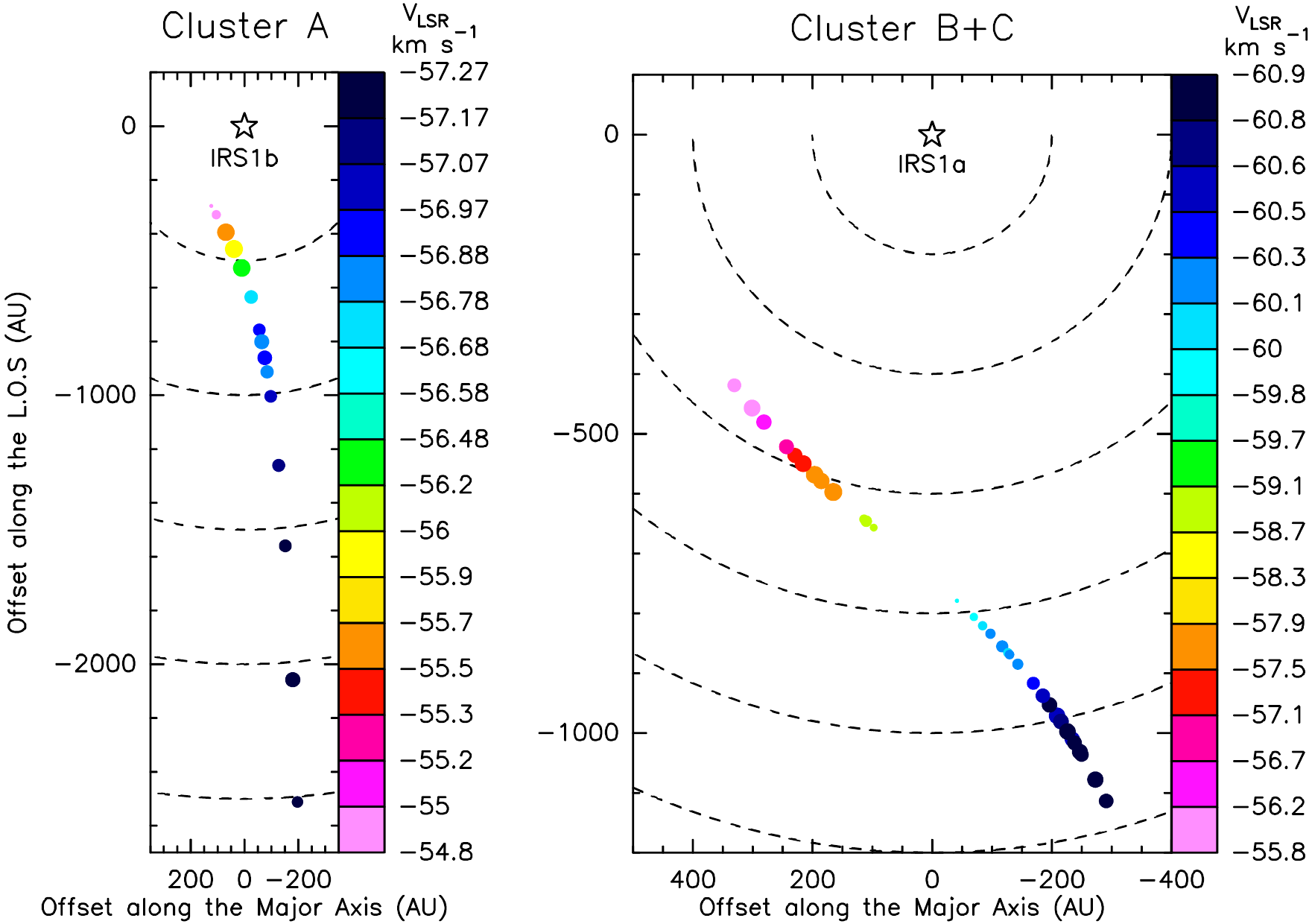}
\hspace{.8cm}\includegraphics[angle=0,width=4cm]{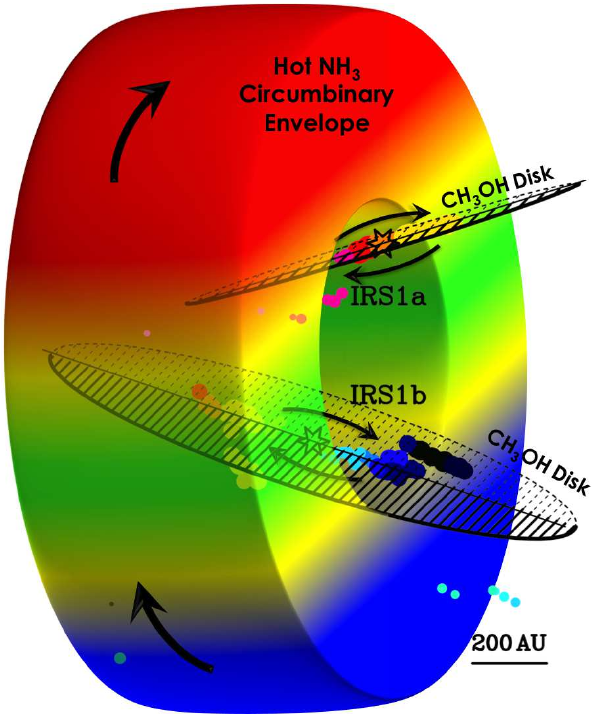}}
\caption{{\it Left panels:} Estimated positions of individual 6.7~GHz 
CH$_3$OH maser features towards the YSOs IRS1b and
IRS1a in the high-mass star-forming region NGC7538 IRS1 derived from an edge-on disk model. The colours of the dots
denote the maser velocity. Dashed arcs indicate
circular orbits with radii increasing by steps of 500 and 200\,au for IRS1b ({\it
left}) and
IRS1a ({\it right}), respectively. From \citet{Moscadelli2014}.  {\it Right panel:} Schematic model showing the
nearly edge-on circumstellar rotating disks around  IRS1a and IRS1b traced by the methanol maser
emission surrounded by a circumbinary envelope traced in NH$_3$. From \citet{Goddi2015}}
\label{fig-masers}
\end{figure*}

Alternatively, the inner disk can be investigated by VLBI observations
of maser emission.  In particular those of CH$_3$OH at a frequency of
6.7~GHz are suitable. In this way, an angular resolution of
$\lesssim$10~mas can be reached which might probe the inner disk gas
dynamics on linear scales of
$\lesssim$20~au \citep[e.g.][]{Moscadelli2011}. The information that
instantaneous maser observations provide on the shape and morphology of
the inner disk is however limited. Observing the masers at multiple
epochs allows to derive the 3-D velocity of the maser spots. The
velocity field derived from such observations has been successfully
modelled with a disk in Keplerian rotation at spatial scales of a few 100~au
(e.g., Sanna et al., 
2010\nocite{Sanna2010}; Moscadelli et al.,
2011\nocite{Moscadelli2011}). The methanol maser spots themselves are
not distributed randomly over the circumstellar disk but in a narrow
ring of radius a few 100 to $\sim$1000~au
\citep[e.g.][]{Bartkiewicz2005,Bartkiewicz2009,Torstensson2011,Sanna2010,Moscadelli2011}.
The ring-like distribution of the methanol maser spots reflects the
sensitive excitation conditions. The masers are not seen in the inner
disk because the local temperature or density probably exceed the upper
excitation limits of 250~K and
$10^9$\,cm$^{-3}$, values predicted by the model calculations of \citet{Cragg2005}. This is 
demonstrated in
Fig.\,\ref{fig-masers} that shows the VLBI observations and a 
corresponding cartoon of the velocity resolved methanol masers in the embedded
late O-type source NGC7538\,IRS1 (Moscadelli \& Goddi, 2014\nocite{Moscadelli2014}). The masers in the left two panels of this figure
are clearly restricted  to a linear distribution on the sky. They are
associated with two high-mass YSOs, named IRS1a, IRS1b,  which are separated 
by $0\farcs2$
(or $\sim$500~au). According to the authors, the masers trace two
quasi-Keplerian accretion disks on linear scales of   
$\sim$400--500~au. This young binary system is embedded in a
larger ($R$$\sim$1000~au) rotating circumbinary envelope 
as depicted in the cartoon of Fig.~\ref{fig-masers} (see also Goddi et
al., 2015\nocite{Goddi2015}).  

The angular resolution and sensitivity achieved by (sub)millimeter
interferometers prior to the advent of ALMA were insufficient to
detect clear disk structure around most HM (proto)stars. 
For the few cases that could be observed with angular resolutions
of $<0\farcs5$, the disk appears as a slightly elongated structure
oriented perpendicularly to molecular outflows or bipolar nebulae. To list a 
few, good examples are  AFGL~2591~VLA3 (Wang et al.,
2012a\nocite{WangK2012}), G35.20$-$0.74~N core B (S\'anchez-Monge et
al., 2013\nocite{Sanchez-Monge2013}),  IRAS~20126+4104 (Cesaroni et al.,
2014\nocite{Cesaroni2014}), and G35.03+0.35 core A (Beltr\'an et al.,
2014\nocite{Beltran2014}). Evaluating the hydrostatic scale height of
these structures by means of Eq.~\ref{height} we find that the
majority of the disks has a scale height of approximately $>$30--40\%
the disk radius. The only disk candidate that 
appears to be geometrically thinner is AFGL~2591~VLA3, for which $H$
is about 10\% the disk radius.  

The embedded circumstellar disks of the stars in this spectral type interval
have typical gas masses that range from 4 to a few 10s of $M_\odot$. Disk masses
as low as $<$1~$M_\odot$ have been estimated for sources observed at
high ($<0\farcs5$) angular resolution (see Table~\ref{OBprop}). 
The most likely explanation for such small disk masses is that the
interferometer has resolved out the outer part of the disk. In a few cases, circumstellar structures with masses as
high as 70--200~$M_\odot$ have been reported for stars with $L_{\rm
bol}\lesssim10^5$~$L_\odot$ (see Table~\ref{OBprop}). Such high masses
correspond to sources located at $\gtrsim$4~kpc. This strongly suggests
that the angular resolution of the observations is not enough to
properly separate the contribution by the envelope from that of the
disk.
An illustrative example is the rotating structure in G24.78+0.08 A1, studied
in \citet{Beltran2004,Beltran2005}. The authors estimated a mass of
130~$M_\odot$ and a radius  of 4600~au for this structure using PdBI
observations at an angular resolution of $\sim$$0\farcs8$. When
increasing the resolution to $\sim$$0\farcs5$, the emission of source
A1 fragments into three separate 
sources \citep{Beltran2011b}. Consequently, the mass of the
rotating structure is reduced by a factor ten, while the radius of the
rotating structure reduces by almost a factor 3 to 1800\,au. In this
way, the new values are brought into consistency with the ones
regularly found for the protostellar disk of early B-type objects.
Moreover, the decrease of the 
mass with size is what is expected from the correlation between the $M_{\rm
gas}/M_{\star}$ ratio and $R$ (see Fig.~\ref{fig-rad} and
Sect.\,\ref{disks-IM-prop}). 

The fact that $M_{\rm disk}$ is mostly $\lesssim M_{\star}$, like in the IM protostar
case (Fig.~\ref{fig-rad}), allows us to argue that the disks of early B-type
and late O-type are in Keplerian rotation. The actual kinematics of the 
embedded disks can be delineated
from high-density tracers. They reveal systematically linear velocity
gradients along well-defined directions, usually perpendicular to the direction
of molecular outflows. When the 
velocity field is probed at high-angular resolution
($\leq0\farcs5$), a clear signature of Keplerian rotation may be determined.
The prime examples for which this has been achieved are:
IRAS~20126+4104 (Cesaroni et al., 2005\nocite{Cesaroni2005}), Cepheus
A HW2 (Patel et al., 2005\nocite{Patel2005}), G35.20$-$0.74~N
(S\'anchez-Monge et al., 2013\nocite{Sanchez-Monge2013}, see also
Fig.~\ref{fig-g35}), and G35.03+0.35
(Beltr\'an et al., 2014\nocite{Beltran2014}). However, material in Keplerian rotation around
the central source is not systematically seen. For some sources, the
observations reveal non-Keplerian rotation in both
varieties: either sub-Keplerian \citep[e.g., 
AFGL~2591~VLA3:][]{WangK2012} or super-Keplerian (IRAS 18151\-$-$1208: Beuther
\& Walsh 2008\nocite{Beuther2008}). Sub-Keplerian motions suggest a
role for magnetic fields that could slow down the rotation below pure Keplerian
(i.e., magnetic braking, e.g., Galli 2006\nocite{Galli2006}). On the other hand,
super-Keplerian velocities could be produced if the inner disk contributes
significantly to the gravitational potential of the protostar-disk system
(Beuther \& Walsh 2008\nocite{Beuther2008}). In such a case, the outer part of
the disk needs to rotate faster in order for it to be centrifugally
supported. Another feature of protostellar disks of early B-type
and late O-type has been revealed by 
 position-velocity diagrams along
the major axis of the rotating structure. In a few cases, these diagrams have
exposed the presence of asymmetries
and inhomogeneities (e.g., Cesaroni 
et al., 2014\nocite{Cesaroni2014}; Beltr\'an et al., 2014\nocite{Beltran2014}).
These disk asymmetries could be produced by the presence of spiral arms or
infalling filaments accreting material onto the disk, or by the interaction with
nearby companions \citep{Cesaroni2014}.

In order to study the inner parts of the disks around young massive
stars, the increase in resolution could be obtained at shorter wavelengths.
In a small fraction of cases the
orientation of the sources provides a line of sight with
relatively low extinction. Such objects can be studied in the
near-IR in the same way as is done for the PMS sources. The only
source that has been studied in exquisite detail is IRAS~13481$-$6124
($\sim$$20\,M_{\odot}$). Using data from the near-IR AMBER beam combiner of the
VLT-I supplied with NTT speckle images, \citet{Kraus2010} obtained a
synthesized image at a spatial resolution of 2.4 milli-arcseconds ($\sim$8.4~au) (see
Fig.~\ref{fig-kraus}).
This is the highest resolution image of an embedded late
O-type star created to date. The image
shows a structure with a smooth spatial profile that is centrally
peaked. The morphology strongly suggests a disk geometry for the
continuum emitting material. This interpretation is re-enforced by the
fact that the disk orientation is perpendicular to that of a
large-scale, bipolar, CO outflow. In addition, this outflow may be
driven by an accretion induced jet, which is the most plausible interpretation of a
parsec-scale collimated H$_2$ flow \citep{Stecklum2010}.  However,
confrontation with a $20\,\mu$m, spatially resolved, single-dish image
makes clear the limitations of the RT disk model used to interpret
the synthesis image \citep[see][]{Wheelwright2012b}, while observations on milli-arcsecond
spatial resolution at 10\,$\mu$m also show that a fraction of the emission
could be due to warmed up cavity walls rather than solely originating in a
disk (see Boley et al.~2013)\nocite{Boley2013}. This is consistent with
the results of the sources W33\,A and AFGL\,2136 studied with optical
interferometry \citep{DeWit2010, DeWit2011}, and from single-dish
mid-IR resolved imaging \citep{DeBuizer2007}.

\begin{figure*}
\centerline{\includegraphics[angle=0,width=12.5cm,height=4.3cm]{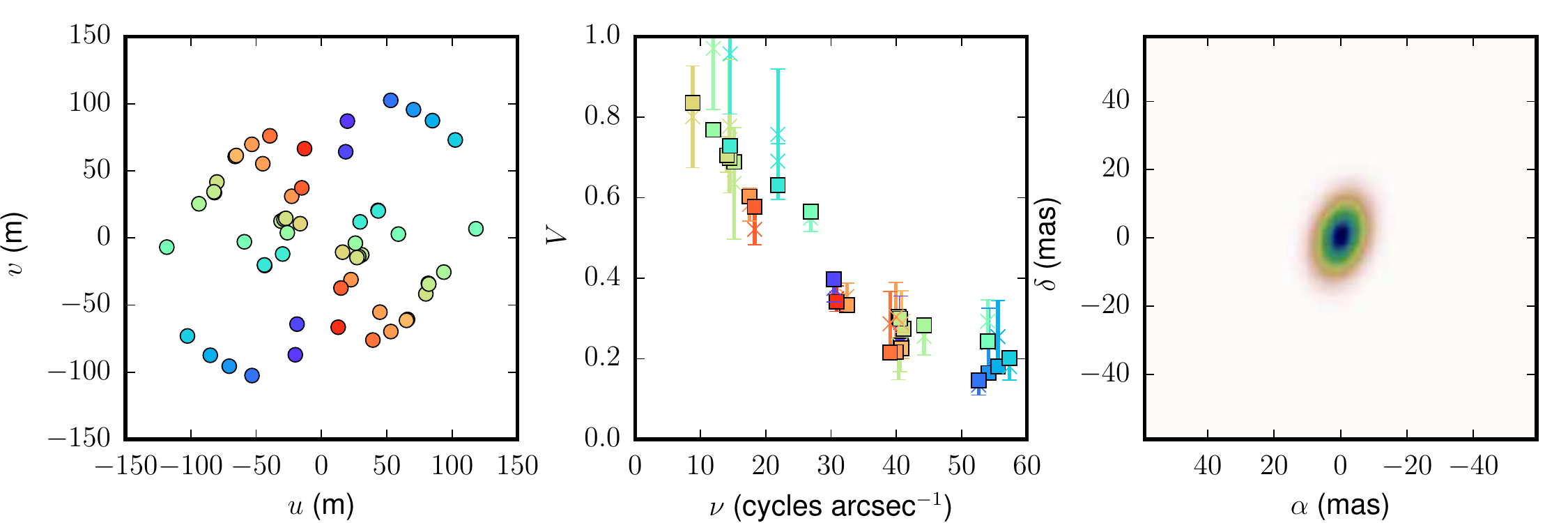}}
\caption{The mid-IR geometry of the high-mass YSO V\,921\,Sco. {\it
    Left panel:} The uv-plane coverage of the VLTI observations  
  with the MIDI instrument. {\it Middle panel:} The visibility
  amplitude as a function of spatial frequency. The colour of the
  points corresponds with the position angle of the measurement. {\it Right panel:}
  Geometrical representation of the visibility function which consists of a
  compact 2D Gaussian combined with 1D Gaussian halo. Adapted from \citet{Boley2013}.
\label{fig-midi}}
\end{figure*}

At  near-IR wavelengths, scattering and envelope extinction hamper
the direct detection of disk emission. At longer, mid-IR wavelengths,
the contribution by the warm envelope dust ramps up, slowly drowning
the number of disk photons. There is a 'sweet spot' in terms of wavelength between K and N-band
where the circumstellar disk is prominent and may even dominate the total
light. Interferometry of embedded early B and late O-type (proto)stars using the MIDI
beam-combiner in the $N$-band probing $\sim$100\,au clearly
demonstrates the absence of circular symmetry (see Fig.\,\ref{fig-midi}), which could be
interpreted as due to a disk or due to outflow cavity wall
emission \citep{Boley2013}. Subsequent RT modeling in
a few studies suggests that the
geometry of the $N$-band emission is consistent with that of an equatorially
flattened structure \citep{Follert2010,Grellmann2011}. The power of
interferometry is that the various contributions to the correlated fluxes is
the flux weighted by the size; at milli-arcsecond resolution the envelope
emission is almost resolved out, and any contribution by an unresolved
structure (e.g., a disk) will become evident. Such an interpretation of MIDI
observations of the well-known embedded late O-type star AFGL\,2136
is presented by \citet{DeWit2011}. An increase in
visibilities at the blue edge of the N-band (see their Fig.~1)
suggests the presence of a geometrically thin, optically thick viscous accretion disk located
within ($\rm <170$\,au) the dusty envelope. The disk would be accreting at a rate of $\rm 3 \times
10^{-3}\,M_{\odot}\,yr^{-1}$ in order to provide sufficient photons
from a compact object.  A remarkable detail in relation to the 
HAeBe star luminosity-size relation revealed by $N$-band,
short-spacing Keck data \citep{Monnier2009}, is
that the rim of the dust envelope is found at about 7 times the dust
sublimation radius. The central dust-free zone could have been evacuated by
the ionized stellar wind as reported in \citet{Menten2004}.

\subsubsection{Disks around early O-type (proto)stars}
\label{toroids}
Observational evidence of circumstellar structures in the most massive
(proto)stars, i.e., those with $L_{\rm bol}>$$10^5$~$L_\odot$, is
growing thanks to the increase in sensitivity and angular resolution of 
radio/mm interferometers.  Young stellar objects with derived masses of over
30\,$M_\odot$ are surrounded by large structures for which a
rotational velocity field could be established.
These structures are characterized by a much higher mass and larger
size than the rotationally supported disks encountered in the lower
mass YSOs, i.e., the objects discussed in the previous subsections.
In order to make a clear distinction, these large, dense, and massive structures
are referred to as ``toroids'' \citep[e.g.,][]{Beltran2005}.  Taken
together, the high masses and relatively large sizes suggest that the
toroids are hosting not just a single massive star but rather a
(proto)cluster of stars \citep[e.g., G29.96$-$0.02:][]{Beuther2007}.

The toroids have radii of a few 1000~au to up 10$^4$~au for the most luminous
sources \citep[e.g.,W51 North:][]{Zapata2008} and are found at typical distances
of $d>$3--4~kpc. Because of that, spatial scales of $<$1000~au can only be
traced through VLBI observations of maser emission  at milli-arcsecond
resolution. However, only for the O-type (proto)star 
W51e2--E the methanol maser features appear located at $<$1000\,au from the central star
(Surcis et al., 2012\nocite{Surcis2012}). What is more, in most cases the proper motions 
of the methanol maser spots do not indicate rotation but expansion
\citep[e.g.,][]{Li2012,Moscadelli2013}.

The hydrostatic scale height of these toroids, estimated following
Eq.~\ref{height} in Sect.~\ref{density-height}, is $>$50\% of the radius,
suggesting that these structures are geometrically thick. Indeed, they are much
thicker than any of the objects discussed so far. The masses of the toroids 
$M_{\rm toroid}$ are a few 100~$M_\odot$ (see
Table~\ref{OBprop}), and because $M_{\rm toroid}$ is much higher than the mass
of the central star, the toroids are not expected to be in Keplerian rotation.  In
fact, they 
are probably self-gravitating structures in
solid-body rotation or, in case the toroids are infalling, pseudo-disks \citep[see e.g.,][]{Allen2003}. Like for the disks around IM
and B-type (proto)stars, the velocity field of the toroids are detected through
observations of high-density tracers at centimeter and (sub)millimeter
wavelengths and the velocity gradients are found to be perpendicular to
molecular outflows (e.g, G29.96$-$0.02: Beltr\'an et al.,
2011a\nocite{Beltran2011a}; W51 North: Zapata et al., 2009\nocite{Zapata2009};
G20.08$-$0.14N: Yu \& Wang 2013\nocite{Yu2013}). Attempts to characterize the
velocity field has proven to be impossible hitherto. For G31.41+0.31,
\citet{Beltran2005} attempted to fit the velocity field with a constant rotation
velocity, $V_{\rm rot}$=constant, (self-gravitating in the absence of magnetic
field) and a constant angular velocity, $\Omega$=constant, which implies $V_{\rm rot}\propto R$, (solid-body
rotation). These authors were not able to discriminate between them  at the
available spatial resolution of the observations. 

Some insight into what the toroids constitute can be obtained from
evaluating the dynamical mass. The dynamical mass, estimated as
$M_{\rm dyn}=V_{\rm rot}^2\,R/G \sin^2 i$ (where $i$ is the
inclination angle, which is 90$\degr$ for an edge-on toroid), are
smaller than $M_{\rm toroid}$. This suggests that toroids might be
dynamically unstable to fragmentation and gravitational collapse.
The toroids
could therefore be transient structures 
with lifetimes of the order of the free-fall time, which, taking into account that typical
densities of toroids are $\sim$10$^7$~cm$^{-3}$, would be
$\sim$10$^4$~yr.

\begin{figure*}
\centerline{\includegraphics[angle=0,width=11cm]{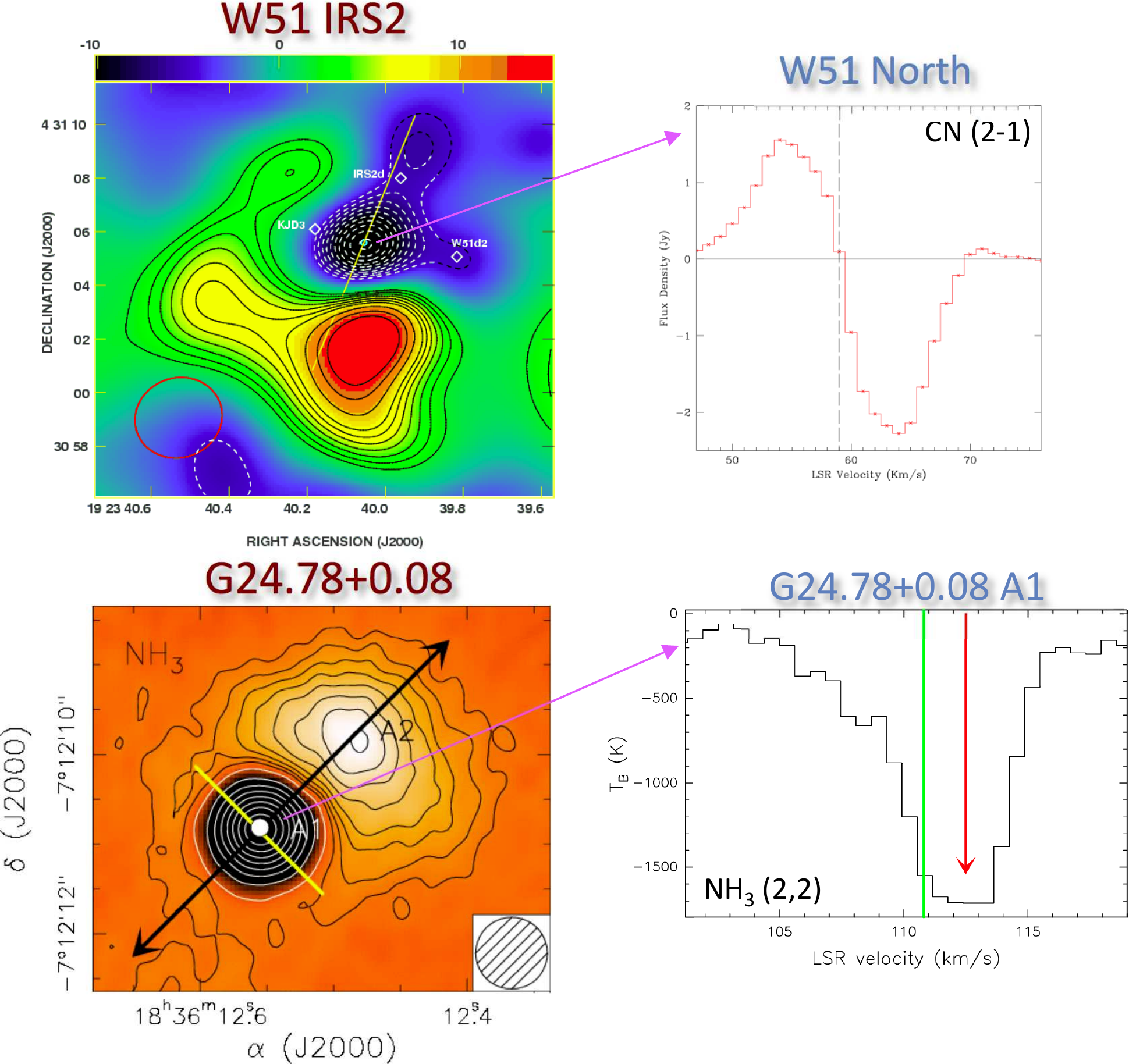}}
\caption{{\it Top panels:} Map of the CN (2--1) integrated line emission of
the high-mass star-forming region W51 IRS2. Black contours indicate positive 
intensity (emission), while white contours are negative (absorption). 
The green circle indicates the position of the dust continuum
emission source W51 North, while the 
yellow  line outlines the orientation of the bipolar 
molecular outflow. The spectrum of CN (2--1) towards the center of W51 
North is shown in the right panel.  The absorption peak is red-shifted with respect to 
the systemic
velocity (dashed vertical line) suggesting infall. From \citet{Zapata2008}.
{\it Bottom panels:} Map of  the NH$_3$ (2,2) integrated line emission of the
high-mass star-forming region G24.78+0.08. Black contours 
indicate positive intensity (emission), while white contours are negative 
(absorption). The white filled circle marks the position of the hypercompact 
\HII region associated with G24.78+0.08 A1, while the black arrows 
outline the direction of the bipolar outflow in the region. The spectrum of the NH$_3$ (2,2) 
satellite line towards G24.78+0.08 A1 is 
shown in the right panel.  The absorption peak indicated
by the red arrow is 
red-shifted with respect to  the systemic
velocity (green vertical line) suggesting infall. 
Adapted from \citet{Beltran2006a}}
\label{fig-infall}
\end{figure*}
 
This inference is supported by evidence for infall detected
towards some of these toroids. Infall was
identified by means of inverse P-Cygni profiles due to red-shifted
absorption against embedded
\HC\ and \UC\ regions (e.g., G10.62$-$0.02: Keto et al., 1988\nocite{Keto1988}; Sollins et
al., 2005\nocite{Sollins2005}; W51e2 $-$E: Zhang \& Ho
1997\nocite{Zhang1997}) or against bright dust continuum emission
(e.g., W51 North: Zapata et al., 2008\nocite{Zapata2008}; see
Fig.~\ref{fig-infall}; G31.41+0.31:
Girart et al., 2009\nocite{Girart2009}). Red-shifted absorption has
also been observed towards late O-type (proto)stars, like for example
in G24.78+0.08 A1 \citep{Beltran2006a} (see Fig.~\ref{fig-infall}) or
NGC7538\,IRS1 \citep{Beuther2012}. The infall rates estimated are comparatively
high, of the order of 10$^{-3}$--10$^{-2}$~$M_\odot$\,yr$^{-1}$, and
they could be high enough to quench the formation of an \HII
region \citep{Yorke1986,Walmsley1995} or to slow down its expansion. What is
more, according to 
the models of \citet{Keto2002}, accretion could proceed through trapped 
\UC\ regions in the form of ionized accretion flows.

\section{Evolution}
\label{evolution}

In this section we discuss disk evolution and mass accretion rate as a
function of stellar mass and of time. To do 
this we compare the properties of disks and toroids around IM and HM
(proto)stars with those of disks around the more evolved Herbig Ae/Be stars. To
study the similarities and differences of the accretion process, we
evaluate the mass accretion rates estimated for IM and HM (proto)stars
alongside those estimated for low-mass Class~0/I YSOs and for more
evolved T\,Tauri and Herbig Ae/Be stars. 

\subsection{Disk evolution}
\label{disk-evolution}
\begin{figure*}[h]
\centerline{\includegraphics[angle=0,width=9cm]{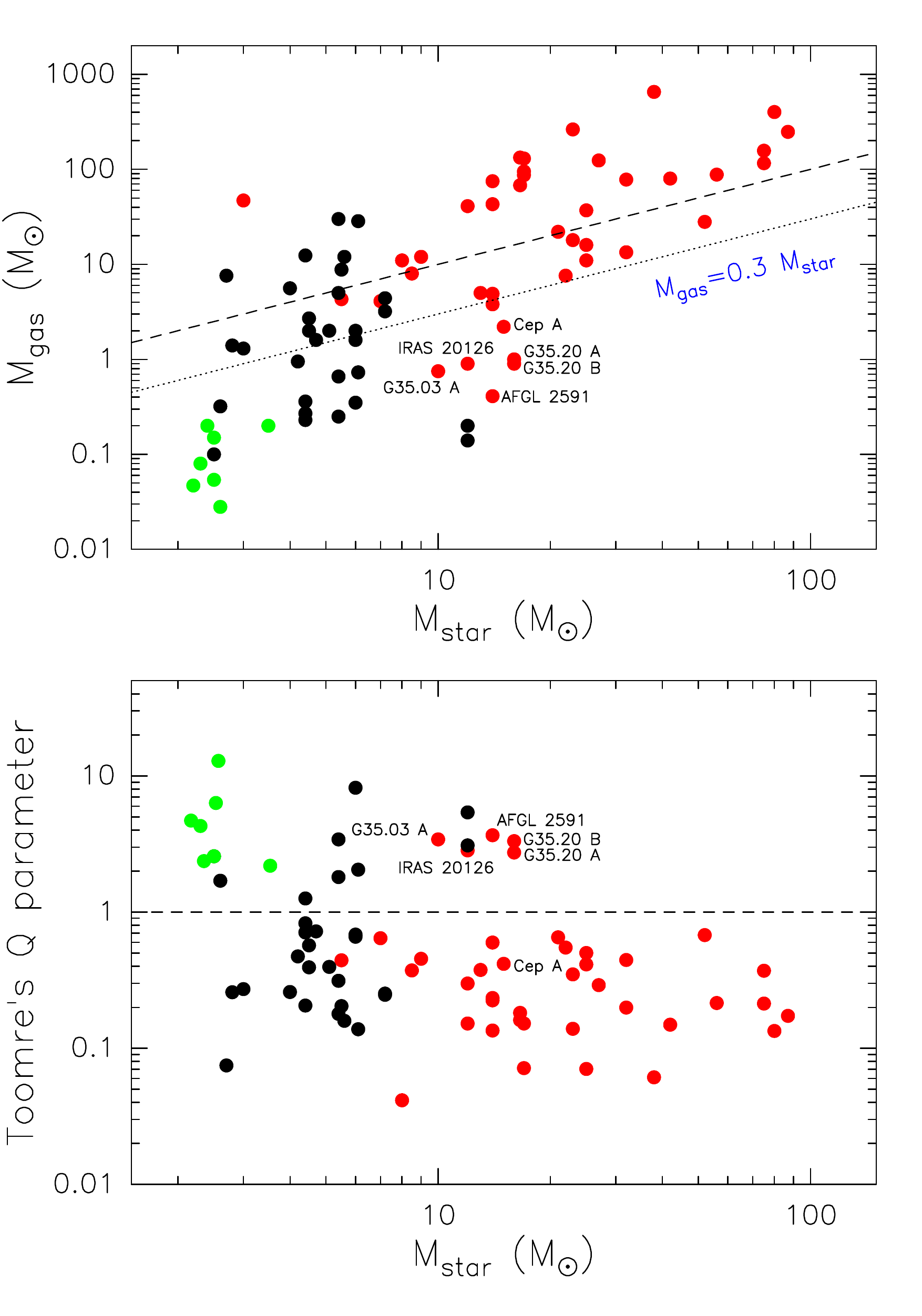}}
\caption{({\it Upper panel)} Gas mass $M_{\rm gas}$ of the circumstellar structures as
a function of the stellar mass $M_\star$ for the IM (black dots)
protostars (Table~\ref{IMprop}) and the HM (red dots) (proto)stars 
(Table~\ref{OBprop}). Green dots correspond to Herbig Ae stars (see
\S~\ref{sect412}). The dotted black line indicates $M_{\rm
gas}=0.3\,M_\star$, the maximum disk mass to allow for disk stability 
\citep{Shu1990}, and the dashed black line 
indicates $M_{\rm
gas}=M_\star$. The labels correspond to the
most likely Keplerian disk candidates around B-type (proto)stars. ({\it Lower
panel)} Toomre's $Q$ parameter \citep{Toomre1964} as a function of 
$M_\star$. The symbols are the same as in the upper panel. The black dashed
line indicates  $Q=1$.} 
\label{fig-stability}
\end{figure*}

A graphical overview of the gas mass $M_{\rm gas}$ as a function of
stellar mass $M_\star$ is provided in Fig.~\ref{fig-stability}. The
stellar mass spans the range from one to 100\,$M_\odot$. The graph
contains the re-derived values for $M_{\rm gas}$ of the previously
discussed embedded objects and also those of the pre-main sequence
Ae star disks, {\it viz.} HD~163296, AB~Aur, MWC~758, HD142527,
MWC~480, HD~100546 ~\citep[data from][]{Alonso-Albi2009}, and HD~34282
\citep{Pietu2003}. For a discussion on the uncertainties in $M_{\rm
  gas}$ see Sect.\,\ref{sect32}. The figure shows clearly that the mass of the circumstellar structure (disk or
toroid) is proportional to the mass of the central object (upper panel
of Fig.~\ref{fig-stability}). Although this trend may be expected a
priori, it is in apparent contrast with what is observed among 
the Herbig~Ae/Be stars. Albeit for a smaller stellar mass range, the disk
mass of the B-type PMS stars are approximately ten times less than that
of A-type  stars \citep[e.g.][]{Fuente2003,Alonso-Albi2009}, with some
noticeable exceptions like the FU\,Ori star Z\,CMa.

The results on the HAeBe star disks may reflect the increasingly shorter disk
dispersal timescale for more massive stars
\citep{Alonso-Albi2009}. Indeed, the ionizing stellar radiation of the
Herbig Be stars could photoevaporate the disk rapidly
\citep{Hollenbach2000}. An alternative explanation involves tidal
interactions in a crowded birth environment leading to disk truncation
\citep{Munoz2015}. The fact that the disks of embedded YSOs display an opposite
trend to that seen among the HAeBe stars could therefore mean that
photo-evaporation is not strong enough to significantly affect the
disk (or toroids) on a timescale of a few 10$^5$~yr, the typical 
formation timescale for high-mass stars \citep{McKee2002}. 
In fact, as pointed out by \citet{Cesaroni2006}, the mass loss rate
due to photo-evaporation would be 100--1000 times lower than the
accretion rate onto the star based on models by \citet{Richling1997}
and \citet{Hollenbach2000}.

\subsubsection{Stability of the disk as a function of mass}

The distribution of points in the upper panel of Fig.~\ref{fig-stability} 
demonstrates that about 50\% of the circumstellar structures around IM and HM
(proto)stars have $M_{\rm gas}$ higher than the mass of the central star. As a
result, these objects are likely to be self-gravitating. These structures cannot
be in Keplerian rotation because the gravitational potential of the system is
dominated by the circumstellar structure itself and not by the stellar mass.
Based on an analytical study of gravitational instabilities in gaseous disks,
\citet{Shu1990} conclude that accretion disks can be stable only if their masses
are less than 0.3\,$M_\star$. For $M_{\rm gas}>0.3\,M_\star$, theory predicts
that gravitational instabilities that induce spiral density waves
appear, leading to a rapid fragmentation of the disk \citep{Laughlin1994,Yorke1995}. The
upper panel of Fig.~\ref{fig-stability} shows that all the circumstellar
structures around the more evolved Herbig~Ae stars have $M_{\rm gas}\ll0.3\,M_\star$,
and therefore should be  gravitationally stable following this
argument. 

On the other hand, only half of
the IM protostars and a handful of HM (proto)stars have $M_{\rm
gas}<0.3\,M_\star$. Although this could be a real effect, the most plausible
explanation is that the angular resolution of
the observations is not high enough to properly separate the disk emission from
that of the inner part of the envelope. In such cases,  $M_{\rm gas}$ has to be
taken as an upper limit. It is expected that higher angular resolution
($\ll 0\farcs5$) observations may reveal  stable circumstellar disks
around most of these sources, given the confirmed presence of outflows powered by
them. 

All the HM circumstellar structures  with $M_{\rm gas} < 0.3\,M_\star$ are around B-type
(proto)\-stars and correspond to the best accretion disk candidates for which
molecular line observations have revealed (quasi-)Keplerian motions
(see Sect.\,\ref{Bprop}). For the most massive early O-type (proto)stars, 
the fact that all circumstellar structures are gravitationally
unstable indicates that the toroids could
be indeed susceptible to gravitational instabilities and fragment in
different cores as observations at increasingly higher angular resolution
suggest \citep[e.g., G29.96$-$0.02:][]{Beuther2007}.

An additional way to assess the local stability of a circumstellar
disk is by evaluating Toomre's parameter $Q$ 
\citep{Toomre1964}. It is defined as
\begin{equation}
Q=\frac{c_s\,\kappa}{\pi\,G\Sigma}, 
\end{equation}
where $c_{\rm s}$ is the sound speed, $\Sigma$ the surface density,
and $\kappa$ is the epicyclic frequency of the disk. The epicyclic frequency is
directly proportional to the angular velocity $\Omega$ and can be expressed as 
$\kappa=f\Omega$, where the proportional factor $f$ depends on the rotation
curve. In case of Keplerian rotation, $f$=1 and therefore, $\kappa\simeq\Omega$.
According to this
stability criterion, a thin disk becomes
unstable against axisymmetric gravitational instabilities if $Q<1$.
 Following \citet{Cesaroni2007}, we estimated $\Omega$ as 
 $\sqrt{G\,M_{\rm total}/R^3}$, where $M_{\rm total}$ is the total
(star plus disk) mass of the system and $R$ is the radius of the structure
(Tables~\ref{IMprop} and \ref{OBprop}). The surface density was estimated 
as $\Sigma=M_{\rm gas}/\pi\,R^2$. For the dust temperature of the
Herbig~Ae stars, needed to estimate $c_{\rm s}$, we used the average
disk temperature per stellar spectral type as reported by
\citet{Natta2000}.  For the IM and HM (proto)stars, the dust
temperature is reported in the references given in Tables~\ref{IMprop}
and \ref{OBprop}. Structures with $Q>1$ should be locally stable.

The lower panel of Fig.~\ref{fig-stability} shows
Toomre's $Q$  parameter as a function of $M_\star$.  The distribution of points 
demonstrate  
that the disks around the Herbig~Ae stars are stable against collapse.
In contrast, only about one third of the circumstellar structures
around the embedded IM protostars appear Toomre stable. For the sources
that are unstable we cannot discard the possibility that $Q$ has been
underestimated as $R$ and $M_{\rm gas}$ might be overestimated at
insufficient angular resolution if the inner envelope emission is not properly resolved
from that of the disk. At higher stellar masses, except for Cepheus
A~HW2, the only circumstellar structures showing stability are the same for which we find that
$M_{\rm gas}<0.3\,M_\star$. They correspond to the bona-fide protostellar disk
candidates around B-type (proto)stars, discussed previously. According to
\citet{Laughlin1994}, these disks are gravitationally stable (see however Cesaroni
et al., 2007\nocite{Cesaroni2007} about the role of heating and cooling on the
stability of these disks).  Nonetheless, Fig.~\ref{fig-stability} shows
that the majority of the disks and 
toroids detected around embedded HM (proto)stars would be unstable
against axisymmetric instabilities.  

\subsubsection{Dynamical status as a function of mass}

\begin{figure*}
\centerline{\includegraphics[angle=0,width=9cm]{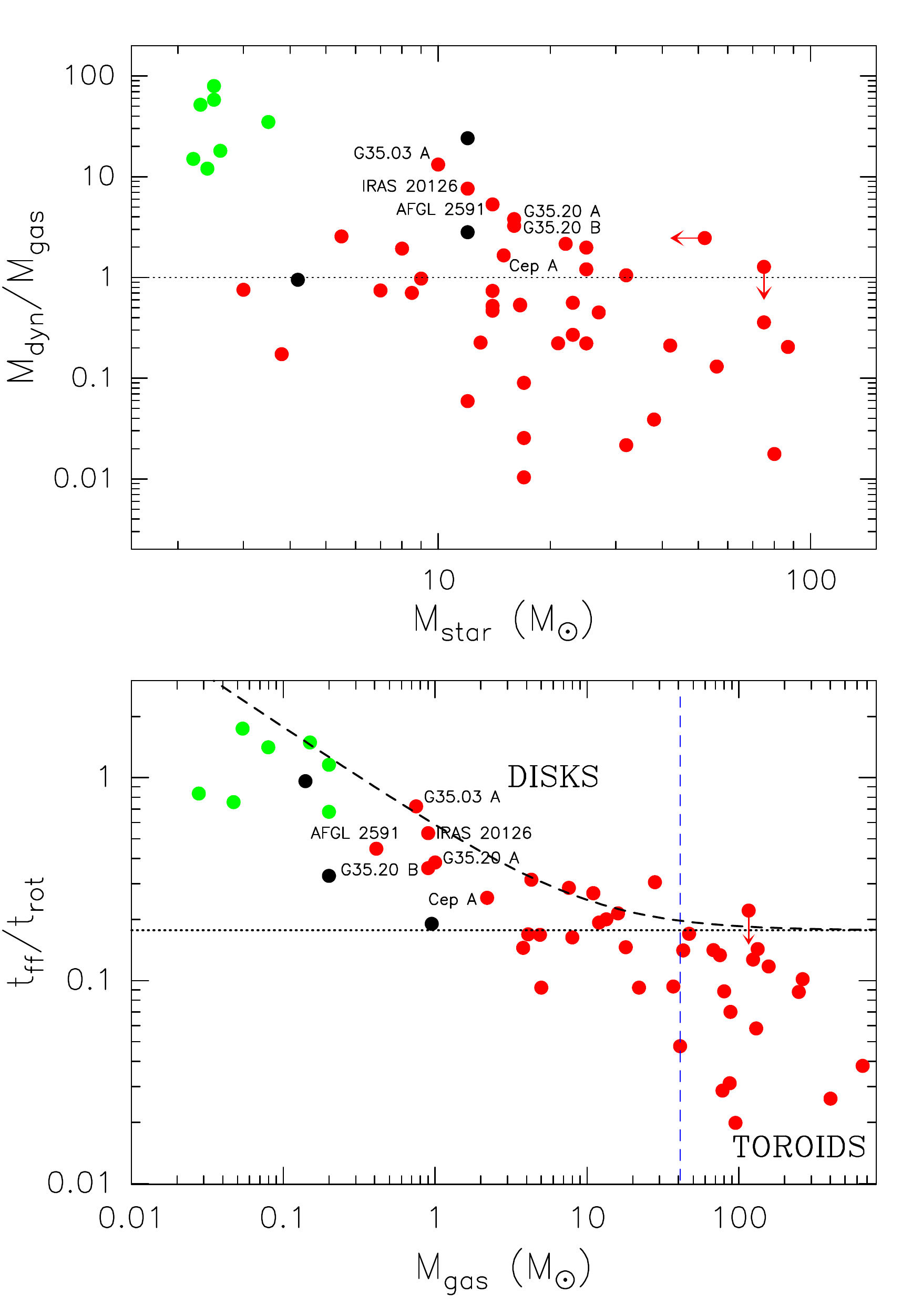}}
\caption{({\it Upper panel}) Dynamical mass to gas mass ratio $M_{\rm dyn}/M_{\rm
gas}$ as a function of $M_\star$ for the rotating structures around 
high-mass (red dots) (proto)stars (Table~\ref{OBprop}) and for the few
intermediate-mass (black dots) protostars 
(Table~\ref{IMprop}) with rotation velocity information available.  Green dots correspond to Herbig Ae stars (see
\S~\ref{sect412}). The dotted black line indicates $M_{\rm dyn}=M_{\rm
gas}$. The labels correspond to the
most likely Keplerian disk candidates around B-type (proto)stars. ({\it Lower
panel}) Free-fall timescale to rotational period ratio, $t_{\rm ff}/t_{\rm
rot}$, versus $M_{\rm gas}$, of rotating disks or
toroids around high-mass (proto)stars (red dots), intermediate-mass protostars
(black dots), and Herbig Ae (green dots). The labels correspond to the
most likely Keplerian disk candidates around B-type (proto)stars. Black dotted
and dashed lines correspond to the 
theoretical values of $t_{\rm ff}/t_{\rm
rot}$ for spherical clouds of mass $M_{\rm gas}$ containing a star 
of mass $M_\star$ at the center, in which the gas is rotationally supported against
the gravity of both the gas plus the star (see $\S$~\ref{disk-evolution}). 
 The dotted black line correspond to  
$M_\star$=0~$M_\odot$ and the dashed to 10~$M_\odot$. The blue dashed line 
indicates a mass of 40~$M_\odot$. 
The rotating structures with masses higher than this value are toroids.}
\label{fig-dynamics}
\end{figure*}

One way of assessing the dynamics of the massive circumstellar
structures surrounding young stars is by evaluating the dynamical
mass $M_{\rm dyn}$ needed for equilibrium (see Sect\,\ref{toroids}). 
$M_{\rm dyn}$ is estimated assuming equilibrium between
centrifugal and gravitational forces.  In the upper panel of
Fig.~\ref{fig-dynamics} we plot the ratio of $M_{\rm dyn}$ to the gas 
mass $M_{\rm gas}$ as a function of $M_\star$ for the same sample of
(proto)stars as in Fig.~\ref{fig-stability}. To evaluate $M_{\rm
dyn}$, the rotation velocity $V_{\rm rot}$ is required. This parameter
is available only for three sources of the sample of IM protostars in
Table~\ref{IMprop}, but for all forty-one HM (proto)stars of the
compiled sample (see Table~\ref{OBprop}).

One can differentiate between two regions in the upper panel of
Fig.~\ref{fig-dynamics}. One for YSOs with
$M_\star\lesssim20\,M_\odot$ that have mostly $M_{\rm dyn}/M_{\rm 
gas}>$1. Such objects are expected to be centrifugally supported. The
other region corresponds to the sources with 
$M_\star\gtrsim 20\,M_\odot$ and $M_{\rm dyn}/M_{\rm gas}<$1. The
Herbig Ae disks are characterized by $M_{\rm dyn}/M_{\rm gas} \gg$1, as
expected.  In fact, the velocity
field of the disk of HD~163296 traced in $c$-C$_3$H$_2$ and in CO~(3--2) with 
ALMA Science Verification
observations was successfully modelled with Keplerian rotation by \citet{Qi2013} and 
\citet{DeGregorio-Monsalvo2013} (see Fig.~\ref{fig-hd163296}). The
three IM protostars that populate this graph have a ratio $\gtrsim$1, and therefore,
could be centrifugally supported. The two IM sources with the highest masses are OMCS-1
139--409 and 134--411, which were discussed in detail in Sect.\,\ref{disks-IM-prop}. 
Their bolometric luminosities, and therefore the estimated
$M_\star$, should be taken as an upper limit because they correspond to the
luminosity of the
whole OMC1-S region \citep{Zapata2007}. Unfortunately, the line
observations towards these three sources do not have enough angular
resolution and sensitivity to properly constrain the nature of the velocity field. 

Moving to the high-mass regime,
one finds that for early-B and late-O type (proto)stars
($M_\star\lesssim$ 25--30~$M_\odot$), the rotating structures could be
centrifugally supported. In fact, the structures with a higher $M_{\rm
dyn}/M_{\rm gas}$ ratio correspond to the best accretion disk candidates
around B-type (proto)stars for which molecular line observations have revealed
(quasi-)Keplerian motions (see Sect.\,\ref{Bprop}; Fig.~\ref{fig-g35}). For 
the early-B and late-O type
sources that have $M_{\rm dyn}/M_{\rm gas}<$1, the most plausible explanation
is that the observations do not have enough angular resolution to properly resolve
the disk emission. Finally, for the highest stellar mass regime, the
early-O type (proto)stars ($M_\star\gtrsim$ 30~$M_\odot$), 
most rotating structures have $M_{\rm dyn}/M_{\rm gas}\ll$1. This
confirms that the massive toroids cannot be centrifugally
supported. These structures are possibly undergoing  gravitational
collapse (see Sect.\,~\ref{toroids}). Some of the sources in the 
figure warrant remarks, for instance the sources with $M_\star>$
30~$M_\odot$ that have $M_{\rm dyn}/M_{\rm gas}>$1, namely W51e8 and G20.08$-$0.14N. 
For W51e8, $M_{\rm dyn}$
should be taken as an upper limit because the rotation velocity of 4~\kms\
estimated from the OCS velocity gradient could be contaminated by outflow emission 
\citep{Klaassen2009}. For G20.08$-$0.14N, the spectral type
estimated from the Lyman-continuum photons of the \HII region is O7.5
\citep{Galvan-Madrid2009} that
corresponds to a stellar mass of $\sim$25~$M_\odot$ \citep{Davies2011,Mottram2011}
instead of 52~$M_\odot$ estimated from cluster
simulations (see Table~\ref{OBprop}). Therefore, $M_{\rm dyn}/M_{\rm
gas}\gtrsim$1 would be consistent with what is found for similar late-O
 type sources.

Following \citet{Beltran2011a}, who made a comparative study of
the stability of the rotating structures around B-type and O-type stars, in the
lower panel of Fig.~\ref{fig-dynamics}  we plotted the ratio of the free-fall
timescale ($t_{\rm ff}$) to the rotational period  ($t_{\rm rot}$) versus $M_{\rm
gas}$ for the same YSOs as in the upper panel plot. $t_{\rm ff}$ is proportional
to the dynamical timescale needed to refresh the material of the rotating
structure and $t_{\rm rot}$ is the rotational period at the outer radius, which is basically the timescale needed by the rotating structure to stabilize 
after incorporating new accreted material. Unlike 
\citet{Beltran2011a}, who use the radius measured from the dust
continuum emission to estimate $t_{\rm ff}$ and $t_{\rm rot}$, we used the radius
estimated from line emission when possible. As \citet{Beltran2014} point out,  
because $t_{\rm ff}$ has been estimated
assuming spherical symmetry, the $t_{\rm ff}/t_{\rm rot}$  ratio for both disks
and toroids could be overestimated. However, because disks and toroids around
embedded (proto)stars are geometrically thick (see \S~\ref{obs-prop}), the
correction factor, which depends on the scale height of the structures, should be
small. The correction factor would be $\sim$2 if the height of the disk is 30\% 
the disk radius and 1.6 if it is 50\%. 
The curves in 
Fig.~\ref{fig-dynamics} (lower panel) are the  theoretical $t_{\rm ff}/t_{\rm rot}$ curves for
spherical clouds of mass $M_{\rm gas}$ containing a star of mass  $M_\star$ at the
center, in which the gas is rotationally supported against the  gravity of the gas
plus the star. These curves can be expressed as $t_{\rm ff}/t_{\rm rot}=[(M_{\rm
gas}+M_\star)/32\,M_{\rm gas}]^{1/2}$. The curves plotted correspond to
$M_\star$=0~$M_\odot$ and 10~$M_\odot$.  

The lower panel of Fig.~\ref{fig-dynamics} confirms the results of
\citet{Beltran2011a}, namely, that disks and toroids are kinematically and
dynamically different structures, with a twice as large sample of embedded HM
(proto)stars and including IM protostars and Herbig Ae stars. In fact, this plot
confirms that disks around B-type (proto)stars and toroids around O-type
(proto)stars occupy two distinct regions of the plot, with disks having a higher
$t_{\rm ff}/t_{\rm rot}$ ratio and lower $M_{\rm gas}$. What is more, disks
around B-type (proto)stars are dynamically similar to those found around IM
protostars and more evolved Herbig Ae stars. Therefore, the faster the rotation
of the structure the more similar to a true accretion disk is, because due to
the shorter rotation timescales, the infalling material has enough time to
settle into a centrifugally supported disk. In fact, as shown in
Fig.~\ref{fig-dynamics}, the disks around B-type (proto)stars that have $t_{\rm
ff}/t_{\rm rot}>1$ are the Keplerian disk candidates. On the other hand, the
rotation timescale for the more massive toroids is so large that the infalling
material does not have enough time to reach centrifugal equilibrium. In this
case, the rotating structure is a transient toroid for which rotation plays a
little role in supporting it. As already mentioned, these massive toroids could
be self-gravitating structures probably supported by thermal pressure,
turbulence and/or magnetic fields or pseudo-disks if these toroids are
infalling. 

\subsection{Mass accretion rate evolution}
\label{evolution-macc}

\subsubsection{Infall rates vs accretion rates}

In Sect.\,\ref{macc} we have described different ways to
estimate the mass accretion rate onto IM and HM (proto)stars. We 
noted that most of these parameters actually measure the mass infall
rate $\dot M_{\rm  inf}$. This quantity corresponds to motion of material
belonging to the larger scale toroid or envelope and not to the
disk. The disk material will be accreted onto the central star at rate 
\macc\ . To avoid confusion, we
use $\dot M_{\rm inf}$  when discussing infall rates estimated
with one of these methods. Comparing $\dot M_{\rm inf}$ with \macc\
can provide an idea on how the material infalling onto the disk or
toroid is incorporated into the central accreting (proto)star.

\begin{figure*}
\centerline{\includegraphics[angle=-90,width=11.5cm]{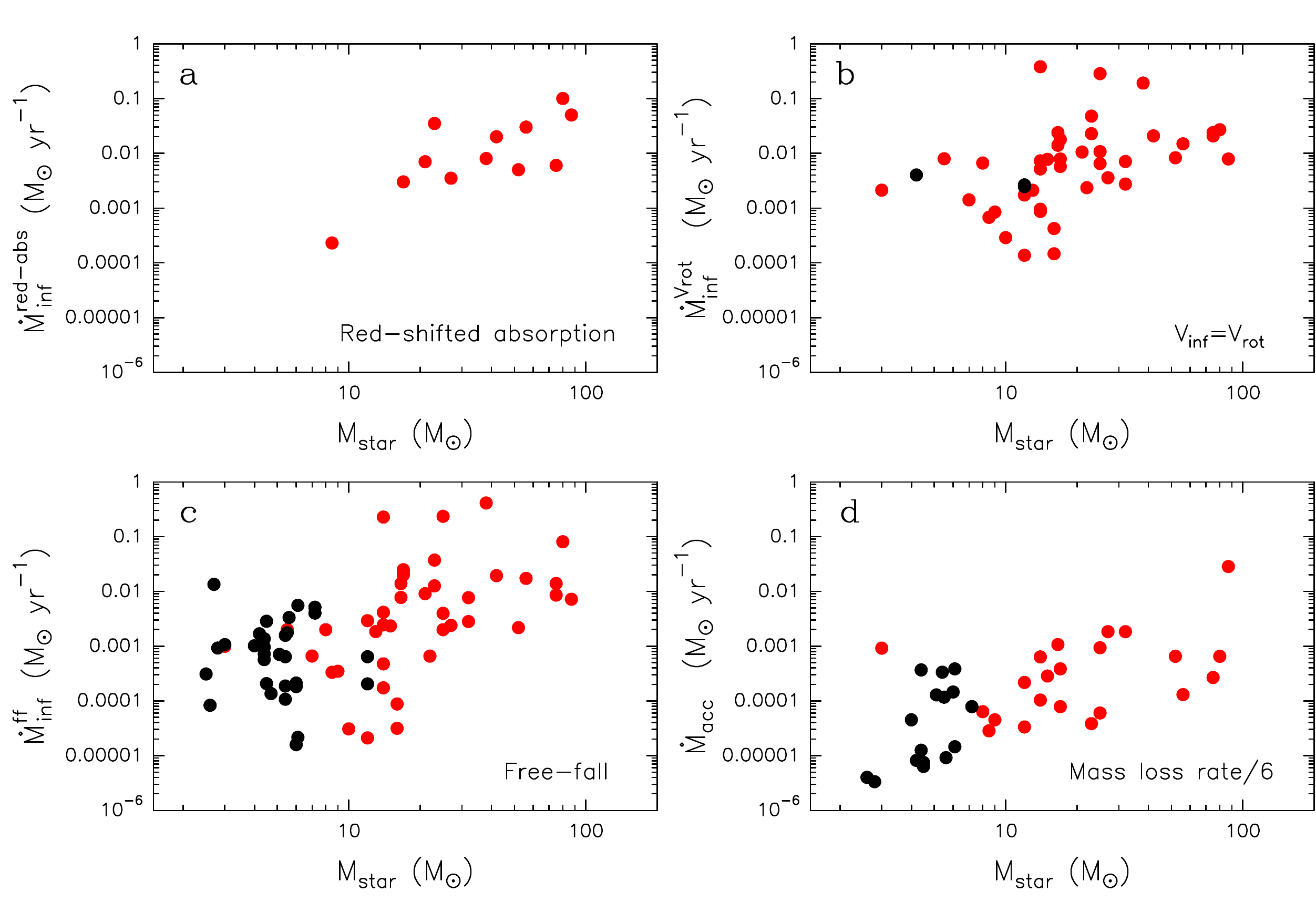}}
\caption{ 
Mass infall and accretion rate as a function of $M_\star$, where the infall
 rate has been estimated ({\it a}) from red-shifted absorbed
line profiles, $\dot M_{\rm inf}^{\rm \, red-abs}$ (see Eq.~\ref{eq-minf}), for the rotating structures
around HM (proto)stars for which red-shifted absorption has been
detected; ({\it b}) assuming that the infall velocity is equal to the
rotation velocity, $\dot M_{\rm inf}^{\rm \, v_{\rm rot}}$ (see Eq.~\ref{eq-mvrot}), for the rotating structures
around HM (red dots) (proto)stars (Table~\ref{OBprop}) and for
the few IM (black dots) protostars (Table~\ref{IMprop})
with rotation velocity information available; and ({\it c}) assuming that
all the material in the rotating structure will collapse in a
free-fall time $t_{ff}$,  $\dot M_{\rm inf}^{\, ff}$ (see Eq.~\ref{eq-mfreefall}), for all the IM
(black dots) and HM (red dots) (proto)stars (Tables~\ref{IMprop}
and \ref{OBprop}). The mass accretion rate $\dot M_{\rm acc}$ ({\it d}) has been
estimated from the outflow mass loss rate $\dot
M_{\rm out}$ (see Eq.~\ref{eq-mout}), for those IM (black dots) and
HM (red dots) (proto)stars for which interferometric observations
of molecular outflows are available (Tables~\ref{IMprop} and
\ref{OBprop}).}
\label{fig-minfall}
\end{figure*}

The four panels of Fig.\,\ref{fig-minfall} show the mass infall rate
estimated by
three different methods and the mass accretion rate estimated from the
outflow mass-loss rate (see Sect.\,\ref{macc}). These quantities are
presented as function of the stellar mass $M_\star$
(Tables~\ref{IMprop} and \ref{OBprop}). Panel\,a depicts the infall
rate $\dot M_{\rm inf}^{\rm \, red-abs}$ derived from observations of red-shifted
absorption features in spectrally resolved molecular line profiles
(see Eq.~\ref{eq-minf}). To generate such absorption components
a background source surrounded by a much colder envelope is required.
The background source can either be a bright (hyper)compact \HII
region or a dust continuum emission source. This requirement renders this
method limited in its application and for only a few
HM (proto)stars a red-shifted absorption component has been observed. 
Panel\,b presents
the mass infall rate $\dot M_{\rm inf}^{\rm \, v_{rot}}$ estimated by assuming that the
infall velocity $V_{\rm inf}$ is equal to the rotation velocity
$V_{\rm rot}$ (Allen et al., 2003\nocite{Allen2003}; see Eq.~\ref{eq-mvrot}). 
This quantity could
be estimated for all the HM (proto)stars and for the three IM
protostars with rotation velocity information available. The quantity
presented in panel\,c is $\dot M_{\rm inf}^{\, ff}$ which
assumes that all the material in the circumstellar 
structure $M_{\rm gas}$ will undergo infall in a free-fall time
$t_{ff}$ (see Eq.~\ref{eq-mfreefall}). Finally, in panel\,d the
mass accretion rate \macc\ is estimated from the outflow mass-loss rate 
$\dot M_{\rm out}$ assuming momentum rate conservation and a
relation between the jet mass-loss rate and the accretion rate (Tomisaka,
1998\nocite{Tomisaka1998}; Shu et al., 1999\nocite{Shu1999}; see
Eq.~\ref{eq-mout}).  This quantity was derived only for those YSOs for
which interferometric observations of 
molecular outflows are available. For a few cases, the direction and
number of molecular outflows associated with a YSO is not evident,
even with interferometric observations \citep[e.g.,
G31.41+0.31:][]{Cesaroni2011}. In such cases, not being sure which
were the powering sources of the molecular outflows, we refrained from
estimating \macc.

\begin{figure*}
\centerline{\includegraphics[angle=-90,width=10cm]{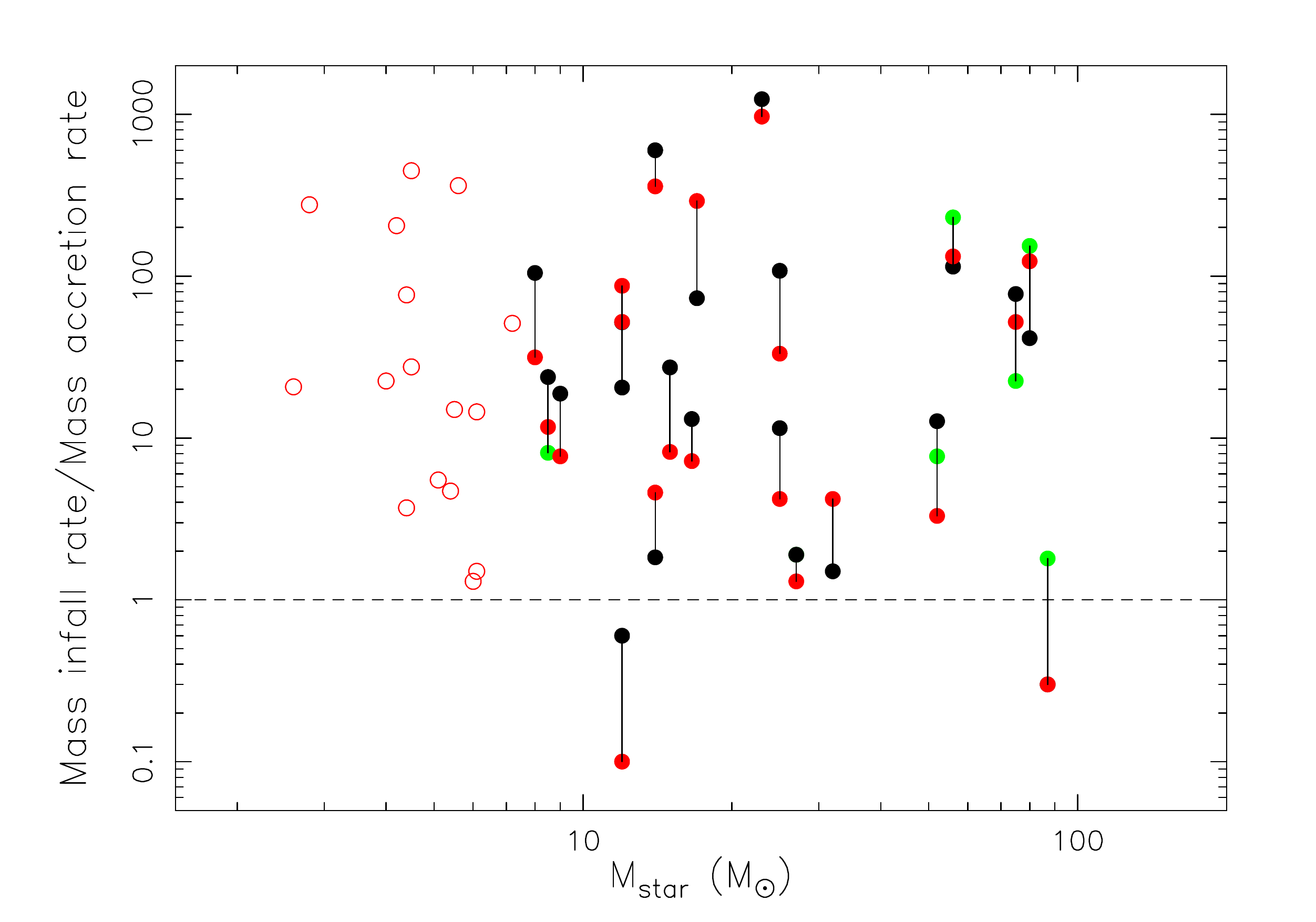}}
\caption{Mass infall rate $\dot M_{\rm inf}$ to mass accretion rate 
\macc\ ratio for all the IM
(open circles) and HM (filled circles) (proto)stars as a 
function of $M_\star$. $\dot M_{\rm inf}$ has been estimated with three
different methods (see Fig.~\ref{fig-minfall}): $\dot M_{\rm inf}^{\rm \, red-abs}$
(green circles), $\dot M_{\rm inf}^{\rm \, v_{\rm rot}}$ (black circles), and 
 $\dot M_{\rm inf}^{\, ff}$ (red circles). Black lines connect the $\dot M_{\rm
inf}/$\macc\ ratio obtained for the same source using different methods. The black
dashed line indicates $\dot M_{\rm inf}$=\macc.}
\label{fig-minfacc}
\end{figure*}

The conclusion that one can draw from Fig.~\ref{fig-minfall} is that
the infall rate increases with the mass of the central star. Moreover,
the infall rate has relatively high values no matter the method used to estimate
it. For those sources for which the infall rate has been
estimated with more than one method, the values of $\dot M_{\rm inf}$ are 
consistent within a
factor of $<$10 (see Fig.~\ref{fig-minfacc}). The
values of $\dot M_{\rm inf}$ are of the order of
$10^{-3}$--$10^{-2}$~\msolyr, being as high as 0.1~\msolyr\ for the
most massive O-type (proto)stars. On the other hand, the values
of \macc\ derived from $\dot M_{\rm out}$ are of the order of
$10^{-4}$-- $10^{-3}$~\msolyr. The infall rates appear therefore to be
much larger that the accretion rate. This observation is further
illustrated in Fig.~\ref{fig-minfacc}. Here we show the
$\dot M_{\rm inf}/$\macc\ ratio for all the IM and HM (proto)stars,
obtained using the three different methods to estimate $\dot M_{\rm
inf}$. Because $V_{\rm rot}$ is available only for three IM
protostars, we decided to use only $\dot M_{\rm inf}^{\, ff}$ for this mass
range sources. No red-shifted
absorption has been detected towards IM protostars, so there are no estimates of 
$\dot M_{\rm inf}^{\rm \, red-abs}$.
As seen in this figure, except for a couple of YSOs, $\dot
M_{\rm inf}$ is always higher than \macc, independently of the method
used to estimate $\dot M_{\rm inf}$. The ratio can be as high as $\sim$200
for $\dot M_{\rm inf}^{\rm \, red-abs}$. 

As already mentioned in Sect.\,\ref{macc}, the main uncertainty in estimating
the infall rate from red-shifted absorbed profiles is the radius at
which $V_{\rm inf}$ is measured. Most authors use the radius of the
 core, and therefore $\dot M_{\rm inf}$ should be taken as an
upper limit. However, this uncertainty in the radius at which
absorption occurs would possibly account only for a factor up to
$\sim$10 (e.g., G24.78+0.08 A: Beltr\'an et al.,
2006c\nocite{Beltran2006c}; W51 North: Zapata et al.,
2008\nocite{Zapata2008}; G31.41+0.31: Girart et al.,
2009\nocite{Girart2009}).  The $\dot M_{\rm inf}/$\macc\ ratio is
as high as 1000 for both $\dot M_{\rm inf}^{\rm \, v_{rot}}$ and 
 $\dot M_{\rm inf}^{\, ff}$.  For the IM protostars, the $\dot M_{\rm inf}^{\,
 ff}/$\macc\ ratio ranges from about $\gtrsim$1
to 450. As seen in Fig.~\ref{fig-minfacc}, there is no correlation
with the mass of the central (proto)star and the result does not seem to depend on 
the method used to estimate $\dot M_{\rm inf}$. This suggests that $\dot M_{\rm inf}$ is
indeed higher than \macc.

A possible explanation for this systematic result is related to the nature of interferometric
observations. As has been mentioned before interferometers could
filter out extended emission. If this extended emission is part of the
molecular outflow then the measured $\dot M_{\rm out}$ is
underestimated as well as $\dot M_{\rm acc}$. However, this observational 
explanation is unlikely,
as \citet{Lopez-Sepulcre2010} obtained a similar result 
for a sample of high-mass clumps which were
observed with a single-dish telescope. 
In particular, these authors found infall rates 2 to 4
orders of magnitude higher than the accretion rates estimated from
outflow mass loss rates.  Another possible explanation
could be that the \macc$/\dot M_{\rm jet}$ ratio is not $\sim$3 
as theoretically proposed
by \citet{Tomisaka1998} and \citet{Shu1999} 
but higher. This implies that the efficiency of the jet/outflow in
removing material from the disk would be lower than expected. 

What we consider as the most likely explanation for the observed fact that $\dot
M_{\rm inf}\gg\dot M_{\rm acc}$ is stellar multiplicity: the infalling material
is not accreted onto the single star that is responsible for the molecular
outflow, but onto a cluster of stars. This explanation seems plausible for the
most massive O-type (proto)stars, because as already explained, the sizes and
masses of the rotating toroids suggest that they are enshrouding stellar
(proto)clusters \citep[see][]{Cesaroni2007}. The fact that the result found by 
\citet{Lopez-Sepulcre2010} at parsec scales, that is, $\dot M_{\rm inf}\gg \dot
M_{\rm acc}$, still holds at much smaller scales suggests that the embedded
clusters would be  very concentrated towards the center of the clump. However,
this explanation cannot solve the problem for the IM protostars and probably
neither for the B-type (proto)stars. As shown in Fig.~\ref{fig-minfacc}, for IM
protostars with $M_\star\simeq$2--3~$M_\odot$, the $\dot M_{\rm inf}/\dot M_{\rm
acc}$ ratio is still 20--300, with $M_{\rm gas}$ of the order of
0.3--1.4~$M_\odot$. Therefore, although these disks could be circumbinary disks,
it seems unlikely that they are circumcluster structures surrounding several
members. What is more, if one takes into account $\dot M_{\rm inf}$ for typical
low-mass Class~0/I protostars like B335 \citep{Zhou1993}, L483
\citep{Tafalla2000}, L1228 \citep{Arce2004}, L1157, BHR~71,  L1527 IRS
\citep{Mottram2013}, L1551 IRS~5 \citep{Chou2014}, L1485 IRS \citep{Yen2014} and
compares it to $\dot M_{\rm acc}$ estimated from the outflow mass loss rate
\citep{Hogerheijde1998,Arce2006,Dunham2014}, the ratio is still $>1$. Actually,
the $\dot M_{\rm inf}/\dot M_{\rm acc}$ ratio is as high as 12 for L1228.

The apparent implication of Fig.~\ref{fig-minfacc} is that infalling material
needs to pile up in the disk and results in disk masses
which are tens to hundreds of solar masses given the observed rates.
This is massive and suggests
a gravitationally unstable disk inducing variable, ``FUOri-like'' accretion
events onto the central object. Comparing this reasoning to observations in
low-mass stars, \citet{Terebey1993} demonstrate from millimeter
observations  that the disk mass during the embedded phase is hardly
different from the one of T Tauri disks. This implies that while the 
infalling envelope is feeding the disk with material, there is no significant
change in mass. The combined mass sinks in the system (the
jet/outflow and the accretion onto the star) are in balance with the
envelope mass infall rate. It seems therefore critical  in IM and HM
(proto)stars that more reliable measurements of mass accretion rates
onto the star are obtained. One such potential \macc\ tracer apt for
young embedded objects is the emission of mid-IR hydrogen recombination
lines \citep[see e.g.][]{Rigliaco2015}.

\subsubsection{Disk accretion rates as a function of stellar mass and time}

\begin{figure}[t]
\centerline{\includegraphics[angle=-90,width=11cm]{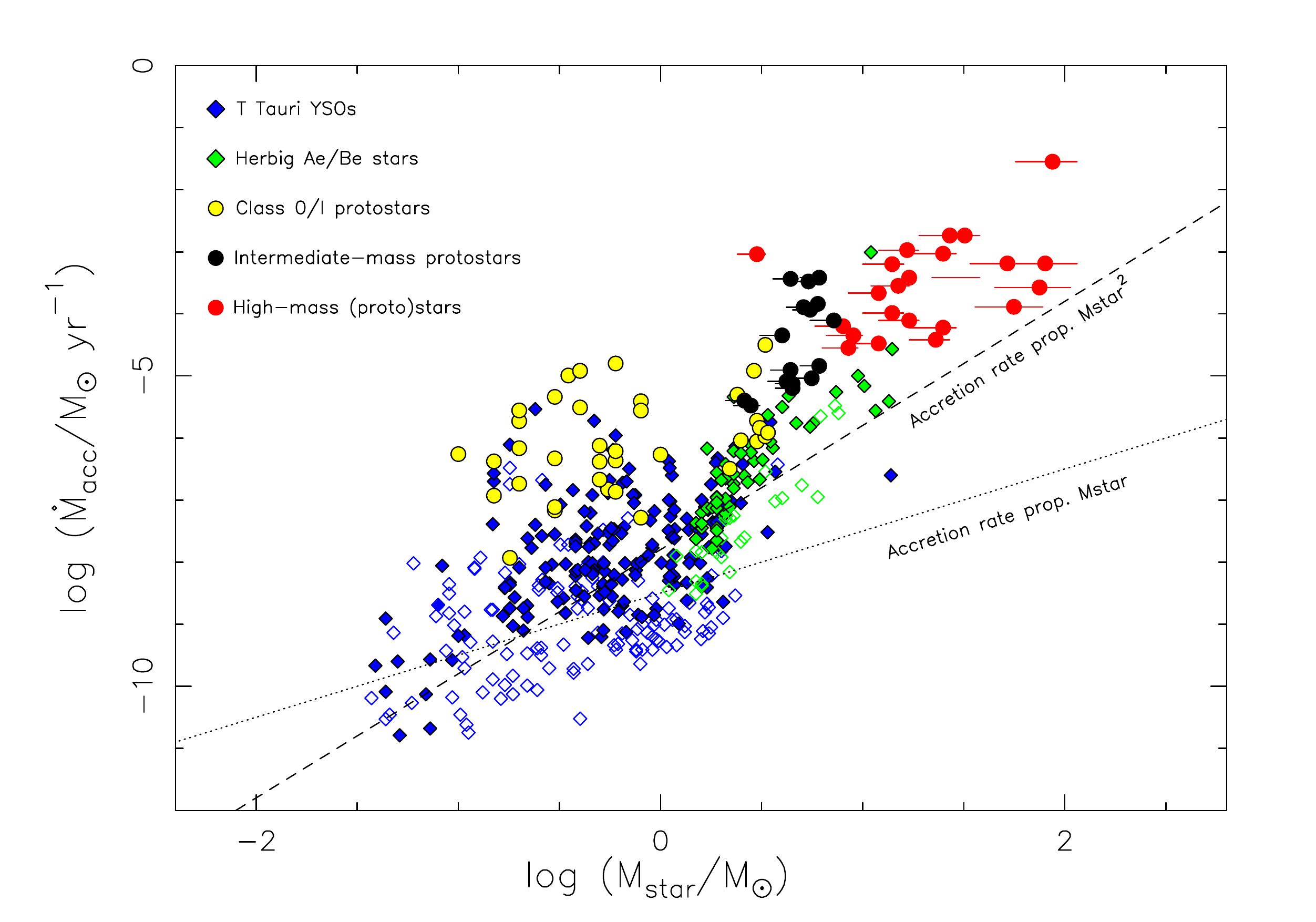}}
\caption{Mass accretion rate $\dot M_{\rm acc}$ as a function of  $M_\star$ for T Tauri
YSOs (blue diamonds), Herbig Ae/Be stars (green diamonds), Class 0/I protostars (yellow
circles), and the IM (black circles) and HM (red circles) (proto)stars 
with available outflow mass loss rate $\dot M_{\rm out}$  (Tables~\ref{IMprop} and \ref{OBprop}). Horizontal black and
red lines indicate the uncertainty in  $M_\star$ for IM and HM (proto)stars,
respectively. Open symbols indicate $\dot M_{\rm acc}$ upper limits. See \S~\ref{evolution-macc} for a
description of the T Tauri, Herbig Ae/Be and Class 0/I samples. }
\label{fig-maccevol}
\end{figure}

The dependence of the accretion rate $\dot M_{\rm acc}$ on the stellar
mass $M_\star$ for low-mass stars has been widely studied through
observational and theoretical works (see Ercolano et al.,
2014\nocite{Ercolano2014} and references therein). These studies
indicate that for low-mass PMS stars $\dot M_{\rm
acc}\propto M_\star^\alpha$, with $\alpha$$\sim$1.5--2.0. The
physical origin of such a steep $\dot M_{\rm acc}$--$M_\star$
correlation is a matter of debate and while some authors propose that
it is the consequence of the initial conditions during disk formation
followed by viscous disk accretion evolution \citep{Alexander2008},
others, on the contrary, suggest that this relationship is the natural
outcome of disk evolution, and in particular, of disk dispersal by
X-ray photoevaporation \citep{Ercolano2014}.

With the data collected in this review, we are in the position to
extend the mass range and to compare the embedded objects
to the optically revealed ones. This is the purpose of
Fig.~\ref{fig-maccevol}. It shows the trend of  
$\dot M_{\rm acc}$ as a function of $M_\star$ 
for different samples of YSOs in different evolutionary phases:
low mass Class~0/I YSOs and T\,Tauri stars, IM protostars 
and Herbig~Ae/Be stars, and HM (proto)stars. The mass accretion rate has been
estimated from the outflow mass-loss rate (Eq.~\ref{eq-mout}) for the most embedded
objects: Class 0/I, IM and HM (proto)stars. The Class 0/I protostars have been
observed by \citet{Yun1994},  \citet{Hogerheijde1998}, \citet{Arce2006},
\citet{Dunham2014}, and Dunham et al.~(in preparation). Like for the IM and HM
(proto)stars of our compiled sample, the mass loss rate has been corrected for
opacity and assuming a mean source inclination of 32.7\degr\ with respect to the
plane of the sky. Following \citet{Hogerheijde1998}, $M_\star$ has been
estimated assuming that all the bolometric luminosity $L_{\rm bol}$ is stellar and
that the object is on the birthline. The pre-main-sequence evolutionary tracks used
to estimate the location of the birthline and $M_\star$ are from
\citet{Stahler2005}.

The sample of T Tauri stars used in Fig.~\ref{fig-maccevol} has been observed
by \citet{Gullbring1998}, \citet{Calvet2004}, \citet{Natta2006}, and
\citet{Antoniucci2014}. \citet{Gullbring1998} estimate $\dot M_{\rm acc}$  for a
sample of T Tauri stars in the Taurus molecular cloud from the excess continuum
emission in the 3200--5200~\AA\ wavelength range. \citet{Calvet2004} estimate $\dot
M_{\rm acc}$ for a sample of IM T Tauri stars, which are a subset of the T Tauri
class with masses $1\,M_\odot \leq M_\star \leq 5\,M_\odot$. The
sources are located in the Taurus-Auriga clouds, 
the Ori OB1c association in the Orion Nebula cluster and the ring around $\lambda$
Ori. The mass accretion rates are estimated from the accretion luminosity $L_{\rm
acc}$, which was obtained from the optical-UV excess, and the stellar parameters.
\citet{Natta2006} estimate $\dot M_{\rm acc}$  for a sample of T Tauri stars in the
$\rho$-Ophiuchi star-forming region from Br$\gamma$ line emission. The accretion and
stellar parameters have been re-calculated by \citet{Manara2013} for a corrected
distance to the region of 125~pc. \citet{Antoniucci2014} estimate $\dot M_{\rm
acc}$  for a sample of T Tauri stars in the Chamaleon, L1641, Serpens, and Lupus
star-forming region from Br$\gamma$ line emission and from the Pa$\beta$ line for
those few sources for which Br$\gamma$ emission was not detected. Finally, the
sample of Herbig Ae/Be stars has been observed by \citet{Fairlamb2015}  and $\dot
M_{\rm acc}$ estimated from the UV Balmer excess.

\begin{figure*}
\centerline{\includegraphics[angle=-90,width=11cm]{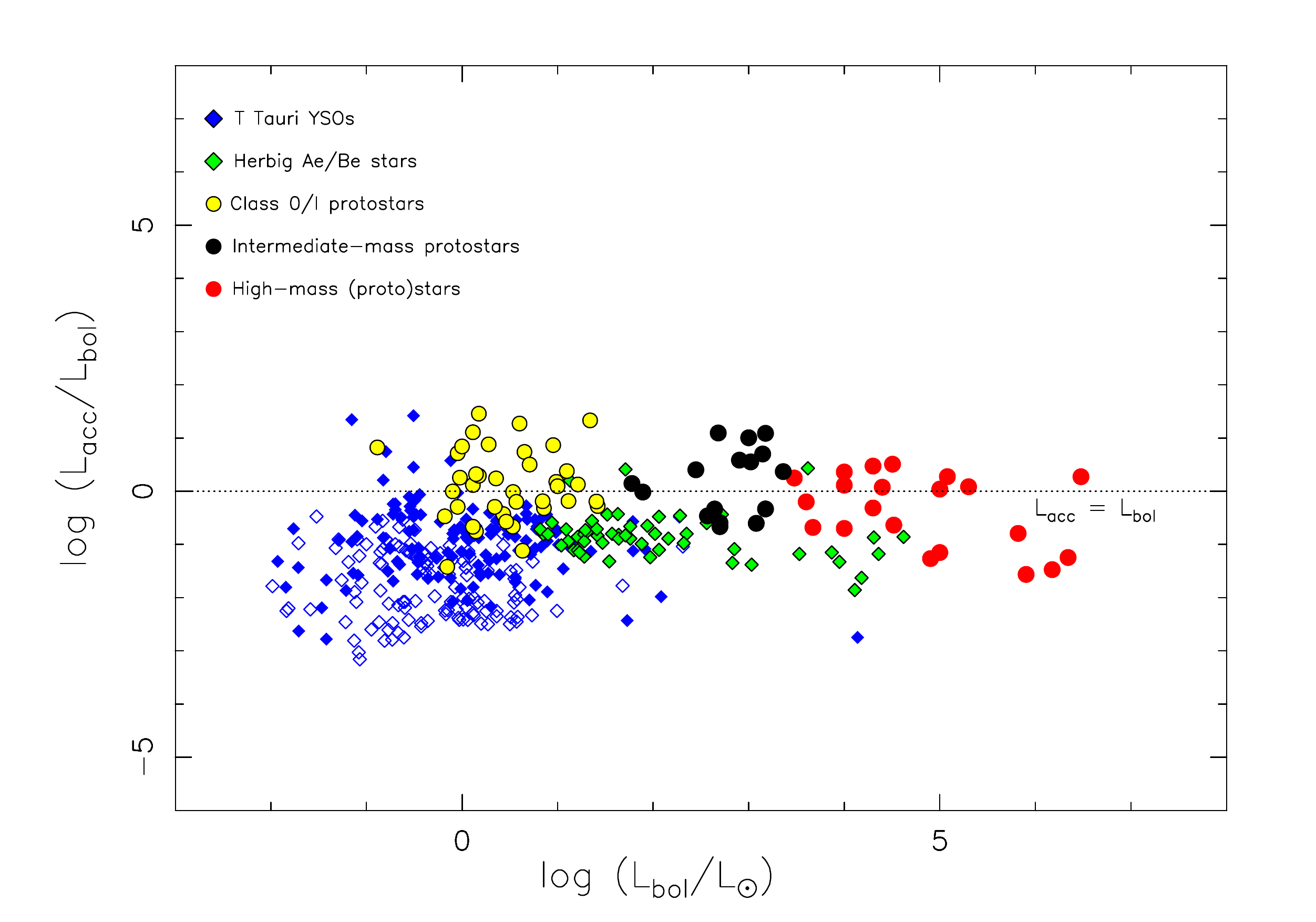}}
\caption{Accretion luminosity $L_{\rm acc}$ to bolometric luminosity $L_{\rm
bol}$ ratio as a function of 
$L_{\rm bol}$  for the
same samples as in Fig.~\ref{fig-maccevol}. Open symbols indicate upper limits.}
\label{fig-laccevol}
\end{figure*}

Figure~\ref{fig-maccevol} shows a general trend of increased \macc\ with larger
stellar masses. To facilitate comparison with previous work we have added a line
showing the relation of \macc\ $\propto M_\star^{2}$. Overall, the
objects follow this relation over the full mass range, but with orders of
magnitude scatter per mass-bin. Figure~\ref{fig-maccevol} shows that
$\dot M_{\rm acc}$ of PMS objects is 1-2 orders of magnitude lower than 
that of embedded YSOs, which suggests a decrease of $\dot M_{\rm
  acc}$ with time. It complies with the expectation that the embedded
objects are in their main accretion phase and accrete material onto the receiving
star at a much higher rate. The {\it average} \macc\ is time dependent and
decreases with age, 
as already established by
studies of disk evolution for T Tauri  \citep[e.g.,][]{Hartmann1998} and HAeBe
\citep[e.g.,][]{Mendigutia2012} stars.  
Figure~\ref{fig-maccevol} also suggests
that both embedded and PMS objects  display a similar dependence of $\dot M_{\rm
acc}$ on $M_\star$ with the relation for the embedded objects offset to higher
$\dot M_{\rm acc}$ values. This inference cannot be tested over the full mass
range as, clearly, no optically revealed high-mass stars with primordial disks
are known to exist and the area for masses $>10~M_{\odot}$ is therefore void in
Fig.~\ref{fig-maccevol}. 

The fact that a relation between \macc\ and $M_\star$  is apparent for
such a broad range of stellar masses (and luminosities) could indicate
that there is continuity in the accretion process from low-mass to
high-mass and that the accreting mechanism is driven by 
similar processes for all luminosities.

By converting \macc\ to $L_{\rm acc}$, we can obtain insight into the
fraction of bolometric luminosity contributed by the accretion process
as function of mass and evolutionary phase. This is presented in
Fig.~\ref{fig-laccevol}, showing the accretion luminosity $L_{\rm
acc}$ as a function of the bolometric luminosity $L_{\rm bol}$ for the
same sample of objects as in Fig.~\ref{fig-maccevol}. $L_{\rm acc}$
was estimated as follows. For the embedded Class~0/I, IM, and HM
(proto)stars, we applied the relation
 $L_{\rm acc}=G\,M_\star\dot M_{\rm acc}/R_\star$, where $R_\star$ is
the stellar radius and \macc\ is estimated from the outflow mass loss
rate. We adopt stellar radii for the HM
(proto)stars under the assumption that the sources are already on the
ZAMS \citep[values taken from the tabulation in][]{Davies2011}. We ignore therefore
the possibility that the objects are swollen up because of high
accretion rates \citep{Hosokawa2010}. The stellar radii of the IM 
protostars are adopted from \citet{Palla1992} for a protostellar mass
accretion rate of $10^{-5}$\msolyr, based on the values estimated 
from the mass loss rate (see Fig.~\ref{fig-maccevol}). For the Class~0/I
protostars, $R_\star$ was estimated assuming that the sources are on
the birthline and following Table~16.1 of
\citet{Stahler2005}, where the estimated values have been computed for a constant
accretion rate of $10^{-5}$~\msolyr. Finally, the bolometric
luminosity of the T\,Tauri and HAeBes stars was assumed to be equal to the
stellar luminosity $L_\star$ tabulated by the different authors. 
The result in
Fig.\,\ref{fig-laccevol} demonstrates that for very few
PMS stars accretion is the dominant source of luminosity. Among the
HAeBe stars there are only a few notable exceptions like FU\,Ori. On the other 
hand, the embedded sources, although scattered around the one-to-one relation, 
have bolometric luminosities comparable to the estimated accretion
luminosities.

\section{Summary and outlook}

\begin{figure*}
\vspace{-1cm}
\centerline{\includegraphics[angle=0,width=12cm]{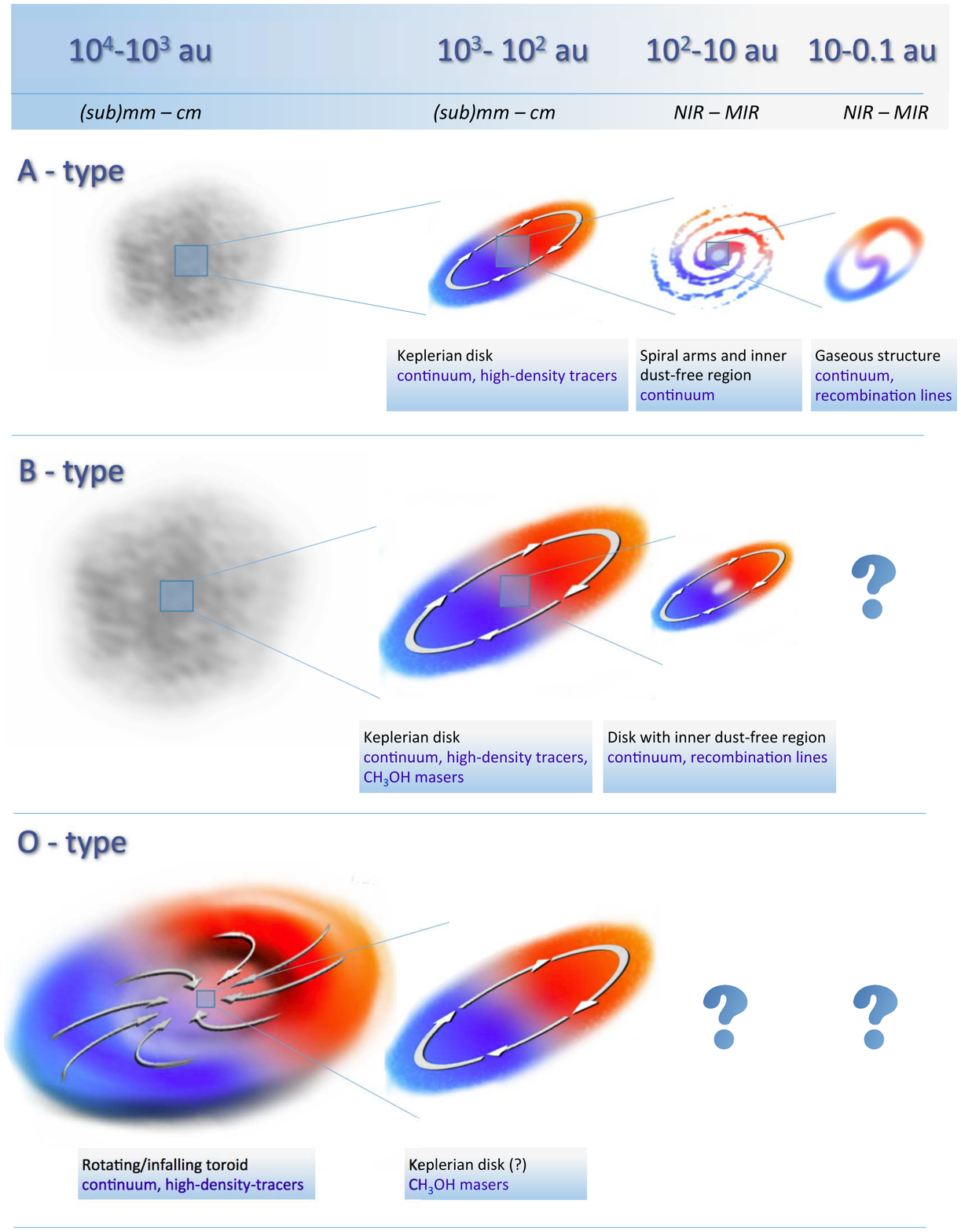}}
\vspace{-.8cm}
\caption{Schematic view of the circumstellar structures around intermediate- (A-type) and
high-mass (B- and O-type) (proto)stars as (mainly) observed by interferometers from near-IR to centimeter
wavelengths. The spatial scales traced by the different observations are
indicated at the top. The tracers used are described in the boxes
below the figures.
The structures observed at spatial scales
of 10$^3$--10$^4$\,au around A- and B-type (proto)stars are simply the surrounding
envelope.}
\label{fig-summary}
\end{figure*}

\subsection{Summary}
We have presented an overview of advancements in the field of
accretion in young high-mass stars based on spatially resolved
observations. Our efforts were aimed at putting into context the 
structures found around these stars by also reviewing the
intermediate-mass objects, both in the embedded as well as in the 
pre-main sequence phase. The focus of the review was to summarize the
basic physical properties of the disks and the toroids alongside the
various methods used in the literature. In Fig.~\ref{fig-summary} we
present a schematical version of our summary showing our knowledge of
accretion disks around intermediate- (A-type) and high-mass (B- and
O-type) young stars as acquired with radio/mm and optical/IR interferometry.

As seen in this schematic, the sizes of the observed circumstellar structures
vary depending on the spectral type of the young star. This is in part  an
effect of the different spatial scales traced by the instruments due to the
different distances of the YSOs. While intermediate-mass A-type young stars are
on average located at distances of $<$1 kpc, high-mass O-type stars are usually
located at  $>$5 kpc. The different sizes are also in part a real effect, in the
sense that higher-mass objects have larger circumstellar structures. 

Radio/mm interferometric observations of embedded IM protostars (A to
late-B spectral type) have revealed circumstellar disks with typical
radii of a few hundreds of au.  Radii of $\lesssim$100~au have been
reported for a few IM protostars observed at an angular resolution
better than $0\farcs4$. These disks are geometrically thick with a
hydrostatic scale height that is more than 20--30\% their radius. The
disk masses turn out to be a few 1\,$M_\odot$ which is less than that 
of the associated central star. These circumstellar disks could
therefore be in Keplerian rotation. Detailed molecular line observations of a few
embedded IM protostars have revealed velocity gradients suggestive of
rotation. However, up to now, the nature of this rotation is
unknown. The fact that Keplerian rotation has been clearly detected
towards the more evolved HAeBe stars, thanks to radio/mm molecular
line interferometric observations, leads us to expect that the disks
around the younger IM protostars are also centrifugally supported. Overall, the circumstellar disks of IM protostars are similar to those
of their lower mass counterparts, at least down to spatial scales of
$\sim$100\,au.

OI observations of the optically revealed IM YSOs, that is the Herbig AeBe stars,
reveal the complexity of the innermost ($<$100\,au) regions of
circumstellar disks around IM stars. Importantly, OI can probe the
structure and nature of the disk interior to the dust sublimation
region. Evidence is building that a large fraction of the emission is
not coming from the region in the disk where a transition takes
  place between gas dominated to dust dominated opacity, i.e. the
  inner rim of the dust disk. Spectroscopic observations are suggesting that the IM PMS
stars are actively accreting objects, possibly through the process of
magneto-spheric accretion. In this sense the IM PMS stars can be quite
similar to the lower mass T\,Tauri stars. Nonetheless, and what is
well known and anticipated, OI has demonstrated the transition in
accretion dynamics that takes place in this mass regime.  The total
disk mass may become much less massive in the B-type stars as a sign
of rapid dispersal, the inner disk ($<10$\,au), nonetheless, may still
be very dense and indeed actively accreting. At least it is clearly
measured that disk erosion takes place at the inner disk. Based on
direct imaging and OI, it is becoming clear that the disk structure of
the IM PMS is varied and there is need to understand better how these
structural differences relate to the disk processes we know.

Evidence for circumstellar disks has been found for HM (proto)stars of early-B
to late-O spectral type, with luminosities up to $\sim$10$^5\,L_\odot$ that
correspond to ZAMS stars of about 25--30\,$M_\odot$. Typical radii of these
disks are a few thousands of au, although radii as small as a few hundreds of au
have been estimated thanks to angularly resolved observations from centimeter to
sub-millimeter wavelengths and to VLBI CH$_3$OH maser emission observations. 
These geometrically thick structures have scale heights of $>$30--40\% of their
radii.  Disk masses range from a few $M_\odot$ to a few tens of $M_\odot$ and in
general are consistent with or smaller than the mass of the central star.
Detailed high-density tracer observations at radio/mm wavelengths have revealed
velocity gradients in these disks consistent with (quasi-)Keplerian rotation,
which has also been suggested by methanol maser emission. These Keplerian disk
candidates are gravitationally stable as suggested by the fact that $M_{\rm
disk}< 0.3 M_\star$ and that Toomre's stability parameter $Q$ is $<1$.
Summarizing, the basic properties of the disks around early-B to
late-O spectral type (proto)stars appear as a scaled-up versions of
those found for disks around low- and intermediate-mass protostars.

Optical/IR interferometry observations of early-B and late-O type
(proto)stars have provided access to the inner regions of their
circumstellar disks by allowing to resolve spatial scales of $\sim$10
to 100\,au. Although this technique applied to embedded sources is
still scratching the surface, we do know now that the disk can persist to a
scale of a couple au. Indeed, an important fraction of NIR and
MIR dust continuum emission comes from the circumstellar
disk and not only from scattered light and/or envelope
emission. Therefore, the near-future NIR and MIR instruments will be
able to provide much better constraints on the disk. 

Early-O type (proto)stars, with luminosities $>$10$^5$\,$L_\odot$, appear to be
surrounded by huge and massive structures called toroids. Their radii of
10$^3$--10$^4$\,au and masses of a few 100\,$M_\odot$ and the fact that their
scale height is $>$50\% of their radius clearly differentiate them from the
circumstellar disks around lower-mass YSOs. Velocity gradients suggestive of
rotation have been observed in these toroids through observations of
high-density tracers from centimeter to (sub)millimeter wavelengths. However,
the fact that usually $M_{\rm toroid}>M_\star$ (in a few cases higher by more
than one order of magnitude) precludes the rotation from being Keplerian.
Toroids are unstable against axisymmetric instabilities as indicated by the fact
that $M_{\rm toroid}> 0.3 M_\star$ and $Q$ is $>1$, and could be susceptible to
gravitational collapse and fragmentation. Instability  against gravitational collapse
is also suggested by the fact that their dynamical mass is $<M_{\rm toroid}$ and
therefore,  toroids cannot be centrifugally supported. Indeed, evidence of
infall has been detected in some of them. Summarizing, toroids around early-O
type (proto)stars are probably pseudo-disks that will never reach equilibrium
and with lifetimes of the order of the free-fall time. Single-epoch VLBI maser
emission  observations of early-O type (proto)stars have revealed CH$_3$OH maser
features in elongated structures of $<$1000\,au in size that could be tracing
circumstellar disks. Only multi-epoch VLBI observations that permit to measure
the proper motions will confirm whether the masers are tracing indeed the
rotation of a disk.

Typical  mass infall rate $\dot M_{\rm inf}$ values for IM and HM YSOs  are of the
order of 10$^{-3}$--10$^{-2}$\,$M_\odot$\,yr$^{-1}$ and can be as high as
0.1\,$M_\odot$\,yr$^{-1}$ for the most massive O-type (proto)stars, while those of
mass accretion rate onto the central (proto)star $\dot M_{\rm acc}$ are  of the order
of 10$^{-4}$--10$^{-3}$\,$M_\odot$\,yr$^{-1}$. Both  $\dot M_{\rm inf}$ and $\dot
M_{\rm acc}$ increase with stellar mass, with $\dot M_{\rm inf}$ usually $> \dot
M_{\rm acc}$ (in a  few cases up to 3 orders of magnitude) for both IM and HM
(proto)stars. The most likely explanation for the discrepancy between infall and
accretion rates for the most massive O-type (proto)stars is that the material
infalling onto the  disk or toroid is not accreted onto a single YSO but a cluster.
On the other hand, for IM and B-type (proto)stars the infalling material could be
piling up in the disk and not being incorporated onto the
central star at the same infalling rate.

The mass accretion rate $\dot M_{\rm acc}$ scales with the stellar mass
$M_\star$ according to a power-law with an exponent close to 2. This dependence of $\dot M_{\rm
acc}$  on the square of $M_\star$ holds for a broad range of luminosities (from
low- to high-mass) and evolutionary stages (from embedded to optically revealed
YSOs). This suggests that there is a continuity in the accretion process from
the low- to the high-mass regime and that the accreting mechanism could be the
same for all luminosities. $\dot M_{\rm acc}$ decreases by 1-2 orders
of magnitude from the embedded phase to the PMS phase, in agreement
with star-formation and disk evolution theory. 

\subsection{Outlook}
New opportunities for revealing the structure and nature of the accretion disks
around IM and HM young stars will present themselves in the near future. Indeed, the
data obtained in the first cycles of the ALMA observatory will significantly advance
our understanding of the initial conditions of star and planet formation and the young star
accretion process. The systematic use of the longest baselines of up to
$\sim$15\,\kms, that will provide maximum angular resolutions of 5--10
milli-arcseconds at the highest frequency bands, and the higher sensitivity achieved
by ALMA will allow us to glimpse into the massive toroids around
O-type young stars and to study the kinematics of the inner regions 
($\lesssim$10\,au) of circumstellar disks around B-type (proto)stars. 
As illustration of the ALMA capabilities, we can mention here the results on the
disk around the low-mass protostar HL\,Tau (see Fig.~\ref{fig-hltau}).
Future
projects will be aimed specifically to achieve an improved view on the exact geometry
of the circumstellar material, the evolution of this geometry and the rates of infall
from the envelope and accretion onto the star.  These aims will lead to a
substantially better understanding of IM and HM star formation and the phenomena
related to it.

As far as OI is concerned, important steps forward in the star-formation arena will be
delivered by the second generation of VLT-I instruments. The two new 
four telescope beam-combiners will be installed and put in operation
in the coming years. \matisse~constitutes an important step forward
in OI as it opens the important mid-IR L- and M-bands. The amount of
young star disks accessible with OI will therefore increase as the
objects are brighter at these wavelengths than at K-band. Importantly, disks at
earlier, embedded evolutionary phases will now be within reach of
OI and image reconstruction techniques. Some results suggest that
the disk emission is strong and possibly dominating the continuum at
$\sim$5\,$\mu$m.  Spectrally resolved OI with \matisse~will also
allow to trace disk kinematics using the line emission of 
\CO. The mid-IR OI data will be supplemented with studies of the
accretion rate. In particular, Humphreys\,$\alpha$ in the atmospheric N-band was recently 
discovered to be an effective accretion tracer in low-mass Class\,0
objects \citep{Rigliaco2015} and holds promise to become a key
accretion tracer for high-mass YSOs. The other second generation VLTI
instrument is Gravity. It operates in K-band and promises to reach
faint ($M_{K}\goa 13^{m}$) objects routinely. The instrument is
equipped with near-IR wavefront sensors, which is of crucial
importance for studies in star-forming regions. The Gravity imaging mode is most
relevant for continuum and line studies of the disks around young
stars.

The 2nd generation \vlti~instruments provide synergy between OI and radio/mm
interferometry. This is especially relevant for the disks of
high-mass stars in their main accretion phase. As the structure of the
disk is directly related to the radial temperature profile, knowledge
of the disk properties necessarily involves observations covering
the full wavelength range. These allow the first characterizations of the
accretion environment around high-mass YSOs from 10 to 1000\,au.

Synergetic studies is what current/near-future facilities allow us to
perform, none\-the\-less the angular resolution may not always be
adequate. It is clear that among the young stars, the most massive
object with a centrifugally supported disk is $\sim 30\,\msol$ so
far. Stepping up in mass, the finest (sub)millimeter observations reveal only
rotating structures at 10,000\,au scales which are {\it not} in
Keplerian rotation. ALMA~will probe scales $\goa 10$\,au and thus not
resolve the crucial star/disk interface. For OI, the current baselines
are not adequate for e.g. resolving stellar surfaces (except for a few
supergiants), and in the context of high-mass star formation this is
something desirable for the optically revealed early-type pre-main
sequence stars. Continuing demands for increased spatial resolution
drive e.g. the Planet Formation Imager concept, and intensity
interferometry at the Cerenkov Telescope Array and set new challenges
beyond 2020. Ideas for innovation are bountiful and are partially
driven by the need to understand the accretion disks around young
stars.


\section*{Acknowledgments}
We would like to thank the following people for their input in
discussion on accretion in young stars: J. Fairlamb, C. Manara,
M. Dunham, H. Arce, T. Alonso-Albi, K. Shiro. We thank L. Carbonaro for his
help on the graphics. We thank T. van Kempen for providing us the NGC 2071 
maps and P. Boley for preparing the figure of V\,921\,Sco. We would
like to thank 
especially D. Galli, R. Cesaroni, and I. Mendigut\'\i a for
reading a preliminary version of the manuscript. M.\ T. Beltr\'an thanks the
ESO Chile Scientific Visitor Programme. This research has
made extensive use of NASA's Astrophysics Data System.

\bibliographystyle{asp2010}
\bibliography{library}   



\end{document}